\documentclass[a4paper,11pt,reqno]{article}
\usepackage{arxiv}

\usepackage[utf8]{inputenc} 
\usepackage[T1]{fontenc}    
\usepackage{hyperref}       
\usepackage{url}            
\usepackage{booktabs}       
\usepackage{amsfonts}       
\usepackage{nicefrac}       
\usepackage{microtype}      
\usepackage{lipsum}
\usepackage{amsmath}
\usepackage{amsfonts}
\usepackage{algorithm}
\usepackage{enumerate}
\usepackage{graphicx}
\usepackage{algpseudocode}
\usepackage{listings}
\usepackage{bm}
\usepackage{color, colortbl}
\usepackage{xcolor}
\usepackage{natbib}

\usepackage[normalem]{ulem} 

\usepackage{adjustbox}
\definecolor{Gray}{gray}{0.8}

\newcommand{\Bgamma}{\boldsymbol{\mathsf{\gamma}}}
\newcommand{\By}{\boldsymbol{Y}}
\newcommand{\Bx}{\boldsymbol{X}}

\usepackage{adjustbox}
\usepackage{array}

\newcolumntype{R}[2]{%
    >{\adjustbox{angle=#1,lap=\width-(#2)}\bgroup}%
    l%
    <{\egroup}%
}

\title{Multilevel regression with poststratification for the national level Viber/Street poll on the 2020 presidential election in Belarus}

\author{
  Ales Zahorski\thanks{This is a collective pseudonym of a group of enthusiasts working on \url{http://narodny-opros.info/} project as well as several external researchers. The group consists of computer and political scientists, sociologists, and mathematicians with backgrounds ranging from BSc to PhD. All-together, the group has 8 members. The author's affiliation mentioned here is fictional. Also, none of the participants is directly affiliated with any kind of political forces. This work is not meant to be manipulative and does not have any political goals. However, the anonymization is vital for the personal security of the participants of the group due to the current political situation in Belarus. Yet, all members of the group hope to be able to deanonymize this piece of research completely when the personal risks vanish.}\\
  Department of Mathematics and Computer Science\\
  Uladzimir Karatkevich National University of Belarus\\
  Miensk,\\
  \texttt{ales.zahorski.1@protonmail.com};
  \texttt{ales.zahorski.2@protonmail.com}\\
  \texttt{ales.zahorski.3@protonmail.com};
  \texttt{ales.zahorski.4@protonmail.com}\\
  \texttt{ales.zahorski.5@protonmail.com};
  \texttt{ales.zahorski.6@protonmail.com}\\
  \texttt{ales.zahorski.7@protonmail.com};
  \texttt{ales.zahorski.8@protonmail.com}\\
}

\begin{document}
\maketitle

\begin{abstract}
Independent  sociological polls are forbidden in Belarus. Online polls performed without sound scientific rigour do not yield representative results. Yet, both inside and outside Belarus it is of great importance to obtain precise estimates of the ratings of all candidates. These ratings could function as reliable proxies for the election's outcomes.  We conduct an independent poll based on the combination of the data collected via Viber and on the streets of Belarus. The Viber and the street data samples consist of almost 45$\,$000 and 1$\,$150 unique observations respectively. Bayesian regressions with poststratification were build to estimate ratings of the candidates and rates of early voting turnout for the population as a whole and within various focus subgroups. We show that both the officially announced results of the election and early voting rates are highly improbable. With a probability of at least 95\%,  Sviatlana Tsikhanouskaya's rating lies between 75\% and 80\%, whereas Aliaksandr Lukashenka’s rating lies between 13\% and 18\% and early voting rate predicted by the method ranges from 9\% to 13\% of those who took part in the election. These results contradict the officially announced outcomes, which are 10.12\%, 80.11\%, and 49.54\% respectively and lie far outside even the 99.9\% credible intervals predicted by our model. The only marginal groups of people where the upper bounds of the 99.9\% credible intervals of the rating of Lukashenka are above 50\% are people older than 60 and uneducated people. For all other marginal subgroups, including rural residents, even the upper bounds of 99.9\% credible intervals for Lukashenka are far below 50\%. The same is true for the population as a whole. Thus, with a probability of at least 99.9\% Lukashenka could not have had enough electoral support to win the 2020 presidential election in Belarus.

\end{abstract}


\section{Introduction}
Analysts often characterize the Belarusian political system as a personalistic autocracy \citep{persdict}. Aliaskandr Lukashenka\footnote{Here and in what follows, we keep to the romanization of personal and geographic names with respect to the phonetics of the Belarusian language \citep{wikibel}. There is ambiguity regarding such a procedure (for example, both Alyaksandr and Aliaksandr can be seen in passports of the citizens of Belarus), however we consistently use the same forms throughout the paper. We emphasize that the romanized names might in some cases not correspond to the transliterated passport names (on which we do not have reliable data) due to this ambiguity.} has served as the country’s president since 1994. In 26 years of his rule, Lukashenka has consolidated and extended his power through a series of referendums. In 1996, he initiated a referendum that gave his decrees the status of law, allowed him to appoint virtually all high-ranking officials instead of the parliament (the chairman of the National Bank and the Constitutional Court included), and provided him with a right to disband the parliament, among other things \citep{nyref, tutref}. In 2004, Lukashenka held another constitutional referendum that resulted in the removal of the two-term limit of serving as president \citep{nyref2}. He has won every election since. The Organisation for Security and Cooperation in Europe (OSCE) did not recognize any of the referendums or elections held since 1994 as free and transparent \citep{osce}.

In principle, presidential elections in Belarus take place in two rounds. If no candidate receives above 50{\%} of the votes in the first round, a second round with the two highest-ranking candidates from the first round is held. This happened only once in the history of Belarus during the  first presidential election in 1994 where Lukashenka won in the second round. There is universal suffrage for all adults above the age of 18 and all votes are weighted equally. It is also important to mention that ballots have an “Against all” option,\footnote{Here, “Against all” means literally expressing explicitly non-support for all of the present candidates.} which is a peculiar leftover of the Soviet electoral system. On a more technical note, the Belarusian electoral law stipulates that the country is divided into about 6000 electoral districts each having a polling station and a Precinct Electoral Committee (PEC) staffing it. The PECs report to a smaller number of District Electoral Committees (DECs) which aggregate their results and pass them to higher-level Committees (city- and regional-level) which in turn report to the Central Electoral Committee (CEC) \citep{electlaw}, the body in charge of elections. Local administrations appoint members of PECs and DECs. The CEC, the body overseeing the elections, is not independent as the president appoints half of the staff of the CEC as well as the chairperson.

 The 2020 presidential campaign proved to be unique for Belarus in many ways: The nonchalant approach of President Aliaksandr Lukashenka to the Covid-19 pandemic caused strong dissatisfaction in the population. At the same time, the actual threat posed by the virus lead to expressions of solidarity in civil society, and many citizens volunteered to help doctors and patients. This voluntary civil engagement in countering the threat of Covid-19 soon led to increased political activity, providing fertile soil for the emergence of new political leaders and eventually presidential candidates \citep{theguardian2020}. These new political leaders did not come from the conventional opposition to the sitting president -- which has been in decline since the repressions after the 2010 presidential election -- and they had no obvious orientation towards Russia or the West. Besides, the new opposition leaders came from different backgrounds and had experience from a wide variety of professional fields. These features of the nascent opposition made them appeal to a broad audience. To mention just a few of the new political leaders, Viktar Babaryka has the image of an experienced manager, who prior to the election was the head of the largest private bank in Belarus. Valery Tsapkala is the founder of the Belarusian analogue of Silicon Valley, a high-tech park populated by the IT community. And Siarhei Tsikhanouski is the author of a very popular political blog about Belarus, "Country for Life". Tsikhanouski appealed to the "ordinary people" by traveling through villages in the regions and talking about local life. These potential candidates quickly formed large initiative groups and were able to efficiently organize and gather the signatures needed to be registered as actual candidates for the presidential election. There were long lines at the gatherings where the signatures were collected, which let people sympathizing with the new opposition meet and discuss. Combined, the factors mentioned above led to a rapid increase in political activity in Belarus, and were also early signs of high support for the potential opposition candidates. 
 
 It is safe to say that the increased political activity, as well as what appeared to be growing support for opposition candidates, did not go unnoticed by the presidential administration. Indirect evidence of this is that all of the three strongest candidates were, in various ways, eliminated from the presidential race. Tsikhanouski and Babaryka were imprisoned for arguably fictional reasons (and were not registered because of this), whilst the CEC alleged that the signatures gathered by Tsapkala were fake, thus preventing him from presenting his candidature  \citep{thewpost2020}.\footnote{One has to submit $100\,000$ signatures to be registered as a candidate.} Among the new political leaders, only Sviatlana Tsikhanouskaya, the wife of Siarhei Tsikhanouski, who was the third most popular candidate according to Internet surveys (see Table \ref{tab:mediapols} for details), managed to register for the presidential election. And largely due to the fact that all three potential candidates had no political history and no former quarrels with each other, their teams could quickly unite around and begin working for a single candidate, namely Sviatlana Tsikhanouskaya. Otherwise, three representatives from the conventional opposition were registered: Siarhei Cherachen, Andrei Dmitriyeu and Hanna Kanapatskaya. None of them had any visible support among the population of the country according to the media polls (see Table~\ref{tab:mediapols}). From the early stages of the 2020 presidential campaign, it was clear that the fairness of the election would be in question. Observers from The Organization for Security and Cooperation in Europe (OSCE) were not granted admission to Belarus, there also were non-admission of independent members to the election commissions,\footnote{In 2020, only 0.009{\%} of PEC{'}s members belonged to the opposition \citep{tutref2}.} and, in fact, non-admission of any independent observers whatsoever — pointing to the possibility of a rigged election. According to the official results, the incumbent Lukashenka won the election by an overwhelming margin \citep{dw2020}. As should be evident, one has ample reason to distrust the official results. The aim of this paper, therefore, is to provide our best estimates of what the results of the $2020$ Belarusian presidential election {\it would have} been, had the election been free and fair.

Independent sociology on political subjects is completely banned in Belarus. For a long time, some sociological studies were conducted illegally, but in 2016 after a fall in Aliaksandr Lukashenka's approval rating, the last independent service - Independent Institute of Socio-Economic and Political Studies (IISEPS) - was suppressed. Performing sociological studies from abroad is practically impossible due to state control over IP telephone service. This makes telephone surveys extremely expensive and de facto impossible for non-state actors. At the same time, online polls performed by the media (which were as of June 1, 2020 also forbidden) can not be trusted, as they lack sound scientific rigour.

The polls that were conducted by the Belarusian media prior to the June 1st ban on polls gave rise to the satiric nickname {``}Sasha-3\%{''} to the incumbent president. {``}Sasha{''} is a common diminutive for Aliaksandr and the $3{\%}$ refers to what was commonly believed to be an upper bound on Lukashenka{'}s rating. The satiric nickname was actively promoted by the opposition during the campaign. We believe, however, that such a low rating cannot be trusted. The pro-governmental polling firm Ecoom announced strikingly different polling results from those reported by the media. The results of media polls are summarized in Table~\ref{tab:mediapols}, while the Ecoom polls are presented in Table~\ref{tab:ecoom}. The absence of independent polling institutes and faced with extremely contradictory results coming from different sources provided the impetus for the current study. We came up with an initiative to carry out a national poll, and based on these data, used the state-of-the-art statistical methodology to estimate the popularity of each candidate. What the election results in Belarus would have been, had they been free and fair, we will never know. With this study, it is our sincere aim to provide a politically unbiased account of what the presidential election results in a counter-factual world -- a Belarus with free and fair elections -- would likely have been.

\begin{table}[!htb]
 \caption{Ratings of different candidates in polls conducted by the Belarusian media in May 2020, and by the messenger bots of the Voice platform (in both Viber and Telegram) and of the official channel of Telegram in Belarus (in Telegram) in August 2020. Here, "*" marks denied, withdrawn or declined candidates. Bold numbers indicate the candidate who took first place in the corresponding poll. Italics - the candidate who took second place. For the early polls, Tsikhanouskaya also includes results for Tsikhanouski since the former was registered as the candidate but it was not clear who of the couple was the actual candidate to be supported.}
  \centering
  \addtolength{\tabcolsep}{-0.2pt}
  \small
  \begin{tabular}{lrrrrrrr}
    \toprule
    Poll&tut.by&nn.by&telegraf.by&onliner.by&svaboda.org&Voice&Telegram\\\hline
Date&20-May-20&20-May-20&22-May-20&26-May-20&26-May-20&20-Aug-20&03-Sep-20\\\hline
Participants&70$\,$029&16$\,$526&35$\,$775&32$\,$128&9$\,$705&1$\,$047$\,$933&1$\,$411$\,$035\\\hline
Babaryka*&\textbf{54.91\%}&\textbf{49.7\%}&\textbf{52.43\%}&\textbf{56\%}&\textit{32\%}&—&-\\
Haidukevich*&1.28\%&0.65\%&3.93\%&0.4\%&—&-&-\\
Hantsevich*&0.34\%&—&0.62\%&—&0\%&-&-\\
Hubarevich*&0.45\%&—&0.23\%&—&0\%&-&-\\
Dmitriyeu&0.58\%&0.32\%&0.33\%&0.29\%&0\%&0.66\%&0.86\%\\
Kanapatskaya&4.14\%&2.5\%&1.7\%&1.4\%&1\%&0.25\%&0.39\%\\
Kisel*&0.63\%&—&0.22\%&—&0\%&-&-\\
Kavalkova*&0.37\%&—&0.21\%&—&0\%&-&-\\
Kazlou*&0.64\%&—&0.61\%&0.32\%&1\%&-&-\\
Lukashenka&6.24\%&2.96\%&3.90\%&3\%&1\%&1\%&\textit{6.39\%}\\
Niapomnyashchykh*&0.4\%&—&0.4\%&—&0\%&-&-\\
Tabolich*&1.37\%&2.3\%&0.25\%&0.9\%&2\%&-&-\\
Tsikhanouskaya&12.71\%&\textit{18.39\%}&\textit{17.37\%}&12\%&\textbf{51\%}&\textbf{95.44\%}&\textbf{85.32\%}\\
Tsapkala*&\textit{15.22\%}&7.5\%&9.8\%&\textit{17\%}&7\%&-&-\\
Cherachen&0.72\%&0.28\%&0.16\%&—&0\%&0.35\%&0.91\%\\
Against all/other&—&15.37\%&7.8\%&7.7\%&4\%&\textit{2.28\%}&6.12\%\\\hline
    \bottomrule
  \end{tabular}
  \label{tab:mediapols}
\end{table}

Our initiative, called {\it Narodny Opros} (National Poll in English, see also our website \url{https://narodny-opros.info/}) is an independent project for estimating pre-election ratings of all the registered candidates and early voting turnout rates of the 2020 presidential election in Belarus. Early turnout rates are extremely important to be addressed as independent observers from the previous elections claimed that the majority of falsifications were done during early voting \citep{falsify2019}.

 \begin{table}[!htb]
 \caption{Ratings of different candidates obtained by Ecoom polling firm in July 2020. The polls held 10–14 July-20 and 23–27 July-20 collected data from 1807 and 1879 respondents correspondingly. Telephone calling was used for sampling but the detail on the methodology is missing. 
Bold numbers indicate the candidate who took first place in the corresponding poll. Italics - the candidate who took second place and "*" marks denied, withdrawn or declined candidates. AA column corresponds to the vote {``}Against all{''} and EV to the positive about early voting responses.}
  \centering
    \addtolength{\tabcolsep}{-2pt}
      \small
  \begin{tabular}{rrrrrrrrrr}
    \toprule
Lukashenka&Tsikhanouskaya&Kanapatskaya&Cherachen&Dmitriyeu&Babaryka*&Tsapkala*&AA&EV\\\hline
\textbf{73.3\%}&\textit{7.5\%}&1.6\%&0.2\%&0.7\%&N/A&N/A&3,7\%&61,4\%\\
\textbf{69.4\%}&2.2\%&1.5\%&0.8\%&0.3\%&\textit{6.7\%}&3.1\%&3,5\%&N/A
\\
    \bottomrule
  \end{tabular}
  \label{tab:ecoom}
\end{table}
We employed two different methods for polling: (1) An online poll using the Viber messenger smartphone/PC application where Viber users responded to a questionnaire; and (2) several street polls taking place at different locations across the country where random people were stopped on the street by one of our data gatherers and questioned. The questionnaires were almost identical in both cases, containing questions about what candidate the respondents intended to vote for, as well as questions about socio-economic and demographic status. However, on the streets, we additionally asked what kind of media communication the respondent uses. The reason for choosing Viber as the vehicle for our digital polling is that Viber is the most popular messenger application in Belarus \citep{belMedia2020}, and is used by citizens from most social strata~\citep{messenger2019}, holding different political views. Of course, a poll conducted via Viber will only reach people owning a smartphone/PC using this particular messaging app, and we are well aware that such people might not be representative of the population as a whole. This is a potential source of bias that we try to adjust for in the statistical analyses. The street polls were conducted to cover a wider array of social groups and the street respondents were also asked whether or not they were Viber users, thus providing data about how representative the Viber users might be. For the street poll, we surveyed groups in different localities ranging from Minsk (the capital of  Belarus) to small villages in the regions. We used the official 2009 census data from Belstat (the official statistics agency of the Republic of Belarus) to calculate the representative size of the statistical group for each category to be surveyed. We then merged the data from the Viber and street polls and used Bayesian regression with poststratification methods to correct for potential biases in our data. 

In short, the methodology we employ involves building a statistical model that attempts to atone for the fact that our survey respondents are not representative of the population as a whole. By properly weighting the predictions of our regression model, we generalise from the sample to the entire population.  The procedure is called multilevel regression with poststratification (MRP). This methodology has previously proven to give precise results in numerous empirical applications in sociology \citep{gelman2016using, park2004bayesian, claassen2018improving}, political science \citep{kiewiet2018predicting, hanretty2019introduction}, epidemiology \citep{zhang2014multilevel, downes2018multilevel, loux2019using}, and marketing \citep{chen2010bayesian}. Among the cited empirical applications, the perhaps most impressive one is the study by \citet{wang2015forecasting}. In that study, the authors collected data from Xbox users that, perhaps not surprisingly, were extremely biased with respect to sex, age, and level of education. After applying statistical methods similar to the ones we use in this paper, the authors obtained estimates of the outcome of the 2012 US presidential election, that, ex-post, turned out to be on the mark. According to \citet{wikimrp2020}, similar statistical methods were also able to successfully predict outcomes of the 2016 US presidential election \citep{trangucci2018voting, wyatt2016estimating}, and the 2017 general election in the UK \citep{johnston2018}. 

The rest of the paper is organized as follows: In Section \ref{sec:modeling}, we provide a full mathematical background for the paper. In particular, we describe the suggested hierarchical Bayesian regression, provide model selection and evaluation criteria, and explain how poststratification works. Further, in Section \ref{sec:data}, we describe how the data was collected, processed and merged. Section \ref{sec:inference} explains the algorithm used to obtain posterior distributions of interest for the parameters of the suggested model. Section \ref{sec:results} describes the main findings of the paper, whilst Section \ref{sec:discuss} proceeds with the conclusions, general discussions of the results in the context of other studies, and gives directions for future research. Finally, Appendix \ref{sec:dataappendix} provides further detail on data collection and a back-end of the IT solution behind it and Appendix \ref{sec:resultsappendix} includes additional results that did not appear in the main part of the article. 

\section{Statistical modelling}\label{sec:modeling}
In this section, we first provide the mathematical details on a multilevel regression model and describe all the prior we use for the unknown parameters. Subsequently, we give an overview of the model selection criteria we employ to assess the fit of our model. Finally, we describe the poststratification procedure we use to adjust for demographic biases in our sample.
\subsection{The multilevel regression model}
Each respondent in our dataset indicated one, and only one, of the alternatives Lukashenka, Tsikhanouskaya, Kanapatskaya, Cherachen, Dmitriyeu, or {``}Against all{''}, and also whether or not they planned to participate in early voting. We marginally model the binary responses of the events of interest, which are indicators of votes for a given candidate (Lukashenka, Tsikhanouskaya, Kanapatskaya, Cherachen, Dmitriyeu, {``}Against all{''}) or indicators of the plans to participate in the early voting, $Y_i \in \{0,1\}, i \in \{1,...,M\}$ to be Bernoulli distributed with the corresponding probability of success $p_i \in \mathbb{R}_{[0,1]}$. The first model has $Y_i = 1$ if the $i$th respondent intended to vote for Lukashenka, and $Y_i = 0$ otherwise; the second one has $Y_i = 1$ if the $i$th respondent intended to vote for Tsikhanouskaya, and $Y_i = 0$ otherwise; with no loss of generality, the same idea is employed to define the responses in the remaining five models. We used the same structure of the logistic model for the probability of success $p_i$ in each of the seven models.\footnote{Strictly speaking, the fact that we use marginal and independent Bernoulli regressions for the events of iterest does not impose any constraint on the probabilities of success to sum up to one across the events, although in practise they are very close to it.} Also, after prepossessing and cleaning the data, the sample size $M = 32108$ was the same for each of the models we fit. Thus, in each model the binary outcomes $Y_1,\ldots,Y_m$ are assumed conditionally independent given the covariates and latent Gaussian structures:
\begin{align}
  \Pr(Y_i = y|  p_i) = & \text{Bernoulli}(p_i), \quad i \in \{1,...,M\},\label{eq:themodel} \\
  \text{logit}(p_i) = & \beta_0 + \sum_{j=1}^{p} \beta_{j}X_{ij} + \delta_{{\rm reg}_i},\label{eq:themodel2}
\end{align}
where $p_i = p_i(x) = 1/\{1 + \exp(-x)\}$, while $\text{logit}(u) = \log(u/(1 - u))$ is the logistic link function, $X_{i1},...,X_{ip}$ are binary 
covariates and $\beta_0,\ldots,\beta_p$ and $\delta_{{\rm reg}_i}$ are random effects of the model. The  covariates ${\bm X_i} = \{X_{i1},...,X_{ip}\}$ are indicators of the various social subgroups and categories the respondents belong to: \textit{gender} (male, female), \textit{region} (Minsk city, Minsk, Mahiliou, Homel, Hrodna, Vitsebsk and Brest regions), \textit{age} (18-30, 31-40, 41-50, 51-60, more than 60), \textit{education} level (elementary school or uneducated, primary or secondary school, professional technical institution, professional college, higher education), and the type of \textit{area of residence} (rural, urban). Thus, we had in total $p = 21$ covariates in our study. The random effects corresponding to the ordinal categorical predictors (age, education) are assumed to have a latent $\textit{AR}_1$ structure between the categories, whilst other factors as well as the intercept term have an $\textit{i.i.d.}$ latent structure. Additionally, a latent Gaussian Besag-York-Mollie ($\textit{BYM}_2$) field \citep{morris2019bayesian, riebler2016intuitive} is included into the model as $\delta_{{\rm reg}}$ in order to account for the spatial dependence of the probabilities between the regions \footnote{Here, Minsk is included into the Minsk region as it is geographically located inside it} and the variance which is neither explained by the covariates nor by the common latent factors included into the random intercept $\beta_0$. Such structural random effects are shown to reduce the bias of the estimates of the model \footnote{We have also compared the model against several alternative models without some or all of the structural effects. The selection criteria are described in Subsection \ref{sec:criteria}, whilst the procedure and empirical results are described in Subsection \ref{sec:modselectres}. This configuration was found optimal in terms of the overall agreement of the addressed criteria.} in a recent study by \citet{gao2020improving}.  
To complete the model specification for \eqref{eq:themodel}, we put the following priors for the parameters of the model:
\begin{align*}
\beta_{j_{{\rm age}_l}} =& \rho_{{\rm age}}\beta_{j_{{\rm age}_{l-1}}} + \varepsilon_{{\rm age}_l},\quad l \in \{1,...,5\},\\ 
\beta_{j_{{\rm edu}_l}} =& \rho_{{\rm edu}}\beta_{j_{{\rm edu}_{l-1}}} + \varepsilon_{{\rm edu}_l},\quad l \in \{1,...,5\},\\ 
\varepsilon_{{\rm age}_l} \sim & \mathcal{N}(0,\tau^{-1}_{{{\rm age}}}),\quad \varepsilon_{{\rm edu}_l}\sim  \mathcal{N}(0,\tau^{-1}_{{{\rm edu}}}),\\
 \tfrac{\tau_{{\rm age}}^{-1}}{(1-\rho_{{\rm age}}^2)}\sim&\text{PC-Prior}(1,0.1),\quad \tfrac{\tau_{{\rm edu}}^{-1}}{(1-\rho_{{\rm edu}}^2)}\sim\text{PC-Prior}(1,0.1),\\
\log\tfrac{1+\rho_{{\rm age}}}{1-\rho_{{\rm age}}}\sim&\mathcal{N}(0,0.15^{-1}),\quad \log\tfrac{1+\rho_{{\rm edu}}}{1-\rho_{{\rm edu}}}\sim\mathcal{N}(0,0.15^{-1}),\\
\beta_j \sim & \mathcal{N}(0,\tau^{-1}_{{\rm  pr}}),\quad
\tau_{{\rm pr}}^{-1} \sim  \text{PC-Prior}(1,0.1), \quad j \in J_{\rm  iid}.\\
\intertext{Here, $J_{\rm  iid} = \{0,...,p\} \setminus \{J_{{\rm age}} \cup J_{{\rm edu}}\}$ are the indeces of the independent random effects and  $J_{{\rm age}} = \{j_{{\rm age}_1},...,j_{{\rm age}_5}\}, J_{\rm edu}  = \{j_{{\rm edu}_1},...,j_{{\rm edu}_5}\}$ are the indices of the effects corresponding to the ordinal predictors \textit{age} and \textit{education}. These priors are available out of the box in INLA \citep{rue2009inla, lindgren2015bayesian}, where the details can be found. Finally, for the the $\textit{BYM}_2$ process we follow the prior spatial structure of regional effects $\{\delta_j\}_{1\leq j\leq 6}$ from \citet{gao2020improving} as specified below:}
 \delta_j = & \frac{1}{\sqrt{\tau_{\rm reg}}} (\omega_j\sqrt{1 - \rho_{\rm reg}}  + \phi_j\sqrt{{\rho_{\rm reg}}} ), \\
    \phi_j \sim& \mathcal{N}({d_j}^{-1}{\sum_{k \sim A_j} \phi_{k}}, {d_j}^{-1} ),\quad j \in \{{\rm reg}_1,...,{\rm reg}_6\}, \\
     \tau_{{\rm reg}}^{-1} \sim & \text{PC-Prior}(1,0.1),\quad \boldsymbol\omega \sim \mathcal{N}_6(0, I_{6}).
\end{align*}
Here, $A_j$ is the set of first-degree neighbours for \textit{region} $j$ (e.g. Hrodna region's first degree neighbours $A_j$ are Minsk, Brest, and Vitsebsk regions) and $d_j$ is the size of set $A_j$ (e.g.~$d_j$ for Hrodna region is $3$). Finally, $I_{6}$ is the identity matrix with 6 dimensions. Further, the marginal precision is $\tau_{\rm reg}$ and the proportion of the marginal variance explained by the spatial effect ($\omega_j$) is $\rho_{\rm reg}$, the latter has the default specification from INLA \citep{rue2009inla, lindgren2015bayesian}.

\subsection{Model evaluation criteria}\label{sec:criteria}

The suggested model can be justified by means of comparing to some competitive model configurations. The set of criteria to be addressed are described in this subsection. They are marginal log likelihood (MLIK), Watanabe–Akaike information criterion (WAIC) and the modified quantile Brier score (MBRIER), which is introduced in this article to evaluate the predictive uncertainty.

For data $\{\By,\Bx\}$ and a model, which includes some unknown parameters $\bm\theta$, the log marginal likelihood is given by
\begin{equation}
\text{MLIK} = \log p(\By|\Bx,\Bgamma)=\log \int_{\Omega}p(\By|\Bx,\bm\theta)p(\theta)d\bm\theta\label{mlikeq}.
\end{equation}
Here, $p(\bm\theta)$ is the prior for $\bm\theta$ while $p(\By|\Bx,\bm\theta)$ is the likelihood function conditional on $\bm\theta$ and covariates $\Bx$. Marginalization in Equation \eqref{mlikeq} is an alternative to penalties on the efficient number of parameters. In particular, greater dimensions of $\boldsymbol\theta$ correspond not only to greater likelihoods, but  typically also to smaller prior probabilities $p(\boldsymbol\theta)$. 

WAIC in turn is defined as follows:
\begin{equation*}
\text{WAIC} = -\mathbb{E}\left\{\log p(\widetilde{\By}|\widetilde{\Bx},\By,\Bx)\right\}.  \label{wai0}
\end{equation*}
Here, $\{\By,\Bx\}$ and $\{\widetilde{\By},\widetilde{\Bx}\}$ are the old and new data sets, respectively. Thus, WAIC is averaging over the posterior distribution of the parameters, which is relevant for prediction purposes, as WAIC is essentially evaluating the predictions based on the unobserved new data in a Bayesian context. Hence, WAIC is a fully Bayesian approach for estimating the out of sample expectation. This gives a computationally convenient approximation to cross-validation.

The modified quantile based Brier score is defined through an expectation with respect to the sampling distribution of the unobserved data, namely: 
\begin{equation*}
\text{MBRIER}^{\alpha} = \mathbb{E}\left\{\tfrac{1}{1-\left(\hat{q}( Y|\widetilde{X})_{\alpha}-\hat{q}( Y|\widetilde{X})_{1-\alpha}\right)^{2}}\times{\min}^2\left\{\left|\widetilde{Y}-\hat{q}( Y|\widetilde{X})_{\alpha}\right|,\left|\widetilde{Y}-\hat {q}(Y|\widetilde{X})_{1-\alpha}\right|\right\}\right\}.
\end{equation*}
Here, $\alpha$ and $1-\alpha$ are the quantile levels of the posterior predictive distribution of the data. Further, the expression ${\min}^2\{|\widetilde{Y}-\hat{q}( Y|\widetilde{X})_{\alpha}|,|\widetilde{Y}-\hat {q}(Y|\widetilde{X})_{1-\alpha}|\}$ measures the distance of the new observation $\widetilde{Y}$ to the nearest predicted by the model quantile ($\{\alpha\}$'s or $\{1-\alpha\}$'s) of the posterior distribution of $Y$ given the newly observed $\widetilde{X}$, i.e. $\hat{q}( Y|\widetilde{X})_{\alpha}$ and $\hat{q}( Y|\widetilde{X})_{1-\alpha}$ respectively. Lastly, $1/({1-(\hat{q}( Y|\widetilde{X})_{\alpha}-\hat{q}( Y|\widetilde{X})_{1-\alpha})^{2}})$ penalizes unreasonably broad credible intervals, in such a way that the intervals that are simultaneously close to the boundaries on both sides are being heavily up-weighted.

\subsection{Poststratification to the whole population}

Poststratification is the method of generalization to the whole population of the estimates obtained from the Bayesian regression model. The method essentially relies on a weighted average of estimates from all possible combinations of factors (sex, age, education, region, type of area of residence).   Each such combination is referred to as a "cell". In our application, we have $P = 700$ cells. Our Bayesian regression model is used to smooth noisy estimates in the cells with too little data by means of using overall or nearby averages. Following \citet{gao2020improving}, with the use of cell sizes obtained from the national population census, $N_j$ of each cell $j$, we poststratify the predictions of the model to get posterior probabilities of the focus parameter at the subpopulation of interest level or at the level of the whole population. Hence, the method fixes issues of non-response in the population by accounting for the size of a given cell $j$ relative to the total population size $N= \sum_{j=1}^J N_j$. Another interpretation of poststratification is understanding it as the weighted average of posterior views on the cell level, where the weights are defined by the size of each cell in the population. Thus, smaller cells get down-weighted and larger - up-weighted. The more observations of each unique cell we have the more reliable our results will be. Additionally, poststratification allows us to estimate the views within a specific subpopulation based on a poll taken across a wider area. Such subpopulations are also allowed to include relatively few respondents from the subpopulation in question, or even to have highly unrepresentative samples from these subpopulations. On the population level, the poststratified $\Theta_{\rm pop}$ parameter of interest becomes:
\begin{align}
    \Theta_{\rm pop} &:= \frac{\sum_{j = 1}^{P}N_j {\Theta}_j}{\sum_{j = 1}^{P} N_j}.
\end{align}
For the subpopulation  $S$, the poststratified parameter is:
\begin{align}
   \Theta_{\rm S} &:= \frac{\sum_{j\in \rm S}N_j {\Theta}_j}{\sum_{j \in \rm S} N_j}.\label{eq:postsub}
\end{align}

For example, $\rm S$ could correspond to the youngest age category among female individuals without higher education. $\Theta_{\rm S}$ then corresponds to the proportion of people in this subpopulation that would, for example, support a given candidate. 


\section{Data collection and filtration}\label{sec:data}

In this section, we describe how the data for our study was collected and processed. Firstly, a set of questions was prepared. Most importantly, we included the questions we could use to further poststratify the estimates from the model using the official statistics from Belstat~\citep{eduCensus2009} (age, education, region, gender, type of area of residence). We further added questions of common research interest. There were two additional questions in both Viber and street surveys about the family's total monthly income and whether the respondent is willing to participate in early voting. There was also a question that was only asked in the street survey. In this question, we wondered about the means of communication and information sources of the respondent. See the full list of questions in Table \ref{tab:qlist} in Appendix~\ref{sec:dataappendix}.

\begin{figure}[!htb]
  \centering
  \includegraphics[scale=0.3]{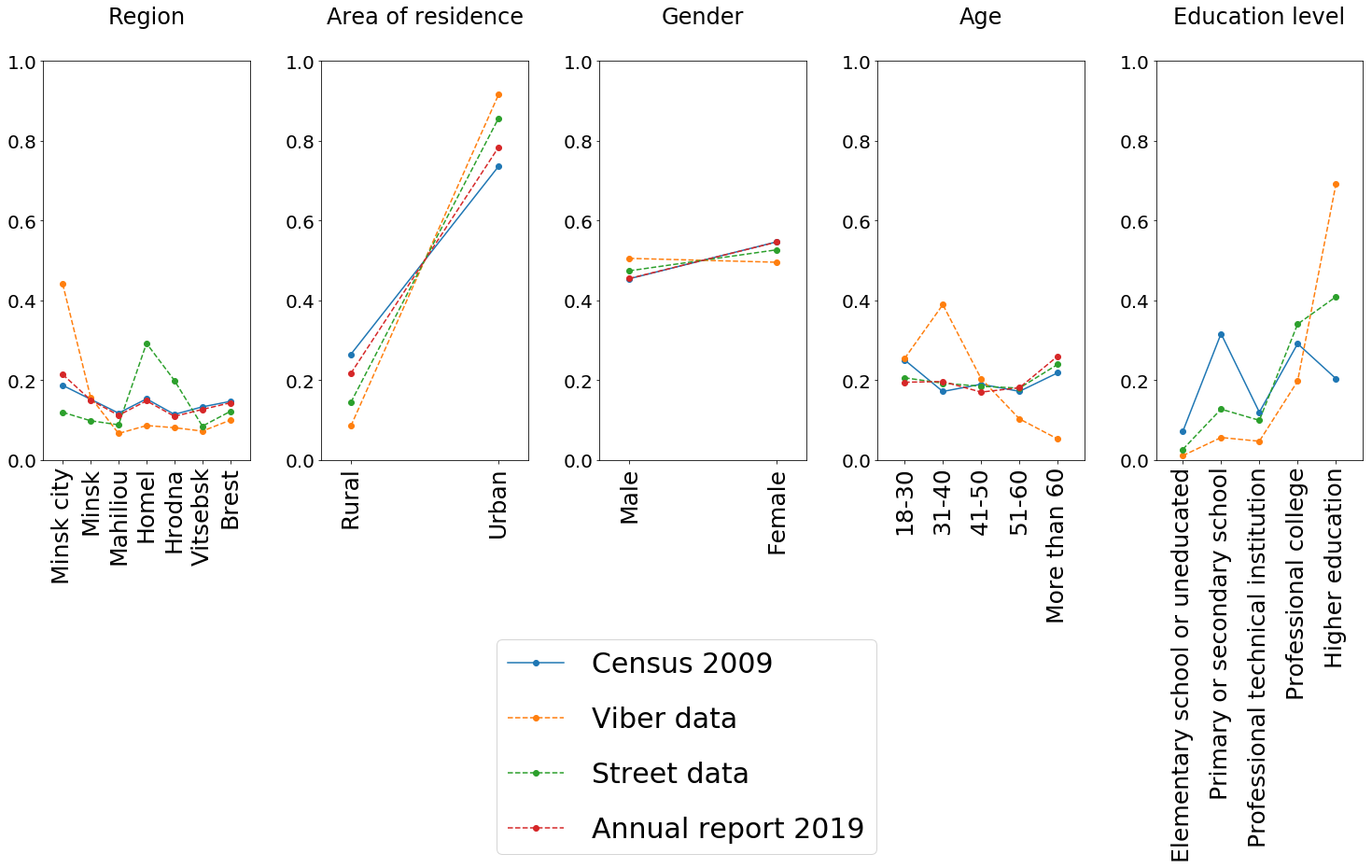}
  \caption{Biases in the Viber and street samples with respect to the addressed demographic variables.}
  \label{fig:biases}
\end{figure}

In our study, Viber was picked as the most popular messenger app in Belarus \citep{belMedia2020}. Yet, the Viber poll can only cover users of this particular app, which is likely to cause a bias towards a certain audience. Therefore, a street poll was also conducted. Properly held street polls allow us to cover all categories of citizens. This is expected to result in significantly smaller biases in terms of demographic characteristics in the street polls. Details on the distributions of the social characteristics of the respondents from these two samples, the 2009 census, and annual report-2019 are provided in Figure \ref{fig:biases}. There, we clearly see strong biases in the Viber sample and somewhat smaller biases in the data gathered on the street. In particular, the Viber data has an extremely over-represented number of participants from Minsk city, the number of participants in the age category from 31 to 40, and participants with higher education. On the one hand, the street data is almost unbiased with respect to the distributions of categories of gender and age. But it still has somewhat a strong bias with respect to the type of area of residence (though slightly smaller than the one from the Viber sample). Also, people from Homel region and people with higher education are over-represented in the street sample. The latter means that even the street data should not be used directly as an unbiased sample from the population and needs proper statistical adjustments.  

We collected all the responses without any filtering on the application level. The response rate in Viber survey was as follows: 82.35\% for those who answered all the questions and 84.68\% for those who answered at least the first five questions (see Table \ref{tab:qlist}).  This means it was required to filter and preprocess the data before modeling. When it comes to the Viber data, we first filtered out the following groups of respondents: Non-Belarusian citizens; Citizens who live outside of Belarus; Participants with a non-Belarusian country code in Viber; Underaged (younger than 18 years old) persons; People who joined the Viber bot after a certain date and time (6 of August 2020 19:12 UTC).
Only meaningful features  that could be used as covariates for statistical modeling were kept at this stage. These features included age, gender, type of the area of residence, region and education level. Additionally, two response variables of interest, namely support of a candidate indicators and an early voting participation indicator, were kept in the dataset here. As we continuously updated our survey right after the release, we collected a small amount of incorrect data in the beginning. This data was corrected if possible or removed otherwise. Initially, we asked people to indicate the age in the ranges "18-25", "26-40", "41-55", "55-70" (should have been "56-70") and "more than 70". So we corrected values "18-25" to "18-30", "more than 70" to "more than 60" and dropped "26-40", "55-70", "41-55" age ranges as they could not be covered by the correct values. We replaced settlements types with residence types (rural for Agro-town / Village, urban - otherwise) as this grouping is widely used in sociological studies and the national census in Belarus, so we could use the corresponding official census data for the purpose of poststratification.
Finally, we omitted all observations with missing values in the covariates, even though, in principle, we could use them in the model by through introducing NA as the additional category for all of the covariates. All preprocessing scripts were implemented in Python and are publicly available on the GitHub page of the project \citep{github}.

For the street poll, we aimed at collecting at least 500 responses to cover all possible categories of citizens with respect to gender, age, region, and type of area of residence. This, as believed, could improve the representativeness of the collected sample. We found volunteers in all regions of the republic of Belarus. Each volunteer was interviewed on the phone. During these interviews, we gave the most vital instructions and tried to understand if the volunteer was able to do this job in general. The details on the given instructions are described in Appendix \ref{sec:volinstrappendix}. As a result of the interviewing process, 40 volunteers were selected to conduct the street poll. We used the official annual report for 2019 from Belstat\citep{belstatDemogr2020} to calculate the representative size of the statistical group for each category surveyed assuming that we needed 500 responses. Volunteers did not have information about other responses and which quotas were already fulfilled; therefore, we checked the coverage of statistical groups at the end of each day and provided volunteers with instructions which categories of citizens needed to be surveyed at a higher rate.
As a result, we collected 1124 responses. There, almost all quotas were fulfilled and some of the groups were even over-represented. The final distribution of the respondents can be found in Appendix \ref{sec:streetquotas}.

The census data was obtained from Belstat in the following fashion: The most recent available data with all combinations of our explanatory variables were available only from the 2009 census held by \citet{eduCensus2009}. The data consists of 7 PDF files, one for each region. In each file, there is a joint distribution of age, gender, area of residence, education level over all combinations of our explanatory variables.  To prepare these data (see \citep{github}) we used Small PDF - an online converter from PDF to Excel \citep{smallPDF}. After this, we summarized the data by the categories of the National Poll, i.e.~age ranges (18-30, 31-40, 41-50, 51-60, more than 60), education level (combined primary and secondary  school levels and elementary school and uneducated levels), using formulas in Excel, and exported this to a .csv file with the following headers: "region", "location\_type", "gender", "age", "education", "count". Hence, we obtained all cells for poststratification. A more recent census was held by Belstat in Belarus in 2019, however the joint distributions of the categories of interest are not yet published. At the same time, marginals with respect to all categories except the education levels are already available from \citet{belstatDemogr2020}. We show them in Figure \ref{fig:biases}. As one can see, the marginal distributions of sub-populations by regions, types of area of residence, genders, and ages did not change significantly, yet there seem to be slightly more elderly people and a bit higher level of urbanization.  The marginal distribution of the education levels for 2019 is not yet available. 

\begin{figure}[!htb]
  \centering
  \includegraphics[scale=0.45]{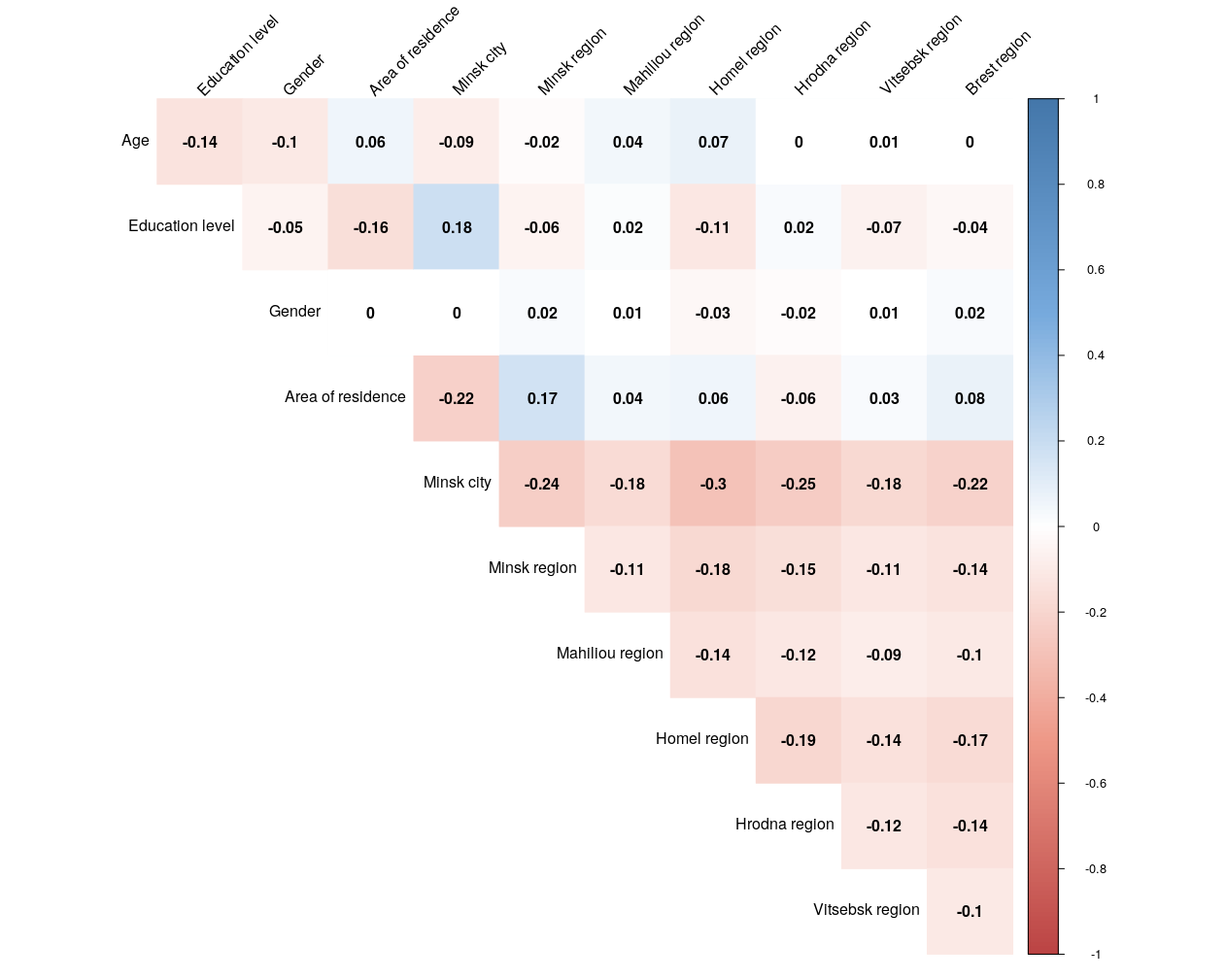}
  \caption{Correlations between categorical covariates in the merged sample (as regions do not have a natural ordering, we provide results separately for all unique regions).}
  \label{fig:corplot}
\end{figure}

After preprocessing the data from the Viber and street polls, we joined the two samples as follows: The filtered Viber sample was randomly divided into two parts consisting of 50\% of the data each. One of these parts was kept as a holdout set for testing the model, whilst the other one was merged with the street sample into a training set, where the street data was uniformly upsampled to the size of 50\% of the whole Viber sample. By means of doing this kind of preprocessing, we equalize the importance of the street and Viber data in the training set, whilst keeping approximately the same amount of information as in the Viber poll data. This kind of equalization is an important stage as it allows to debias the training sample due to the significantly lesser biasedness of the data from the street survey. Additionally, this allowed us to obtain a relatively heterogeneous holdout set that we used to test predictive uncertainty handling by our model. Correlations between demographic variables in the merged set are depicted in Figure \ref{fig:corplot}. There, we clearly see an absence of strong multicollinearities but observe several interesting patterns: Older people tend to live in rural areas and are less educated. People living in urban areas and especially in Minsk are more educated. There are more women among older people. All these preliminary observed patterns make generally sense in the context of the demographics of the country. This gives evidence that we have enough heterogeneity in the merged sample. The scripts used for merging the data are implemented in R as a part of the statistical modelling pipeline. These scripts are also freely available on the GitHub page of the project \citep{github}. Further details on the addressed polling methods and technical solutions including technical pipeline details, methods of distribution of the poll in the population, and other aspects are given in Appendix \ref{sec:dataappendix}.

\begin{figure}[!htb]
  \centering
  \includegraphics[scale=0.4]{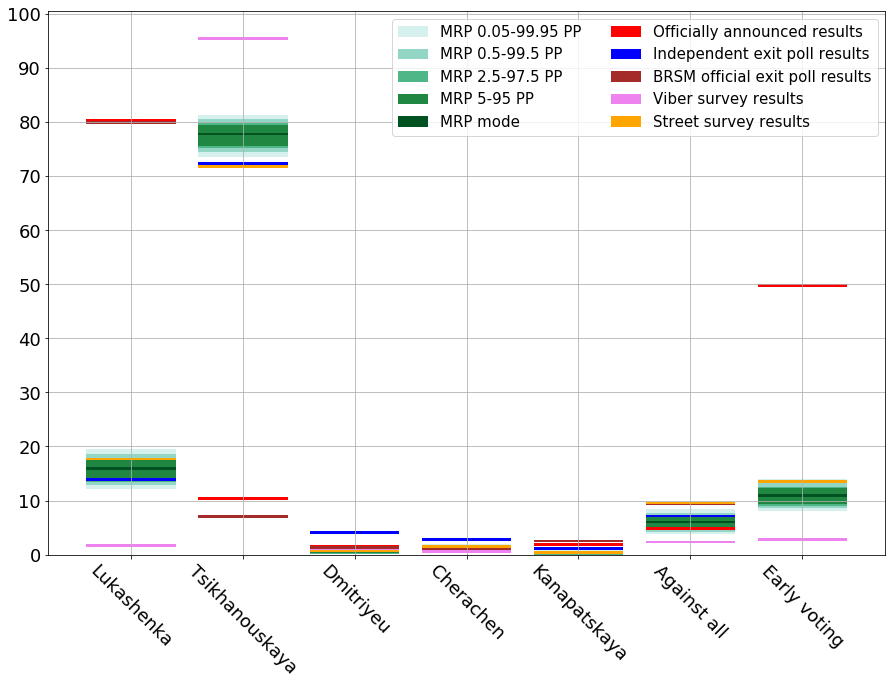}
  \caption{Posterior modes and quantiles for the parameters of interest on the population level. The officially announced results are drawn in red. The independent exit poll results are drawn in blue. PP here means posterior probability. Viber and Street survey results correspond to the sample means of the corresponding polls.}
  \label{fig:ratings}
\end{figure}

\section{Bayesian inference}\label{sec:inference}

For the suggested in Equations \eqref{eq:themodel}-\eqref{eq:themodel2} MRP, we used integrated nested Laplace approximations (INLA) for the inference on the posterior distributions \citep{rue2009inla} of interest. In this section, we briefly describe the INLA approach. Define $\bm \gamma$ as the set of all structured latent Gaussian variables ($\textit{AR}_1, \textit{BYM}_2$ processes) and the $\textit{i.i.d.}$ random effects $\bm\beta$, also define $\bm\lambda$ to include the remaining hyperparameters (a low-dimensional vector).
The INLA approach is based on two steps.  In the first step, the marginal posterior of the hyperparameters is approximated by
\begin{align*}
 p(\boldsymbol\lambda|\By,\Bx) \propto \frac{p({\boldsymbol \gamma},\boldsymbol\lambda,\By,\Bx)}{ p(\boldsymbol \gamma|\boldsymbol\lambda,\By,\Bx)} = 
\frac{p({\boldsymbol \gamma},\boldsymbol\lambda,\By,\Bx)}{ \tilde{p}_G(\boldsymbol \gamma|\boldsymbol\lambda,\By,\Bx)} \Big{|}_{{\boldsymbol \gamma} = {\boldsymbol \gamma}^*(\boldsymbol\lambda)} + \mathcal{O}(M^{-{3}/{2}})\,,
\end{align*}
where  $\tilde{p}_G(\boldsymbol \gamma|\boldsymbol\lambda,\By,\Bx)$ is the Gaussian approximation to $p(\boldsymbol \gamma|\boldsymbol\lambda,\By,\Bx)$, and $\boldsymbol \gamma^*(\boldsymbol\lambda)$ is the mode of the distribution $p({\boldsymbol \gamma}|\boldsymbol\lambda,\By,\Bx)$. The posterior mode of the hyperparameters is found by maximizing the corresponding Laplace approximation employing some gradient descent method. For example, the Newton-Raphson algorithm can be used at this stage. Then, an area with a high posterior density of the hyperparameters is explored with either a grid-based procedure or variational inference \citep{rue2009inla}.  The second step approximates the latent variables for all sets of the explored hyperparameters. At this stage, the computational complexity of the approximation heavily depends on the likelihood type for the data. In case of the Gaussian likelihood, the posterior of the latent variables is Gaussian and the approximation becomes fully tractable. Otherwise, a Gaussian approximation of the latent variables might be inaccurate and a Laplace approximation should be used instead:
\begin{align*}
\tilde p_{\text{LA}}(\gamma_i|\boldsymbol\lambda, {\By,\Bx})  \propto \frac{p({\boldsymbol \gamma},\boldsymbol\lambda, {\By,\Bx})}{\tilde p_{\text{GG}}({\boldsymbol \gamma}_{-i}|\gamma_i,\boldsymbol\lambda,{\By,\Bx})}\Big{|}_{{\boldsymbol \gamma}_{-i} = {\boldsymbol \gamma}^*_{-i}(\gamma_i,\boldsymbol\lambda)},
\end{align*}
where $\tilde p_{\text{GG}}$ is the Gaussian approximation to $p({\boldsymbol \gamma}_{-i}|\gamma_i,\boldsymbol\lambda,{\By,\Bx})$ and ${\boldsymbol \gamma}^*_{-i}(\gamma_i,\boldsymbol\lambda)$ is its posterior mode.\footnote{The error rate of the full Laplace approximation is $\mathcal{O}(M^{-{3}/{2}})$. The full Laplace approximation of the latent fields is rather time-consuming, hence more crude lower order Laplace approximations are often used instead \citep{tierney1986accurate}. This increases the error rate up to $\mathcal{O}(M^{-1})$.} After the posterior distribution of the latent variables given the hyperparameters is approximated, the uncertainty in the hyperparameters is integrated out using numerical integration and the law of total probability \citep{rue2009inla}:
\begin{align*}
\tilde p(\gamma_i|{\By,\Bx} ) = \sum_k \tilde p_{\text{LA}}(\gamma_i|\boldsymbol\lambda_k, {\By,\Bx}) \tilde p(\boldsymbol\lambda_k|{\By,\Bx} ) \Delta_k.
\end{align*}

\begin{figure}[!htb]
  \centering
  \includegraphics[scale=0.38]{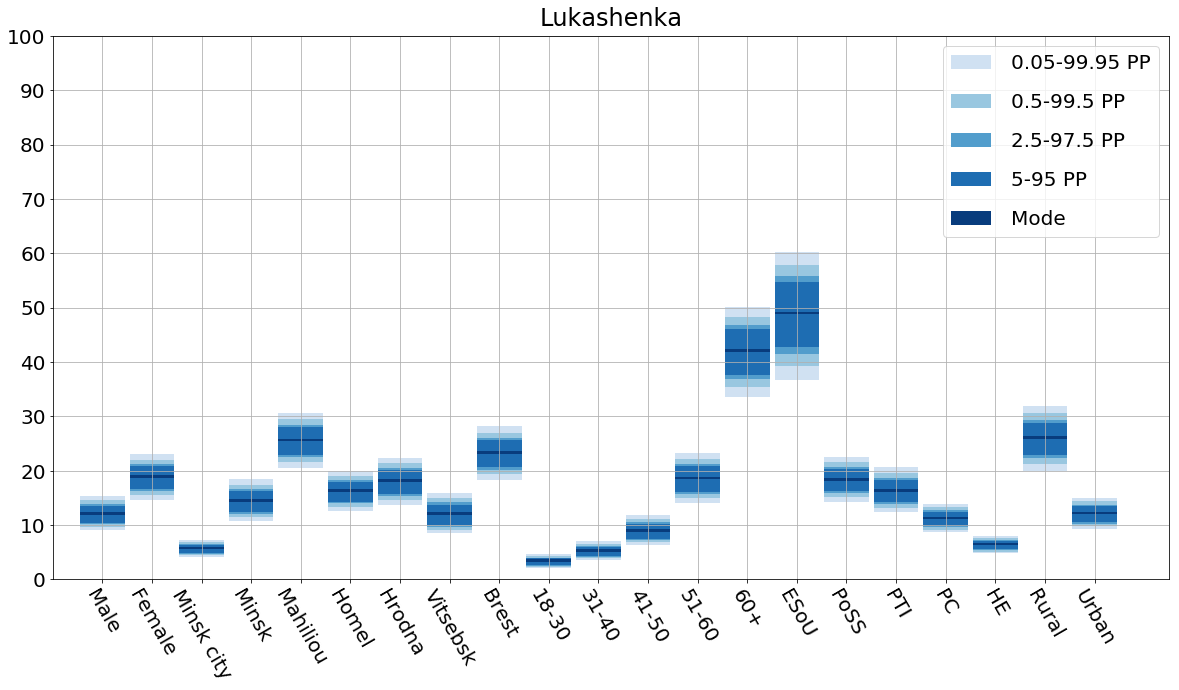}
  \includegraphics[scale=0.38]{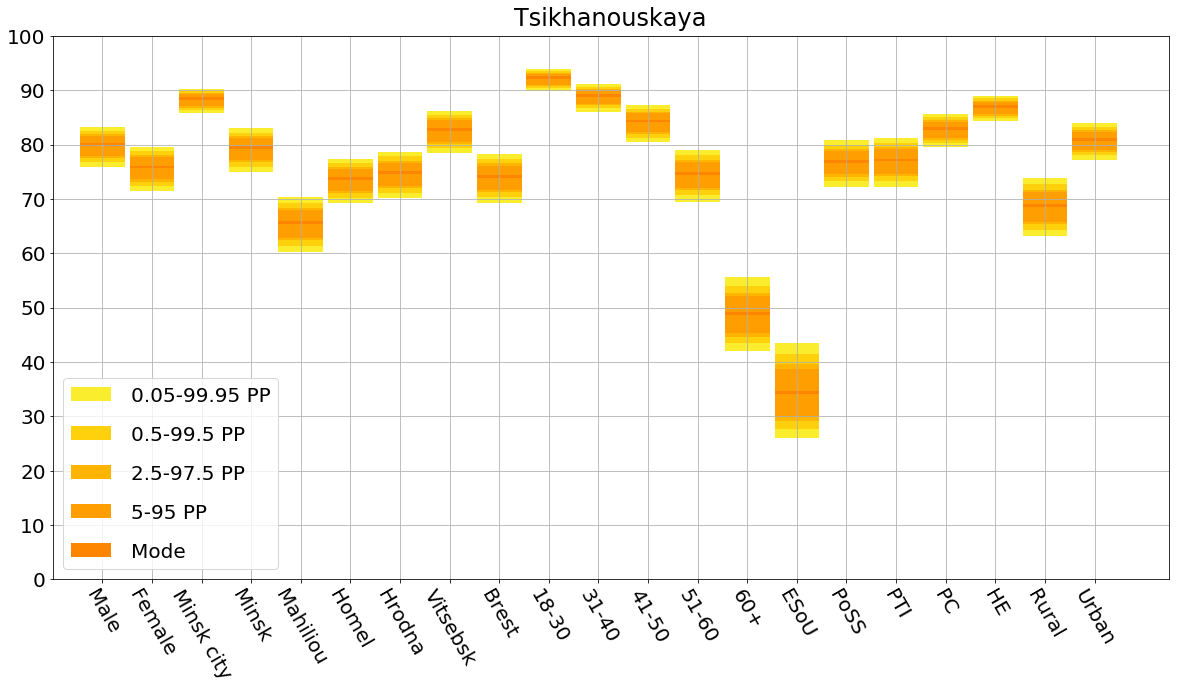}
  \caption{Posterior modes and quantiles for the votes in support of Lukashenka (top) and Tsikhanouskaya (bottom)  across all marginal subpopulations. Here and in what follows, 60+ means older than 60, ESoU - elementary school or uneducated, PTI - professional technical institution education, PC - professional college education, HE - higher education.}
  \label{fig:ratings_lukatiha}
\end{figure}

INLA approximates the marginal likelihoods by
\begin{equation*}
\widetilde{p}(\By|\Bx) = \int_{\boldsymbol \gamma} \left. \frac{p(\By,\boldsymbol \gamma,\boldsymbol{\lambda}|\Bx)}{\tilde{\pi}_G(\boldsymbol{\lambda}|\By,\Bx,\boldsymbol \gamma)} \right \rvert_{\boldsymbol{\lambda} = \boldsymbol{\lambda}^*(\boldsymbol \gamma)} d\boldsymbol \gamma.
\end{equation*}
Here, $\boldsymbol{\lambda}^*(\boldsymbol \gamma)$ is the posterior mode, while $\tilde{\pi}_G(\boldsymbol{\lambda}|\By,\Bx,\boldsymbol \gamma)$ is a Gaussian approximation to $\pi(\boldsymbol{\lambda}|\By,\Bx,\boldsymbol \gamma)$. The integration of $\bm \gamma$ over its support is done by an empirical Bayes (EB) approximation or using numerical integration based on a central composite design (CCD) or a grid. Note that WAIC is computed very similarly within the INLA framework ~\citep[see][for further detail]{rue2009inla}. Finally, the $\text{MBRIER}^\alpha$ is estimated as the sample mean of the scores on realizations of the data from the holdout set, namely:

\begin{equation*}
\widetilde{\text{MBRIER}}^{\alpha} = \tfrac{1}{M_{\text{test}}} \sum_{i=1}^{M_{\text{test}}}{\tfrac{1}{1-\left(\hat{q}( Y_i|\widetilde{X}_i)_{\alpha}-\hat{q}( Y_i|\widetilde{X}_i)_{1-\alpha}\right)^{2}}\times{\min}^2\left\{\left|\widetilde{Y}_i-\hat{q}( Y_i|\widetilde{X}_i)_{\alpha}\right|,\left|\widetilde{Y}_i-\hat {q}(Y_i|\widetilde{X}_i)_{1-\alpha}\right|\right\}},
\end{equation*}
where $M_{\text{test}}$ is the number of observations in the addressed holdout set $\{\boldsymbol{\widetilde{Y},\widetilde{X}}\}$.

\section{Results}\label{sec:results}

\subsection{Model selection and validation}\label{sec:modselectres}
In this section, we provide the results of our study. To begin with, we provide a comparison of the competing models with respect to WAIC, MLIK, and $\text{MBRIER}^{0.975}$ criteria. The two alternative \footnote{Here, (I) is our main model.} model formulations addressed are: (II) The model with $\textit{i.i.d.}$ effects for age and education level instead of the $\textit{AR}_1$ structured effects, and (III) The model with $\textit{i.i.d.}$ effects for age and education level instead of the $\textit{AR}_1$ structured effects and no $\textit{BYM}_2$ spatial latent field. As can be seen in Table \ref{tab:mliks}, model (I) outperforms other candidates almost uniformly in terms of MLIKs (Bayes factors), it is also the best model in terms of WAIC in the majority of the cases. When it comes to the $\text{MBRIER}^{\alpha}$ score at the level $\alpha = 0.975$, models (II) and (III) perform slightly better than the chosen model. However, the results are extremely low (indicating very good predictive uncertainty handling by all of the methods) anyway and the differences in the scores can be (subjectively) seen as almost negligible. Even though we have decided to choose model (I) as the primary model (since the two classical Bayesian model selection criteria choose it), to proceed in this section, we do not drop the predictions of the two other models and report them in Tables \ref{tab:other_models_preds1} and \ref{tab:other_models_preds2} in Appendix B to the paper.  From these tables, one can see that the differences between the predictions of all the compared models are insignificant.

\begin{table}[!htb]
 \caption{Marginal log-likelihoods of the chosen and alternative models. Here, index (I) corresponds to the addressed model, index (II) corresponds to the model with no $\textit{AR}_1$ structural effects for ordinal variables, and index (III) corresponds to the model with no structural $\textit{AR}_1$ effects for ordinal variables and no spatial $\textit{BYM}_2$ process. Clearly, model (I) addressed in this paper is almost uniformly the best model both with respect to MLIK (Bayes factors) and the majority of the responses according to WAIC, whilst the $\text{MBRIER}^{0.975}$ scores are almost identical for the three compared model configurations and are extremely low for all cases.}
  \centering
  \addtolength{\tabcolsep}{-4pt}
  \small
  \begin{tabular}{lrrrrrrrrr}
    \toprule
&$\text{MLIK}_\text{(I)}$&$\text{WAIC}_\text{(I)}$&$\text{MBRIER}^{.975}_\text{(I)}$&$\text{MLIK}_\text{(II)}$&$\text{WAIC}_\text{(II)}$&$\text{MBRIER}^{.975}_\text{(II)}$&$\text{MLIK}_\text{(III)}$&$\text{WAIC}_\text{(III)}$&$\text{MBRIER}^{.975}_\text{(III)}$\\\hline
Lukashenka&\textbf{-6085.51}&\textbf{12102.26}&{1.35E-05}
&{-6086.31}&{12102.41}&\textbf{1.34E-05}&-6091.76&12102.37&\textbf{1.34E-05}\\
Tsikhanouskaya&\textbf{-9999.35}&{19934.09}&6.99E-04&-10000.50&\textbf{19934.02}&{6.98E-04}&-10005.77&19934.06&\textbf{6.97E-04}\\
Dmitriyeu&\textbf{-857.55}&1688.77&4.45E-08&-857.68&\textbf{1688.10}&\textbf{3.34E-08}&-865.15&{1688.43}&{3.35E-08}\\
Cherachen&{-1180.19}&2313.95&1.78E-09&\textbf{-1178.37}&\textbf{2313.04}&{1.43E-09}&-1183.59&{2313.05}&\textbf{1.30E-09}\\
Kanapatskaya&\textbf{-300.24}&\textbf{569.34}&1.24E-11&-300.30&{569.35}&1.39E-11&-306.04&569.71&\textbf{1.21E-11}\\
Against all&\textbf{-5386.98}&\textbf{10716.79}&1.60E-05&-5388.52&10717.04&\textbf{1.55E-05}&-5393.63&10717.15&\textbf{1.55E-05}\\
Early voting&\textbf{-7890.02}&\textbf{15719.36}&\textbf{8.89E-05}&-7891.72&15719.45&{8.93E-05}&-7895.95&15719.43&\textbf{8.89E-05}\\
    \bottomrule
  \end{tabular}
  \label{tab:mliks}
\end{table}

\subsection{National poll results}
We now discuss the overall results of the predicted ratings provided by MRP. The overall national level rankings are provided in Table \ref{tab:ratings}, with posterior means, modes, medians, and multiple quantiles of interest (covering 90\%, 95\%, 99\% and 99.9\% credible intervals). The rankings are also depicted graphically in Figure \ref{fig:ratings}, where the same quantiles and the posterior modes are given. Moreover, in Table \ref{tab:ratings}, we  present the officially announced results, as reported by the CEC. In Table \ref{tab:ratings}, we also indicate whether (and if so on which credible level) the results reported by the CEC can be seen as anomalies with respect to the posterior distributions of the corresponding parameters predicted by our model. The official results are also depicted with red lines in Figure \ref{fig:ratings}. We clearly see that when it comes to the support for Lukashenka or Tsikhanouskaya, the official results are outside of the $99.9{\%}$ credible intervals provided by MRP. The official number for the early election turnout is also well outside our $99.9{\%}$ credible interval. In fact, the officially announced early election turnout is more than $35$ percentage points above the upper bound of the $99.9{\%}$ credible interval provided by our model. This finding is in line with the estimates reported by \citet{schpilkin2020}. There it is claimed that at least one million of the early votes are falsified, a number that, if we are to trust the turn-out rate reported by the Belarusian CEC, makes up at least $20{\%}$ of all the votes cast in the election. Also, even the upper bound of the $99.9{\%}$ credible interval of the support of Lukashenka predicted by MRP is about $61$ percentage points away from his official election results announced by the Belarusian CEC. On the other hand, for Tsikhanouskaya, the officially announced result is about $63$ percentage points below the lower bound of the $99.9{\%}$ credible interval of her rating predicted by our model. In short, judged by our data and our model, the official results are extremely unlikely.

\begin{table}[!htb]
 \caption{Official election results and posterior statistics of interest for the poststratified distribution on the population level. Here, $g_x$ means $x$'s quantile of the focus posterior distribution. For the election results, the notation ${}^{***}$ means outside the 99.9\% credible interval, ${}^{**}$ means outside the 99\% credible interval, ${}^{*}$ means outside the 95\% credible interval, and ${}^{.}$ means outside the 90\% credible interval  of the parameter with respect to its posterior distribution. The anomalies in the official results found on either of the addressed credible levels are bold.}
  \centering
    \addtolength{\tabcolsep}{-3pt}
  \small
  \begin{tabular}{llrrrrrrrrrrr}
    \toprule
Event&Official&Mean&Mode&Median&$q_{0.0500}$&$q_{0.9500}$&$q_{0.0250}$&$q_{0.9750}$&$q_{0.0050}$&$q_{0.9950}$&$q_{0.0005}$&$q_{0.9995}$\\\hline
Lukashenka&\textbf{0.8010}${}^{***}$&0.1560&0.1562&0.1561&0.1379&0.1752&0.1346&0.1790&0.1282&0.1865&0.1209&0.1953\\
Tsikhanouskaya&\textbf{0.1012}${}^{***}$&0.7753&0.7752&0.7753&0.7555&0.7943&0.7516&0.7979&0.7440&0.8048&0.7351&0.8127\\
Dmitriyeu&\textbf{0.0120}${}^{**}$&0.0042&0.0044&0.0043&0.0022&0.0078&0.0019&0.0087&0.0015&0.0108&0.0011&0.0138\\
Cherachen&0.0114&0.0077&0.0079&0.0078&0.0047&0.0127&0.0042&0.0140&0.0035&0.0167&0.0027&0.0204\\
Kanapatskaya&\textbf{0.0167}${}^{***}$&0.0011&0.0013&0.0012&0.0004&0.0033&0.0003&0.0040&0.0002&0.0058&0.0001&0.0090\\
Against  all&\textbf{0.0459}${}^{.}$&0.0578&0.0579&0.0578&0.0473&0.0701&0.0455&0.0726&0.0421&0.0778&0.0384&0.0842\\
Early voting&\textbf{0.4954}${}^{***}$&0.1070&0.1072&0.1071&0.0930&0.1226&0.0905&0.1257&0.0857&0.1320&0.0803&0.1395\\
    \bottomrule
  \end{tabular}
  \label{tab:ratings}
\end{table}

The deviation of the official results for Lukashenka and Tsikhanouskaya from the nearest bound of our credible intervals is about the same no matter what credible interval we consider: For the 99\% credible intervals the upper bound Lukashenka{'}s rating is about $62$ percentage points below the official outcome, whilst for Tsikhanouskaya the official results are $64$ percentage points below the lower bound of her predicted rating. For the 95\% and 90\% credible intervals, these deviations are 62-63 and 64-65 percentage points for Lukashenka and Tsikhanouskaya, respectively. We believe this is a rather suspicious coincidence which might indicate that the votes cast for Tsikhanouskaya may have been systematically subtracted from her and added to the votes cast for Lukashenka. Based on a subset of the officially reported protocols at some of the local election committees, \citet{tutby2020} also suggested such a pattern of systematic voting fraud. We must stress, however, that whilst our results are in line with such a conclusion, they do not exclude other explanations for this pattern. As for the votes cast for the {``}Against all{''} alternative, the official result is inside the $95\%$ credible interval. The results of Kanapatskaya are less than one percentage point of the $99.9\%$ credible interval of her rating predicted by our model. Whilst for Dmitriyeu, the official results are inside the $99.9\%$ credible interval and are just $0.08$ percentage points above the upper bound of the $99\%$ credible interval predicted by the model. The official results for Cherachen do not deviate from the corresponding posterior distribution. The fact that the official results for other candidates than Lukashenka and Tsikhanouskaya are much less deviant, may corroborate the hypothesis of the election authorities having systematically counted a share of Tsikhanouskaya{'}s votes as votes for Lukashenka. 

In Figure \ref{fig:ratings}, we show the results of the independent exit poll \citep{charter2020}  based on the responses of $85\,500$ participants surveyed by 76 volunteers according to \url{tmtr.me/modbelarus20/305}. Unfortunately, it is not yet clear what methodology was used in that study, but its results were updated through the whole early voting period and several times on the main election day. We notice some trends in the results: Lukashenka{'}s support rate gradually declined as the main election day approached, while Tsikhanouskaya{'}s support rate gradually grew. For Tsikhanouskaya this trend continued during the main election day itself. Furthermore, the {`}support{'} for the  {``}Against all{''} alternative declined on the main election day. These trends are consistent with the trends observed in the administrative voting (when people from the public sector are pressed to participate in the early voting), as well as with the calls of Tsikhanouskaya{'}s team to vote, preferably in the evening of the main election day~\citep{pamyatka2020}. The result for the {``}Against all{''} alternative from this poll is inside the $90\%$ credible interval predicted by our model; for Kanapatskaya, it is on the boundary of the $99.9\%$ credible interval; and for Lukashenka, it is inside the $95\%$ credible interval. The results for Cherachen and Tsikhanouskaya are less than one percentage point outside their respective $99.9{\%}$ credible intervals. Only for Dmitriyeu are the results slightly more than two percentage points away from the $99.9{\%}$ credible interval. The deviations of the results for Tsikhanouskaya could be explained by the fact that a great number of people supporting her -- identified by the white bracelets they chose to wear -- were willing to vote in the evening, but could not do so due to large queues\footnote{The queues were artificially created by CEC, which decided to reduce the capacity of the local commissions. Allegedly, this measure was taken to prevent the spread of the Covid-19 virus in the afternoon of the main election day. At the same time, no measures were taken to extend working hours of local commissions \citep{coronaermosh2020,queues2020,meduza2020}.} and commissions that closed on time and refused to wait until all who wanted could vote \citep{queues2020,coronaermosh2020,meduza2020,pamyatka2020}. According to the estimates of the observers at some voting places, these queues consisted of several hundred people. Even though the methodology used for polling is missing, the results of the independent exit poll seem to be in overall agreement with the ratings predicted by our model. The latter gives additional evidence for both the MRP ratings and the results of the poll. 

Returning back to our data and model, Figure \ref{fig:ratings_lukatiha} reveals that Lukashenka is not estimated to hold the majority in any marginal sub-group of the population. In the top graph of Figure \ref{fig:ratings_lukatiha}, we see that the only two marginal sub-groups where the upper bound of the 99.9\% credible interval is above 50\% are either people with no education or people above 60 years of age. According to the bottom graph of Figure \ref{fig:ratings_lukatiha}, for the former group, he actually has a higher ranking than Tsikhanouskaya, but for the latter group, Tsikhanouksaya is still more popular on average. For all other marginal sub-groups, Sviatlana is by far the most popular candidate with a variation of having the lower bound of the 99.9\% credible interval at around 60\% for Mahiliou region to having the upper bound of the same interval at almost 94\% for the youngest group of respondents. Ratings of the other candidates, as well as the popularity of the {``}Against all{''} choice and early voting across the sub-groups are of lesser interest; we have, however, included results for these groups in Figures \ref{fig:ratings_dimtr}-\ref{fig:ratings_earl} in Appendix \ref{sec:resultsappendix}.

 \begin{figure}[!htb]
 \centering
   \includegraphics[scale=0.27]{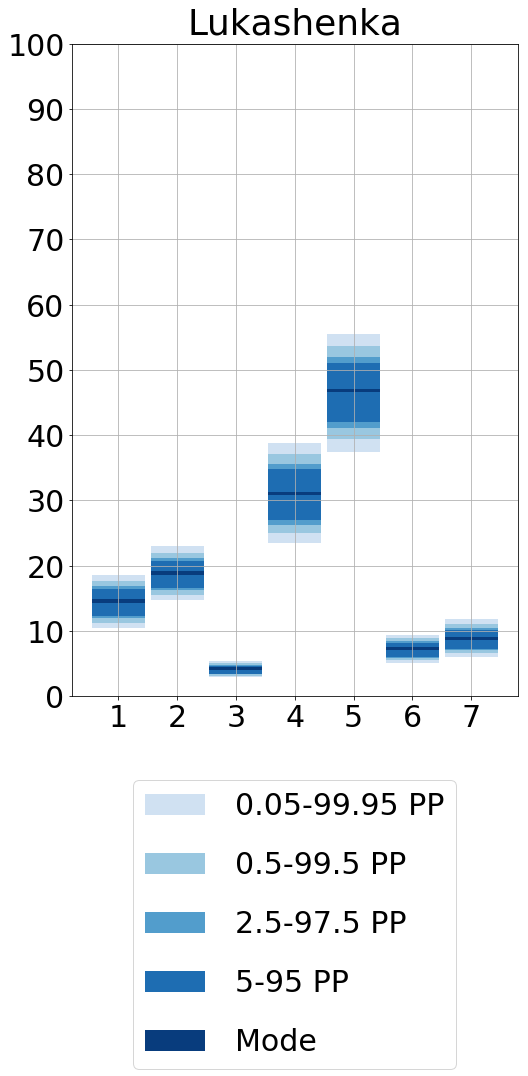}
   \includegraphics[scale = 0.27]{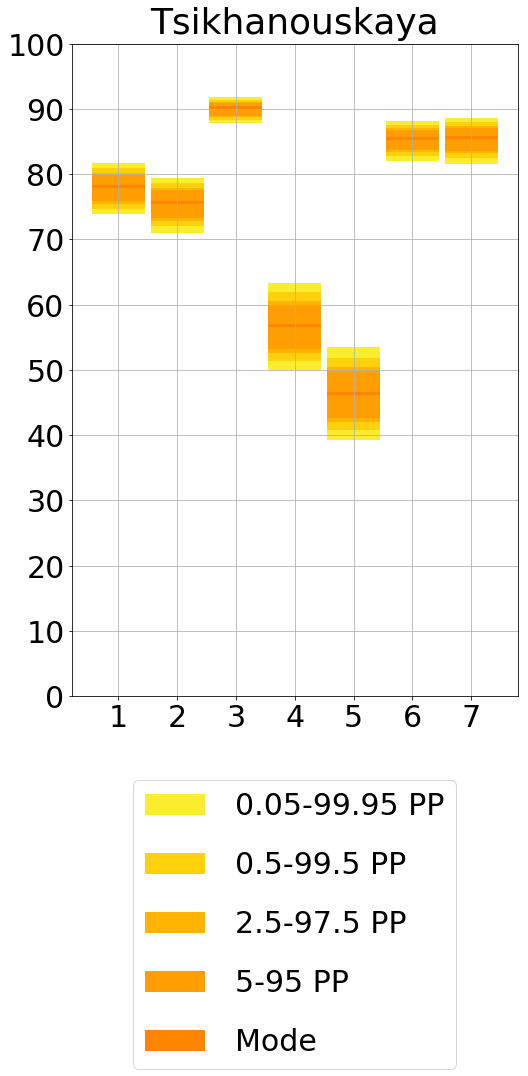}
   \caption{Posterior modes and quantiles for the support of Lukashenka (left) and Tsikhanouskaya (right) on the level on subpopulations corresponding to the 7 found clusters.}
   \label{fig:lukatihaclust}
 \end{figure}

To find other interesting demographic patterns in the data, we performed clustering of the respondents employing mixtures of multinomials with 7 latent states \footnote{The number of latent states was selected by BIC on a set from 1 to 7. We did not allow more than 7 clusters to enhance interpretability of the results and to avoid tiny clusters.} and the observed process consisting of independent multinomially distributed vectors corresponding to the 5 demographic variables of interest (age, gender, region, type of area of residence and education) and a dummy variable of political preferences with three categories (support Lukashenka, support Tsikhanouskaya, support neither of the two). The mixture model was fitted on the merged data by the expectation maximization (EM) algorithm from \textit{poLCA} R-package \citep{linzer2011polca}. Further, distributions of the subgroups inside the clusters and cluster sizes were poststratified to the whole population using Equation \eqref{eq:postsub}. The poststratified distributions of demographic variables are depicted in Figure \ref{fig:democlust} in Appendix \ref{sec:resultsappendix}. The distributions of the support of Lukashenka and Tsikhanouskaya across the 7 clusters are given in Figure \ref{fig:lukatihaclust} in the main part of the article, whilst similar distributions for other candidates, votes {``}Against all{''} and indicators of early voting are deferred to Figure~\ref{fig:othersclust} in Appendix~\ref{sec:resultsappendix}. 

As one can see from the right graph of Figure~\ref{fig:lukatihaclust}, the third segment (a term for cluster used in sociology), amounting to 22\% of all voters, constitutes Tsikhanouskaya’s main electorate where her support reaches 89.9\% (99\% CI  from 88.3\% to 91.4\%). It is also one of the largest segments in our study in general. The most represented age groups in this segment are middle-aged voters (31-41) and the youngest category of voters (18-30). They are likely university educated, urban, and the majority of these probably reside in the capital. In general, this segment has the assumed characteristics of a Belarusian {``}middle class{''} which explains Tsikhanouskaya{’}s country-wide popularity. The first segment, totaling $9{\%}$ of all voters, is another stronghold for Tsikhanouskaya – it consists of university educated people living in small cities in the Minsk region, but not in the capital itself. Tsikhanouskaya{’}s support there reaches $77.9{\%}$ ($99{\%}$ CI  from $74.7{\%}$ to $80.9{\%}$). According to the left graph of Figure~\ref{fig:lukatihaclust}, the fifth segment representing $11{\%}$ of the total electorate is Lukshashenka’s main voter group. This is the only segment where Lukashenka rivals Tsikhanouskaya standing at 46.6\% (99\% CI  from 39.4\% to 53.6\%). It consists of people in their fifties and older that mostly live in rural areas. In terms of education, they are likely high-school or vocational-school graduates. Tellingly, although this is Luskashenka’s most devout voter group, Tsikhanouskaya still has nearly the same level of support within this segment. Interestingly, Lukashenka does not have strong  support among the voters that traditionally have been regarded as his voter-base. For example, $85{\%}$ of pre-retirement aged men without higher education living in rural areas, as well as rural youth (the 7th segment – 15\% of the total) voted for Tskihanoyskaya. The other 15\% percent voted {``}Against all{''} or for other candidates. These results are indeed striking since Lukashenka, who used to be the head of a collective farm before becoming the president of the country, habitually emphasizes the importance of rural parts of the country and agriculture in his rhetoric.  The sixth segment (we called the “working class”)  is also overwhelmingly pro-Tsikhanouskaya with 85.3\% (99\% CI  from 82.7\% to 87.5\%) supporting her on average. This {``}working class{''} segment, amounting to $13{\%}$ of the population, consists of mainly working-aged men in urban areas with vocational school education.  They are likely well-qualified workers and have traditionally been assumed to be one of Lukashenka{'}s strongholds. The dismal support levels for the former president, however, can be explained by the government{’}s subjugation of trade unions, the increase in retirement age, layoffs, and the introduction of the so-called {``}contract system{''} that is extremely employer-centric and has striped employees of most of their rights. The second segment (24\% of the total) consisting mainly of university educated women older than 40 and mainly living outside the capital, also used to be considered in favour of Lukashenka. Despite Lukashenka{'}s efforts to appeal to this group in his rhetoric and personal-branding, our study shows that only $18.6{\%}$ (99\% CI  from 15.5\% to 22\%) of them support him while 75.5\% (99\% CI  from 72\% to 78.6\%) support Tsikhanouskaya. Moreover, the fourth segment (almost 5\% of the population) consisting of university educated retirement and pre-retirement aged people likely living in Minsk is also not convinced by Lukashenka. Despite a relatively high pension as compared with other countries in the Commonwealth of Independent States,\footnote{A regional intergovernmental organisation founded after the dissolution of the Soviet Union in 1991. Its current member states are: Armenia, Azerbaijan, Belarus, Kazakhstan, Kyrgyzstan, Moldova, Russia, Tajikistan, Uzbekistan.} and despite Lukashenka{'}s efforts to promote an image of Belarus as a comfortable and peaceful place to live that would supposedly appeal to them, Lukashenka still did not achieve high support in this segment: 56.6\% (99\% CI  from 51.4\% to 61.9\%) voted for Tsikhanouskaya, 30.8\% (99\% CI  from 24.9\% to 37\%) for Lukashenka, and 10.3\% voted {``}Against all{''} (99\% CI  from 7.4\%-13.9\%). This likely indicates that not all of the members of this segment are ready to vote for a woman without a strong political profile, however, Lukashenka does not appeal to them either.

Thus, our research shows that regular Belarusians have found {``}their{''} candidate – Tsikhanouskaya. Many of those who previously supported Lukashenka voted {``}Against all{''} since they were not convinced by Tsikhanouskaya, but nevertheless disillusioned by their previous choice. This is evident in groups like older urban university educated voters, 11\% of which voted {``}Against all{''} which is higher than in any other segment. On the other hand, Lukashenka does not have a strong voter-base, i.e.~a major social group that unconditionally supports him. He lost in all regions, in all age groups, even in rural areas. There are only small subgroups where he is leading. They are either likely not well-educated (some consisting of mostly secondary school drop-outs) or very small. These groups are very sparse or politically inactive to have them as a credible voter base. Thus, this might be one of the reasons why Lukashenka had to rely on loyal security forces and special police units to stifle dissent. Therefore, we can credibly state that Lukashenka{'}s reign is upheld by the corporate interests of affiliated elites and administrative and forceful methods, and not by popular support. In general, our research suggests that the $2020$ presidential election in Belarus was not simply abounded with fraud, but falsified altogether. It is supported by the conclusions from other monitoring initiatives based on analyzing collected polling stations{’} protocols. Since it is not possible to use ballot staffing to add an extra $60{\%}$ of the vote (which translates to millions of ballots), we state that most probably the results were announced at Lukashenka{’}s discretion only, thus effectively {``}cancelling{''} the actual election. This is supported by the evidence of independent observers who claim that the 2020 election results have been blatantly frauded in favour of Lukashenka \citep{economist1}.

%



\section{Discussion}\label{sec:discuss}

\subsection{Main conclusions}
In this study, we first used a Viber bot and a street poll to collect data on the political preferences of the people of Belarus leading up to the 2020 presidential election. The data from these two sources were then merged, and used as the basis for our statistical analyses. From Belstat, we also obtained aggregate demographic and economic data on the Belarusian population, allowing us to properly adjust and debias the sample obtained via the Viber bot and the street poll. A multilevel Bayesian regression model with poststratification was then used to estimate pre-election ratings, with associated uncertainty estimates, for all the candidates of the 2020 presidential race in Belarus, as well as estimates for the share of voters supporting the {``}Against all{''} alternative. Besides, we estimated the share of citizens who intended to participate in early voting. We further compared the results to the official outcomes of the election, as well as to two polls conducted prior to the election. We found that the results of the election announced by CEC and the results of the pro-governmental BRSM poll strongly disagree with the estimated pre-election ratings of the candidates, whilst the results of the independent poll are much more consistent with our estimated ratings. In particular, we found that both the officially announced results of the election and the officially reported early voting rates are highly improbable according to the estimates we obtained from the merged Viber and street poll data. We showed that with a probability of at least $95{\%}$, the rating of Sviatlana Tsikhanouskaya lies between $75{\%}$ and $80{\%}$, whereas Aliaksandr Lukashenka{'}s rating lies between $13{\%}$ and $18{\%}$ and early voting rates predicted by the method ranges from $9{\%}$ to $13{\%}$ of those who took part in the election. These results highly contradict the official ones, which are $10.12{\%}$, $80.11{\%}$, and $49.54{\%}$, respectively \citep{dw2020}, and all lie far outside even the $99.9{\%}$ credible intervals predicted by our model. The Ratings of other candidates and voting {``}Against all{''} are insignificant and correspond to the official results. The same conclusions are valid when comparing the pre-election ratings to the BRSM poll. Of particular interest is also the fact that the only groups of people where the upper bounds of the $99.9{\%}$ credible intervals for the rating of Lukashenka are above the $50{\%}$ mark are people older than $60$ and uneducated people. For all other groups, including rural residents and residents from all of the regions of the country, even the upper bounds of the $99.9{\%}$ credible intervals for Lukashenka are far below $50\%$. The same is valid for the population as a whole, showing that with a probability of at least $99.9{\%}$, as predicted by our model, Lukashenka could not have had enough electoral support to win the 2020 presidential election in Belarus in the first round.
In summary, our findings strongly suggest that the 2020 Belarusian election was a scam.

\subsection{Connections to other studies}
Our findings are in line with other sources of indirect evidence of the presidential election being rigged. Firstly, the results of the alternative votes verification Voice platform \citep{voice2020}, which joined $1\,049\,344$ votes of which $545\,947$ were confirmed with the verified photo of ballots (9.38\% of $5\,818\,955$ voted totally) from $1\,310$ places (22.7\% of all $5\,767$ polling places), show that the official voting results were systematically falsified. The Voice platform claims that the official records from $1\,310$ polling places account for 75\% of the total
votes for Tsikhanouskaya announced by CEC, i.e.~$471\,709$ out of her total vote count of $588\,619$. Hence, Tsikhanouskaya received only $116\,910$ additional votes from the remaining $4\,457$ polling locations. In other words, one might infer from the available data that the missing voting records from the unaccounted polling places show that Tsikhanouskaya received on average $26$ votes at each of these locations, which seems suspicious as she received on average 360 votes at the polling places where ballot records were available. Also, according to the CEC results, in Minsk $126\,861$ voters at 731 polling places supported Tsikhanouskaya, however the Voice platform shows that based on the official records from 432 polling places in Minsk (accounting for 59\% of all polling places in Minsk), there were at least $132\,941$ votes for Tsikhanovskaya reported, i.e.~$4.8{\%}$ more than the total number of votes reported by CEC for all 731 polling places, which is striking evidence of falsification. Further, according to the CEC, in the Minsk region, $115\,304$ voters at $993$ voting places voted for Tsikhanouskaya. The Voice platform analyzed the results that were reported by 257 voting places (26\% of all voting places in Minsk region) and found that Tsikhanouskaya received $114\,553$ votes there. This is 99.3\% of the votes reported by CEC. These results imply that in the remaining 74\% of the voting places, only 0.7\% of citizens supported Tsikhanouskaya, which seems rather suspicious. Finally, according to the Voice platform, the comparison of official ballot results with the photographs that were reported by the voters to the Voice platform (and hence the lower bound of the number of votes at each voting place was available for every candidate and {``}Against all{''}), showed falsifications in at least 440 polling places out of analyzed $1\,310$. These results support evidence of massive falsifications on all levels of the election committees. It also shows that at least 30\% of the final ballot reports were falsified. Hence the election could not be accepted as legitimate \citep{voice2020}.

\begin{table}[!htb]
 \caption{Official election results and posterior statistics of interest for the poststratified distribution on the population level for the selected model 1 trained on the street data only. Here, $g_x$ means $x$'s quantile of the focus posterior distribution. For the election results, the notation ${}^{***}$ means outside the 99.9\% credible interval, ${}^{**}$ means outside the 99\% credible interval, ${}^{*}$ means outside the 95\% credible interval, and ${}^{.}$ means outside the 90\% credible interval  of the parameter with respect to its posterior distribution. The anomalies in the official results found on either of the addressed credible levels are bold.}
  \centering
    \addtolength{\tabcolsep}{-3pt}
  \small
  \begin{tabular}{llrrrrrrrrrrr}
    \toprule
Event&Official&Mean&Mode&Median&$q_{0.0500}$&$q_{0.9500}$&$q_{0.0250}$&$q_{0.9750}$&$q_{0.0050}$&$q_{0.9950}$&$q_{0.0005}$&$q_{0.9995}$\\\hline
Lukashenka&\textbf{0.8010}${}^{***}$&0.1945&0.1956&0.1949&0.1165&0.2995&0.1044&0.3221&0.0834&0.3681&0.0631&0.4236\\
Tsikhanouskaya&\textbf{0.1012}${}^{***}$&0.7046&0.7050&0.7047&0.5966&0.7948&0.5743&0.8099&0.5296&0.8376&0.4768&0.8666\\
Dmitriyeu&0.0120&0.0046&0.0055&0.0049&0.0009&0.0185&0.0006&0.0239&0.0003&0.0395&0.0001&0.0717\\
Cherachen&0.0114&0.0114&0.0129&0.0119&0.0032&0.0354&0.0023&0.0436&0.0011&0.0656&0.0005&0.1055\\
Kanapatskaya&\textbf{0.0167}${}^{*}$&0.0020&0.0027&0.0022&0.0003&0.0110&0.0001&0.0147&0.0000&0.0269&0.0000&0.0562\\
Against all&0.0459${}^{.}$&0.0890&0.0926&0.0902&0.0461&0.1600&0.0400&0.1776&0.0299&0.2166&0.0210&0.2701\\
Early voting&\textbf{0.4954}${}^{**}$&0.1364&0.1392&0.1373&0.0777&0.2224&0.0690&0.2417&0.0541&0.2819&0.0403&0.3323\\
    \bottomrule
  \end{tabular}
  \label{tab:other_model_street}
\end{table}

Furthermore, the results of the polling initiated by the Telegram messenger \citep{telegapoll2020} as of 13:00 CET, September 3, 2020 collected around $1\,411\,035$ votes confirmed by the unique Belarusian phone numbers (around 25\% of those who voted which is the largest media poll in the history of Belarus), of which $1\,203\,862$ support Tsikhanouskaya (around 85\%), $90\,163$ support Lukashenka  (around 6.4\%), $86\,375$  support {``}Against all{''} (around 6.1\%), $12\,907$ support Cherachen (around 0.9\%), $12\,190$ support Dmitriyeu (around 0.9\%), and $5\,538$ support Kanapatskaya (around 0.4\%). 

Also, statistical analyses conducted by independent sociologists and statisticians from Russia assess the overall level of falsified ballots to constitute about {60\%} of the total number of ballots {`}cast{'}. More specifically, by analyzing $1\,000\,000$ digitized protocols (16\% of the total count), \citet{golgansky2020a} concluded that Lukashenka earned 55\% of the vote and Tsikhanouskaya 30\% without correcting for ballot stuffing. However, the balance between these two candidates becomes 45\% vs 45\% when correcting for the most obvious ballot stuffing. In a later inquiry, to showcase the scale of fraud, \citet{golgansky2020b} graphed the votes for Lukashenka as a function of turnout at a given polling station from the publicly available data. Unlike the graph of Tsikhanouskaya{’}s votes, the distribution of votes for Lukashenka has a specific parallelogram-like form that suggests strong positive linear relations between the turnout and votes for him. In turn, this points to extensive ballot stuffing in favour of Lukashenka. \citet{yudin2020}, a professor at The Moscow School of Social and Economic Sciences, aggregated the results of online and {``}on the ground{''} polls before and on the day of the election. He concludes that, notwithstanding the limitations of the then-available data, Tsikhanouskaya wins the presidency in the first round collecting around $70{\%}$ of the votes. \citet{schpilkin2020}, a physicist and data-scientist known for his research on electoral fraud in Russia, analyzed the fraud levels in the turnout during the early voting period of the 2020 presidential election in Belarus. He aggregated a dataset from $1\,804$ polling stations where the turnout data from the official protocols and independent observers differed by less than $10{\%}$. After analyzing the distribution of voter turnout over the five days of early voting, \citet{schpilkin2020} concluded that the actual turnout during early voting was a half or less than the official voter turnout of $41.7{\%}$.\footnote{Percentage of the total number of voters in the country is considered here.} According to \citet{schpilkin2020}, the upper bound on the early voting turnout is $24{\%}$ and this difference from the official number translates into at least $1.2$ million fraudulent votes.

\begin{figure}[!htb]
  \centering
  \includegraphics[scale=0.35]{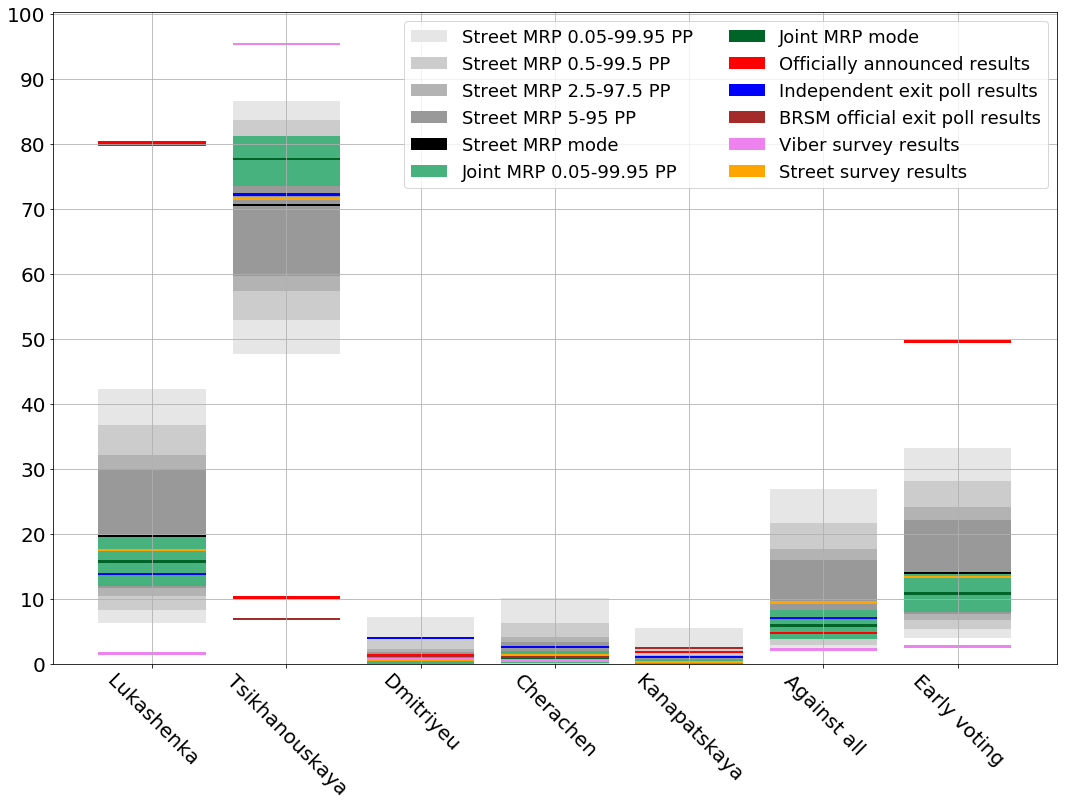}
  \caption{Posterior modes and quantiles for the parameters of interest on the population level for the MRP on the street and merged samples. The officially announced results are drawn in red. The independent exit poll results are drawn in blue. PP here means posterior probability. Viber and Street survey results correspond to the sample means of the corresponding polls.}
  \label{fig:ratings_str}
\end{figure}

\subsection{Criticism and our responses to it}

Whilst MRP is, without doubt, a very useful tool for estimating the distributions of focus parameters on the population level based on a biased sample, \citet{buttice2013does} show that it has to be used with caution. When dealing with samples of the size of typical national surveys, MRP provides the possibility of analyzing many political phenomena previously believed to be outside the bounds of systematic empirical inquiry. Investigating its performance with sample sizes typical of national surveys produce rather optimistic assessments. At the same time, \citet{buttice2013does} examine a larger number of cases and a greater range of opinions than in previous studies and find some variation in MRP{'}s performance. These authors explore such variations through empirical and Monte Carlo analyses. Their findings suggest that the conditions necessary for MRP to perform are not always met.  The main conclusion is that the most significant factor for the success of MRP is the strength of the geographic-level covariates included in the multilevel model of opinion and the ratio of opinion variation across geographic units relative to opinion variation within units. If these values are sizable, MRP will typically produce reliable estimates from national surveys of conventional size. Their empirical analysis based on Monte Carlo simulations, however, suggests that these conditions will not always be satisfied. At the same time, the authors show that even if MRP does not perform well, it still has significantly higher accuracy than a more traditional disaggregated means approach. These conclusions are fully supported by  \citet{warshaw2012should}  and \citet{lax2009should}. Whilst the disaggregated means method requires larger sample sizes, \citet{lax2009should} show that MRP performs equally well with both small and large samples of survey respondents. Additionally, they claim that since MRP can effectively be used with relatively little data and simple demographic typologies, it can also be applied to studies of public opinion over smaller time periods or in smaller geographic units, such as congressional districts or school districts, for which detailed demographic data are not available. \citet{gelman2013} gives a general discussion of the criticism. Further, in \citet{gao2020improving}, it is shown via a set of extensive simulation scenarios that structural priors improve MRP. Other important assumptions to be met are discussed by Daniel Simpson in his scientific blog \citep{simpson2019}: Assumption 1: The composition of the population is known. Assumption 2: The people who did not answer the survey are similar to the people who did answer the survey.

Assumption~1 is important because, in order to reconstruct the population from a smaller sample, we need to know how many people should be in each subgroup. This provides restrictions on how one can stratify the population. For surveys of people, one typically builds out the population information from some trustworthy census data, as well as from smaller official surveys.\footnote{Like the American Community Survey for estimation focus parameters about the US.} Assumption~2 can be formalized as follows: First, the people who were asked to participate in the survey in subgroup $j$ correspond to a random sample of subgroup $j$. Secondly, the people who actually answered the survey in subgroup $j$ correspond to a random sample of the people who were asked.

In response to the criticism from \citet{buttice2013does}, we have adopted advanced structural priors for the random effects recently suggested by \citep{gao2020improving}, where significant improvements were shown in extensive simulation scenarios. We also performed some additional simulations that can be found on the GitHub page of the project \citep{github}. Regarding Assumption~1, we used precise survey data from the 2009 Belstat census. We had to assume, however, that the demographics of Belarus have not changed significantly since then. In Figure~\ref{fig:biases}, we show this to be true at least marginally for four groups of the addressed demographic variables (when compared to the 2019 annual report), but the data on the fifth group (education levels) from 2019 is not yet available. Assumption~1 will be additionally checked when the results of the 2019 census in Belarus are published. Then, we will have the possibility to restratify the results if some significant changes in the demographics appear. Finally, regarding Assumption~2, we agree with \citet{simpson2019} that this sort of missing at random assumption is almost impossible to verify in practice. There are various things one can do to relax this assumption, but generally this is \textit{the assumption} that we are making. Whilst this assumption is likely met for the street survey,\footnote{Nevertheless, there is room for three sources of bias: (1) selection of respondents by interviewers - tendency to select more approachable/friendly-looking people, although we gave the explicit instructions to select random people; (2) response/refusal of respondents when approached (those in a hurry, those afraid to answer because of their pro-opposition views, possibly pro-governmental respondents who are not eager to answer due to their distrust in polls and other activities around the election); (3) item non-response, i.e. respondents not answering specific questions (some respondents did not want to report their income levels). The net effect of (1)-(3) is unknown, but street surveys typically assume it to be negligible.} validating it in the Viber poll is much more difficult. According to \citet{simpson2019}, one option is to assess how well the prediction works on some left out data in each subgroup. This is useful because poststratification explicitly estimates the response in the unobserved population. This viewpoint suggests that our goal is not necessarily unbiasedness but rather a good prediction of the population. It also means that if we can accept a reasonable bias, we will get the benefit of much tighter credible bounds of the population quantity than the survey weights can give. Hence, we return to the famous bias against variance trade-off. We have tried to approach this assumption from several perspectives. First of all, in the Viber poll, we used sampling of random phone numbers to invite respondents and advertised at different venues frequented by people with various demographic and political backgrounds (see Appendix \ref{sec:viber}). Secondly, in the attempt to obtain better results in the bias-variance trade-off sense and to assess predictive properties of the underlying Bayesian regression, we uniformly upsampled the street data to the size of $50{\%}$ of the Viber data and randomly divided the Viber data into two halves: One half was merged with the upsampled street data to form the training sample. The other one was left as a hold-out set to test predictive uncertainty handling by the introduced modified Brier score. Here, we aimed at reducing the variance by possibly introducing some bias and at testing predictive qualities (which appeared excellent according to Table~\ref{tab:mliks}) of the model.  Lastly, to assess and confirm our findings on the joint sample, we performed the same analysis based on MRP fitted on the street data only. This analysis is likely to have no violations of Assumption~2 above, however the sample is much smaller, and in the sense of a bias-variance trade-off, we are likely to have a significantly increased variation in the posterior distributions of the focus parameters. At the same time, we can validate the results obtained by MRP on the joint sample. As a result, equivalents to Table~\ref{tab:ratings} and Figure~\ref{fig:ratings} for the MRP fitted to the street data only are presented in Table~\ref{tab:other_model_street} and Figure~\ref{fig:ratings_str}. Moreover, the distributions of the focus parameters across all marginal subpopulations are presented in Figures~\ref{fig:ratings_luka_str}-\ref{fig:ratings_earl_str} in Appendix~\ref{sec:resultsappendix}. In short, one can see that even though the level of uncertainty is significantly increased due to the reduced sample size, ultimately all of the conclusions are equivalent to those presented above for the MRP trained on the joint sample. Though for some important conclusions the level of significance drops from 99.9\% to 99\% or 95\%. Moreover, the 99.9\%, 99\%, 95\%, and 90\% credible intervals of the MRP trained on the joint sample are almost always inside the corresponding credible intervals obtained on the street data. This allows us to conclude that we have obtained a very reasonable bias-variance trade-off on the joint data, corroborating the conclusions we have drawn from the joint sample. 

\subsection{Directions for future research and applications}

We also believe that obtaining the estimates of the actual ratings of the presidential candidates is not just interesting as a mathematical problem, but also as a tool for civil society under any authoritarian regime. It is common practice for autocrats to limit independent sociological research. Our solution is openly available and can, with a moderate level of effort, be applied in other countries with similar problems. Thus, our research could serve as a starting point to inform civil society action in similar non-democratic regimes. 

In the case of Belarus, we applied our method to estimate ratings of the candidates for the 2020 presidential election. However, the questions of interest do not have to be limited to politics. Our methodology could be applied to obtain other data that governments are trying to conceal or simply do not have enough resources to study. For example, in countries where the government refuses to reveal the Covid-19 mortality/spread statistics, our method could potentially be of help for the civil society to estimate the overall death/spread levels. In democratic countries like the United States, it could be possible to obtain independent estimates of the levels of police violence on different racial communities and minorities.

From a statistical standpoint, it might be of interest to obtain a model that can classify the respondents from the Viber bot to be similar to the respondents from the street survey. That might help to further improve checking Assumption 2 for MRP from \citet{simpson2019}. At the same time, it is far from obvious how to tackle this problem optimally. Also, statistically extending the cells for poststratification could be of interest. Given the five currently used categorical variables, one can build a model for estimating conditional probabilities of the new ones (for example the native language and the confession of a respondent). Then one automatically gets an approximation for the joint distribution of the original and predicted categories. This distribution can further be used in poststratification. Although the idea seems relatively simple, proper uncertainty handling under such an approach might be rather challenging. 

Finally, we are working on new polls, that could be of interest for the Belarusian civil society (see Figure~\ref{fig:timeline}). We hope their results could help to learn the mood of the population which in turn could be used to smoothly overcome the political, legal and social crises in the country.  

\section*{Acknowledgements}

We would like to thank Dr B and Ms A for proofreading the article and giving useful comments on the description of the model, the language, and the presentation of the results. We would also like to thank Dr K and Dr K for their critical remarks. Furthermore, we express our most sincere gratitude to Mr P for the help with legal details on authorship under anonymization.\footnote{The names are also anonymized here for the time being.} Last but not least, we cannot be more thankful to all the volunteers that participated in gathering the answers on the streets of the Belarusian cities, towns, and villages, continuously having at stake their health and freedom. This project would never have succeeded without the help of our data gatherers. 


\appendix

\begin{figure}[!htb]
  \centering
  \includegraphics[scale=0.465]{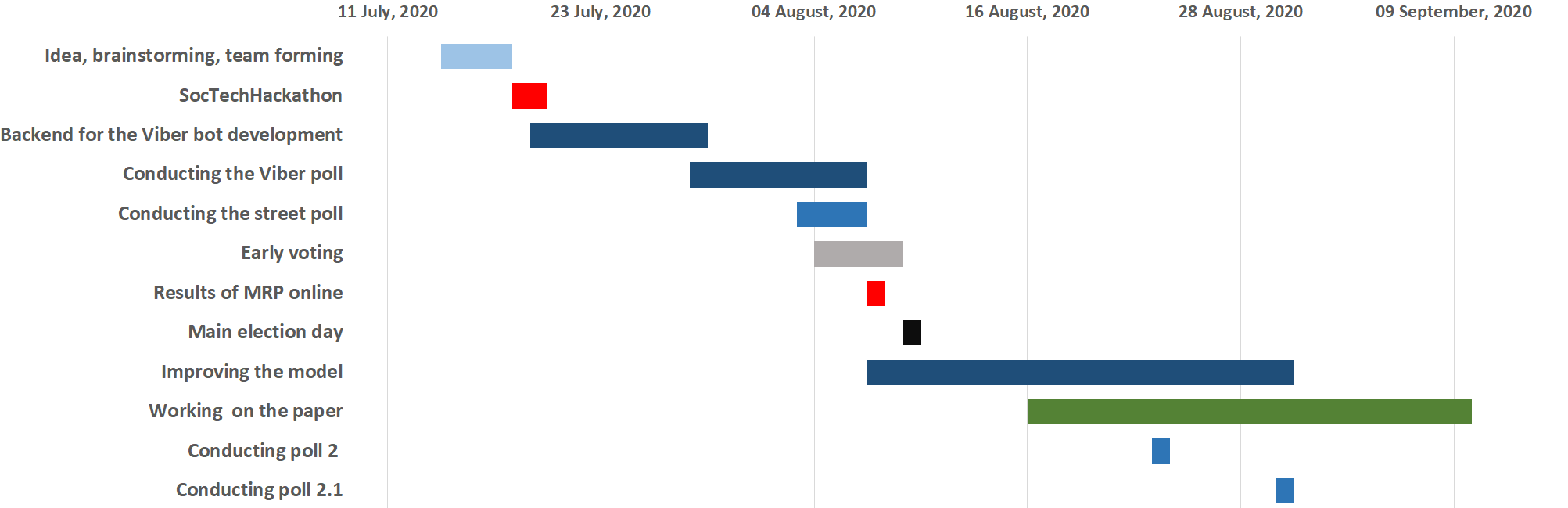}
  \caption{Project timeline.}
  \label{fig:timeline}
\end{figure}

\begin{table}[!htb]\footnote{Question 10 was asked only in street survey} 
 \caption{The full list of questions that we asked respondents before the elections. The mark $^*$ means including pensions, scholarships, allowances, etc.}
  \centering
    \addtolength{\tabcolsep}{-2pt}
      \small
  \begin{tabular}{lll}
    \toprule
\#&Question&Answer options\\\hline
1&Are you a citizen of the Republic of Belarus?&
    \begin{tabular}
    {@{}l@{}}
    1. Yes \\ 
    2. No
    \end{tabular}\\\hline
2&Please indicate your age&
    \begin{tabular}
    {@{}l@{}l@{}l@{}l@{}l@{}}
    1. Under 18 \\
    2. 18-30 \\
    3. 31-40 \\
    4. 41-50 \\
    5. 51-60 \\
    6. More than 60
    \end{tabular}\\\hline
3&What is the type of settlement you live in?&
    \begin{tabular}
    {@{}l@{}l@{}l@{}}
    1. Regional center / Minsk \\
    2. City or urban village \\
    3. Agro-town / Village \\
    4. I live outside the Republic of Belarus 
    \end{tabular}\\\hline
4&Select the region you live in. If you live in Minsk, choose Minsk&
    \begin{tabular}
    {@{}l@{}l@{}l@{}l@{}l@{}l@{}l@{}}
    1. Brest region \\
    2. Vitsebsk region \\
    3. Homel region \\
    4. Hrodna region \\
    5. Minsk region \\
    6. Mahiliou region \\
    7. Minsk city \\
    8. I live outside the Republic of Belarus
    \end{tabular}\\\hline
5&Who are you planning to vote for?&
    \begin{tabular}
    {@{}l@{}l@{}l@{}l@{}l@{}l@{}l@{}}
    1. Dmitriyeu \\
    2. Kanapatskaya \\
    3. Lukashenka \\
    4. Tsikhanouskaya \\
    5. Cherachen \\
    6. Against all \\
    7. Difficult to answer \\
    8. I will not go to vote
    \end{tabular}\\\hline
6&Please indicate your gender&
    \begin{tabular}
    {@{}l@{}}
    1. Male \\
    2. Female 
    \end{tabular}\\\hline
7&What is your educational level?&
    \begin{tabular}
    {@{}l@{}l@{}l@{}l@{}}
    1. Primary or secondary school \\
    2. Professional technical institution \\
    3. Professional college \\
    4. Higher education \\
    5. Other (Elementary school or uneducated)
    \end{tabular}\\\hline
8&Please indicate your family's total monthly income*&
    \begin{tabular}
    {@{}l@{}l@{}l@{}l@{}}
    1. Up to 500 BYN \\
    2. 500 - 1000 BYN \\
    3. 1000 - 2000 BYN \\
    4. Above 2000 BYN \\
    5. I don't want to answer this question
    \end{tabular}\\\hline
9&When do you plan to vote?&
    \begin{tabular}
    {@{}l@{}l@{}}
    1. Early (from 4 to 8 of August) \\
    2. On the main voting day (August 9) \\
    3. I will not go to vote
    \end{tabular}\\\hline
10&Which means of communication and information sources do you use? &
    \begin{tabular}
    {@{}l@{}l@{}@{}l@{}l@{}l@{}@{}l@{}l@{}}
    1. vk.com \\
    2. ok.ru \\
    3. Viber \\
    4. Telegram \\
    5. Youtube \\
    6. TV \\
    7. Radio \\
    8. Paper media \\
    9. Other (open option)
    \end{tabular}\\
    \bottomrule
  \end{tabular}
  \label{tab:qlist}
\end{table}

\section{Further detail on data collection and processing}\label{sec:dataappendix}
\subsection{Viber survey}\label{sec:viber}
\subsubsection{Technical pipeline details}
\begin{figure}[!htb]
  \centering
  \includegraphics[scale=0.45]{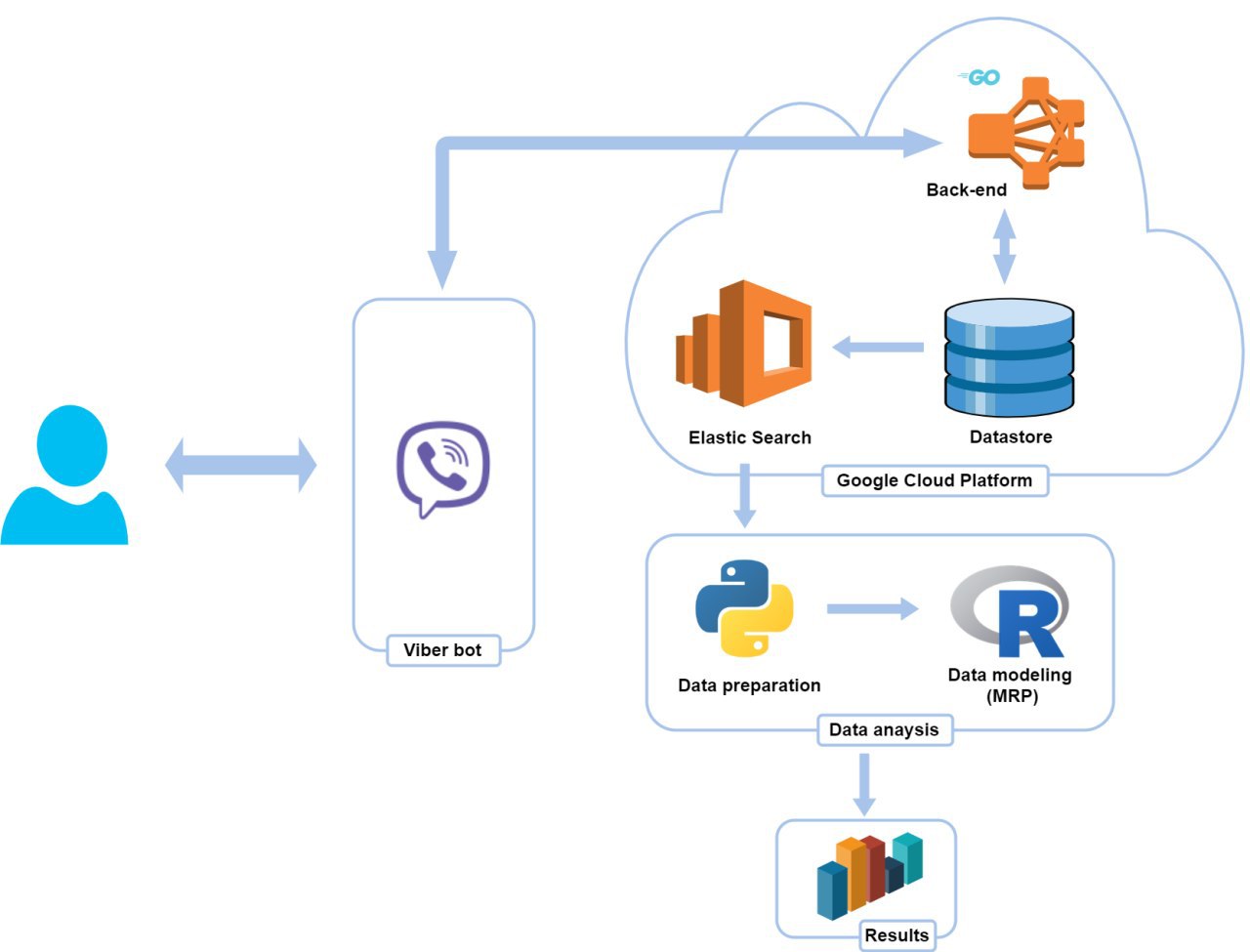}
  \caption{A high-level architecture of our solution.}
  \label{fig:pipeline}
\end{figure}

The implemented technical solution consists of two main parts: the Viber bot itself and a server-side application which sends questions to the users and receives their answers. Firstly, we created a chat bot using Viber Admin Panel \citep{viberAdmin}. Then a back-end solution based on Viber Http API \citep{viberAPI} was implemented. There, Go programming language was used for business logic and Google Cloud Platform developments for deployment and storing the data. The data was continuously transferred to ElasticSearch, from where it was retrieved for preprocessing with Python. Postprocessing, statistical modelling, and analysis was conducted in R, see Figure \ref{fig:pipeline}.  All data were stored without any corrections. Thus, for example, if a respondent pointed out that he is underaged, we allowed him to pass further. However, a warning that his answer could not be counted was sent. We further asked such a respondent to share the survey with his relatives or friends who were at least 18 years old. All the inappropriate data was filtered after collection in a Python preprocessing pipeline.

\subsubsection{Distribution of the poll in the population}
Since online surveys in Belarus typically attract the young and progressive audience, we decided to spread the bot within more general communities including the venues popular among supporters of Aliaksandr Lukashenka. We also published the advertisement for the survey in the most popular Telegram channels dedicated to politics in Belarus.  The list of venues is present in Tables \ref{tab:communitiesMSG}, and \ref{tab:communitiesOK}.

\begin{table}
 \caption{The list of venues our bot was spread within (Messengers)}
  \centering
  \begin{tabular}{lp{3cm}llll}
    \toprule
Channel&Name&Audience&Link&Accepted\\\hline
Viber&WestMotors&18$\,$764&-&No \\
Viber&The magical world of books&436&-&Yes \\
Viber&PORA center - helping children&115&-&Yes \\
Viber&Coronavirus defeated&1$\,$043&-&Yes \\
Viber&BREST Buy \& Sell&2$\,$833&-&Yes \\
Viber&Stepyanka&1$\,$550&-&Yes \\
Viber&Flea market in Baranovichi and Belarus&1$\,$402&-&Yes \\
Viber&Part-time job for women of Belarus&3$\,$500&-&Yes \\
Viber&Flea market Minsk&2$\,$183&-&Yes \\
Viber&Khaltura Uruchye Minsk&10$\,$023&-&Yes \\
Telegram & Home Belsat & 4$\,$554 & \url {t.me/belsat\_chat} & - \\
Telegram & Tea with raspberry jam & 191$\,$111 & \url {t.me/belteanews} & - \\
Telegram & Real   Belarus & 2$\,$435 & \url {t.me/strana\_by} & - \\
Telegram & Tut.by News$\,$ Chat & 14$\,$152 & \url {t.me/tutby\_news\_chat} & No \\
Telegram & Commments room & 1$\,$058 & \url {t.me/joinchat/LptSM1dPAC3ro5\_GRi8E1Q} & - \\
Telegram & My country Belarus & 14$\,$247 & \url {t.me/majakrainablr} & Yes \\
Telegram & Tractor   cures &  34$\,$509 & \url{t.me/traktorlechit} & - \\
Telegram & Nexta Live & 358$\,$466 & \url {t.me/nexta\_tv} & Yes \\
Telegram & Nexta   TV & 310$\,$453 & \url {t.me/nexta\_live} & Yes \\
Telegram & Belarus   of brain & 184$\,$709 & \url {t.me/belamova} & - \\
Telegram & Viktar Babaryka official & 48$\,$608 & \url {t.me/viktarbabarykaofficial} & - \\
Telegram & Onliner & 45$\,$008 & \url {t.me/onlinerby} & No \\
\bottomrule
    \end{tabular}
  \label{tab:communitiesMSG}
\end{table}

\begin{table}
 \caption{The list of venues our bot was spread within (ok.ru)}
  \centering
  \begin{tabular}{p{5cm}p{1cm}ll}
    \toprule
   Name&Audience&Link&Accepted\\\hline
     My free and   great Belarus!!! & 7$\,$872 & \url {ok.ru/mayakranav} & Yes \\
 Minsk$\,$   Belarus & 550 & \url {ok.ru/group52104198684871} & Yes \\
 Belarus.   Minsk. News. Jobs. & 744 & \url {ok.ru/group53644681085141} & Yes \\
 Belarus.   Minsk. News. Jobs & 5$\,$304 & \url {ok.ru/myprotivd} & Yes \\
 News.   Belarus. Minsk & 3$\,$322 & \url {ok.ru/zamirzamir} & Yes \\
 Minsk.   Belarus. News. Events & 3$\,$914 & \url {ok.ru/minsk.bela} & Yes \\
 Belarus & 6$\,$362 & \url {ok.ru/group51288141463705} & On moderation \\
 Belarus & 8$\,$794 & \url {ok.ru/belorusiy} & On moderation \\
 Belarus is us & 22$\,$285 & \url {ok.ru/belaruseto} & Yes \\
 For a prosperous   Belarus! We no longer believe Lukashenka. & 12$\,$386 & \url {ok.ru/zaprotsvet} & On moderation \\
 Let's   repost$\,$ Belarus & 20$\,$215 & \url {ok.ru/postimby} & On moderation \\
 My   native corner$\,$ how lovely you are! & 44$\,$086 & \url {ok.ru/moyrodkut} & No \\
 Biarazino   chronicle & 163 & \url {ok.ru/group/56156556296222} & On moderation \\
 What's   happening in Belarus & 156$\,$060 & \url {ok.ru/cpbelarus} & On moderation \\
 Give   away & 23$\,$152 & \url {ok.ru/v0x0v/topics} & Yes \\
 I'm   against this regime & 19$\,$673 & \url {ok.ru/group/52914981306543/topics} & Yes \\
 Homel   junk market & 127$\,$472 & \url {ok.ru/group53645952811238/topics} & Yes \\
 Free   junk market By & 20$\,$552 & \url {ok.ru/besplatkaby/topics} & Yes \\
 Belarus.   Everything about everything & 25$\,$960 & \url {ok.ru/prostointeresnotut} & On moderation \\
 My   beloved old Vitsebsk & 19$\,$090 & \url {ok.ru/retrovit} & On moderation \\
 ePramova & 17$\,$921 & \url {ok.ru/epramova} & On moderation \\
 Your   Minsk & 17$\,$004 & \url {ok.ru/etominsk} & On moderation \\
 Belarus   bulletin board & 15$\,$887 & \url {ok.ru/internetmagazinvitebsk/topics} & Yes \\
 Hrodna & 14$\,$509 & \url {ok.ru/grodnosimplythebest} & Yes \\
 Internet   is against zombobox & 14$\,$137 & \url {ok.ru/oo.85} & Yes \\
 Underwater   hunting$\,$ fishing in Belarus & 12$\,$232 & \url {ok.ru/podvodnaya.ohota} & Yes \\
 Belarus.   News. Announcements. All cities & 9$\,$897 & \url {ok.ru/o0v0o} & Yes \\
 Everything   about Belarus and Belarusians & 9$\,$448 & \url {ok.ru/vseobelaru51885784760525} & Yes \\
 We   are Belarusians! & 9$\,$231 & \url {ok.ru/mibelarusi} & Yes \\
 Minsk   and Minsk people & 8$\,$283 & \url {ok.ru/nashminsk} & Yes \\
 Give Free |  Belarus & 8$\,$137 & \url {ok.ru/darombelarus} & On moderation \\
 Advertising   panel & 8$\,$043 & \url {ok.ru/ads.ob} & Yes \\
 Junk   market "White Rus" & 7$\,$674 & \url {ok.ru/group53197935935704} & On moderation \\
 The   power is in truth! & 34$\,$336 & \url {ok.ru/group53901106151576} & On moderation \\
 Give   away free & 6$\,$615 & \url {ok.ru/doska.rusk} & Yes \\
 Belarus   is our motherland! & 5$\,$620 & \url {ok.ru/group52860189540462} & On moderation \\
 Grape   growing in Belarus$\,$ Sviatlahorsk & 5$\,$495 & \url {ok.ru/vinogrador} & On moderation \\
 Belarusian   mothers & 5$\,$305 & \url {ok.ru/mamybelarusi} & Yes \\
 Second   by & 3$\,$522 & \url {ok.ru/sekondby} & Yes \\
 Advertising   panel. Brest region & 2$\,$956 & \url {ok.ru/brest2016} & On moderation \\
 Belarus   plus & 2$\,$849 & \url {ok.ru/belplus} & Yes \\
 Realty   By & 2$\,$556 & \url {ok.ru/club500} & Yes \\
 Homel.   Ads. News & 2$\,$410 & \url {ok.ru/gomelobyav} & Yes \\
 Zialiony   Luh$\,$ Minsk & 2$\,$330 & \url {ok.ru/zelenylugm} & Yes \\
 People   connections in Belarus & 2$\,$160 & \url {ok.ru/znakomstvavbelarusi} & Yes \\
 Slonim.   Ads. News & 1$\,$989 & \url {ok.ru/slonimobya} & Yes \\
    \bottomrule
    \end{tabular}
  \label{tab:communitiesOK}
\end{table}

In addition, to further make the data more representative, we sent about 1500 SMS to randomly chosen telephone numbers of the residents of Belarus. We also tried to advertise our survey on social networks. In particular, we created a 1 minute video about our project that was put on YouTube \citep{promoVideo}. Furthermore, an Instagram page \citep{instaPage} was created. Both Youtube and Instagram were artificially targeted to the audience outside the Minsk region. We also tried to spread the survey among influencers who could actively distribute the information about the project on social networks and in their entourage. Finally, we appealed to the people who had already participated in our survey to invite their relatives and friends with different political opinions, emphasizing that it was especially important to invite people living in the rural areas. This had a good effect on coverage and representativeness. We believe, all these steps allowed us to collect relatively heterogeneous data. \footnote{We also made an attempt to advertise via the most popular Belarusian web portals (\url{www.tut.by/},  \url{www.onliner.by/}, \url{kyky.org/}, \url{citydog.by/}, \url{www.svaboda.org/}, \url{euroradio.fm}). But they refused to publish the advertisement since sociological studies made by non-accredited organizations are not allowed in Belarus.}

\subsection{Street survey}

\subsubsection{Respondents quotas for street poll }\label{sec:streetquotas}
The quotas of all combinations of age, sex, type of residence of area, and region are shown in Table \ref{tab:quotastab}. Each row consists of the number of collected answers and is supplemented with the correspondent quotas, which were calculated based on a minimum of 500 respondents and distributions from age and sex structure of the average annual population in Belarus for 2019 \citep{belstatDemogr2020}. We collected these data using the same technique as described in Section \ref{sec:data} for the census data. The numbers of respondents belonging to the groups that we were not able to fulfill are highlighted in red.

\begin{table}[!htb]
\small
\caption{Respondents quotas per region, age and types of area of residence based on a sample size of 500.}
\begin{tabular}{lrrrrllrrll}
\hline
\cellcolor{lightgray}\textbf{Brest   region} & \multicolumn{2}{c}{\cellcolor{lightgray}\textbf{Urban}} & \multicolumn{2}{c}{\cellcolor{lightgray}\textbf{Rural}} &  & \cellcolor{lightgray}\textbf{Hrodna   region} & \multicolumn{2}{c}{\cellcolor{lightgray}\textbf{Urban}} & \multicolumn{2}{c}{\cellcolor{lightgray}\textbf{Rural}} \\
\multicolumn{1}{l}{\cellcolor{lightgray}\textbf{Age}} & \cellcolor{lightgray}\textbf{Male} & \cellcolor{lightgray}\textbf{Female} & \cellcolor{lightgray}\textbf{Male} & \cellcolor{lightgray}\textbf{Female} &  & \cellcolor{lightgray}\textbf{Age} & \cellcolor{lightgray}\textbf{Male} & \cellcolor{lightgray}\textbf{Female} & \multicolumn{1}{r}{\cellcolor{lightgray}\textbf{Male}} & \multicolumn{1}{r}{\cellcolor{lightgray}\textbf{Female}} \\
\textbf{18-30   quotas} & 5 & 5 & 1 & 1 &  & \textbf{18-30   quotas} & 4 & 4 & \multicolumn{1}{r}{1} & \multicolumn{1}{r}{1} \\
\textbf{18-30   responses} & 10 & 10 & 5 & 2 &  & \textbf{18-30   responses} & 28 & 18 & \multicolumn{1}{r}{1} & \multicolumn{1}{r}{1} \\
\textbf{31-40   quotas} & 5 & 5 & 2 & 1 &  & \textbf{31-40   quotas} & 4 & 4 & \multicolumn{1}{r}{1} & \multicolumn{1}{r}{1} \\
\textbf{31-40   responses} & 9 & 10 & 6 & 3 &  & \textbf{31-40   responses} & 19 & 16 & \multicolumn{1}{r}{\cellcolor{red}0} & \multicolumn{1}{r}{1} \\
\textbf{41-50   quotas} & 4 & 5 & 2 & 2 &  & \textbf{41-50   quotas} & 3 & 4 & \multicolumn{1}{r}{1} & \multicolumn{1}{r}{1} \\
\textbf{41-50   responses} & 10 & 9 & 9 & 7 &  & \textbf{41-50   responses} & 16 & 19 & \multicolumn{1}{r}{2} & \multicolumn{1}{r}{5} \\
\textbf{51-60   quotas} & 4 & 5 & 2 & 2 &  & \textbf{51-60   quotas} & 3 & 4 & \multicolumn{1}{r}{2} & \multicolumn{1}{r}{1} \\
\textbf{51-60   responses} & 4 & 12 & 6 & 5 &  & \textbf{51-60   responses} & 9 & 16 & \multicolumn{1}{r}{2} & \multicolumn{1}{r}{2} \\
\textbf{60+   quotas} & 5 & 8 & 3 & 5 &  & \textbf{60+   quotas} & 4 & 6 & \multicolumn{1}{r}{2} & \multicolumn{1}{r}{3} \\
\textbf{60+   responses} & 8 & 12 & 3 & 5 &  & \textbf{60+   responses} & 10 & 21 & \multicolumn{1}{r}{4} & \multicolumn{1}{r}{1} \\
 & \multicolumn{1}{l}{} & \multicolumn{1}{l}{} & \multicolumn{1}{l}{} & \multicolumn{1}{l}{} &  &  & \multicolumn{1}{l}{} & \multicolumn{1}{l}{} &  &  \\
\cellcolor{lightgray}\textbf{Vitsebsk   region} & \multicolumn{2}{c}{\cellcolor{lightgray}\textbf{Urban}} & \multicolumn{2}{c}{\cellcolor{lightgray}\textbf{Rural}} &  & \cellcolor{lightgray}\textbf{Minsk   region} & \multicolumn{2}{c}{\cellcolor{lightgray}\textbf{Urban}} & \multicolumn{2}{c}{\cellcolor{lightgray}\textbf{Rural}} \\
\cellcolor{lightgray}\textbf{Age} & \cellcolor{lightgray}\textbf{Male} & \cellcolor{lightgray}\textbf{Female} & \cellcolor{lightgray}\textbf{Male} & \cellcolor{lightgray}\textbf{Female} &  & \cellcolor{lightgray}\textbf{Age} & \cellcolor{lightgray}\textbf{Male} & \cellcolor{lightgray}\textbf{Female} & \multicolumn{1}{r}{\cellcolor{lightgray}\textbf{Male}} & \multicolumn{1}{r}{\cellcolor{lightgray}\textbf{Female}} \\
\textbf{18-30   quotas} & 4 & 5 & 1 & 1 &  & \textbf{18-30   quotas} & 4 & 3 & \multicolumn{1}{r}{2} & \multicolumn{1}{r}{2} \\
\textbf{18-30   responses} & 8 & 8 & \cellcolor{red}0 & 3 &  & \textbf{18-30   responses} & 12 & 6 & \multicolumn{1}{r}{4} & \multicolumn{1}{r}{2} \\
\textbf{31-40   quotas} & 5 & 5 & 1 & 1 &  & \textbf{31-40   quotas} & 5 & 4 & \multicolumn{1}{r}{3} & \multicolumn{1}{r}{3} \\
\textbf{31-40   responses} & \cellcolor{red}2 & 6 & 1 & 1 &  & \textbf{31-40   responses} & 14 & 6 & \multicolumn{1}{r}{3} & \multicolumn{1}{r}{\cellcolor{red}1} \\
\textbf{41-50   quotas} & 4 & 4 & 1 & 1 &  & \textbf{41-50   quotas} & 4 & 4 & \multicolumn{1}{r}{3} & \multicolumn{1}{r}{3} \\
\textbf{41-50   responses} & 7 & 5 & 1 & 1 &  & \textbf{41-50   responses} & 12 & 9 & \multicolumn{1}{r}{3} & \multicolumn{1}{r}{\cellcolor{red}2} \\
\textbf{51-60   quotas} & 4 & 5 & 2 & 2 &  & \textbf{51-60   quotas} & 3 & 4 & \multicolumn{1}{r}{3} & \multicolumn{1}{r}{3} \\
\textbf{51-60   responses} & \cellcolor{red}3 & \cellcolor{red}3 & 4 & 2 &  & \textbf{51-60   responses} & 4 & 7 & \multicolumn{1}{r}{\cellcolor{red}1} & \multicolumn{1}{r}{4} \\
\textbf{60+   quotas} & 5 & 9 & 2 & 3 &  & \textbf{60+   quotas} & 4 & 7 & \multicolumn{1}{r}{4} & \multicolumn{1}{r}{7} \\
\textbf{60+   responses} & 8 & 11 & \cellcolor{red}1 & \cellcolor{red}2 &  & \textbf{60+   responses} & 9 & 10 & \multicolumn{1}{r}{5} & \multicolumn{1}{r}{7} \\
 & \multicolumn{1}{l}{} & \multicolumn{1}{l}{} & \multicolumn{1}{l}{} & \multicolumn{1}{l}{} &  &  & \multicolumn{1}{l}{} & \multicolumn{1}{l}{} &  &  \\
\cellcolor{lightgray}\textbf{Homel   region} & \multicolumn{2}{c}{\cellcolor{lightgray}\textbf{Urban}} & \multicolumn{2}{c}{\cellcolor{lightgray}\textbf{Rural}} &  & \cellcolor{lightgray}\textbf{Mahiliou   region} & \multicolumn{2}{c}{\cellcolor{lightgray}\textbf{Urban}} & \multicolumn{2}{c}{\cellcolor{lightgray}\textbf{Rural}} \\
\cellcolor{lightgray}\textbf{Age} & \cellcolor{lightgray}\textbf{Male} & \cellcolor{lightgray}\textbf{Female} & \cellcolor{lightgray}\textbf{Male} & \cellcolor{lightgray}\textbf{Female} &  & \cellcolor{lightgray}\textbf{Age} & \cellcolor{lightgray}\textbf{Male} & \cellcolor{lightgray}\textbf{Female} & \multicolumn{1}{r}{\cellcolor{lightgray}\textbf{Male}} & \multicolumn{1}{r}{\cellcolor{lightgray}\textbf{Female}} \\
\textbf{18-30   quotas} & { 5} & { 5} & { 1} & { 1} &  & \textbf{18-30   quotas} & 4 & 3 & \multicolumn{1}{r}{1} & \multicolumn{1}{r}{1} \\
\textbf{18-30   responses} & { 18} & { 26} & { 2} & { 6} &  & \textbf{18-30   responses} & 9 & 10 & \multicolumn{1}{r}{2} & \multicolumn{1}{r}{2} \\
\textbf{31-40   quotas} & { 6} & { 6} & { 1} & { 1} &  & \textbf{31-40   quotas} & 5 & 5 & \multicolumn{1}{r}{1} & \multicolumn{1}{r}{1} \\
\textbf{31-40   responses} & { 23} & { 18} & { 5} & { 4} &  & \textbf{31-40   responses} & 9 & 5 & \multicolumn{1}{r}{3} & \multicolumn{1}{r}{1} \\
\textbf{41-50   quotas} & { 5} & { 5} & { 2} & { 1} &  & \textbf{41-50   quotas} & 4 & 4 & \multicolumn{1}{r}{1} & \multicolumn{1}{r}{1} \\
\textbf{41-50   responses} & { 17} & { 26} & { 2} & { 5} &  & \textbf{41-50   responses} & \cellcolor{red}3 & 6 & \multicolumn{1}{r}{4} & \multicolumn{1}{r}{2} \\
\textbf{51-60   quotas} & { 4} & { 6} & { 2} & { 2} &  & \textbf{51-60   quotas} & 3 & 4 & \multicolumn{1}{r}{1} & \multicolumn{1}{r}{1} \\
\textbf{51-60   responses} & { 18} & { 20} & { 7} & { 3} &  & \textbf{51-60   responses} & 8 & 12 & \multicolumn{1}{r}{7} & \multicolumn{1}{r}{8} \\
\textbf{60+   quotas} & { 5} & { 10} & { 2} & { 4} &  & \textbf{60+   quotas} & 4 & 8 & \multicolumn{1}{r}{2} & \multicolumn{1}{r}{3} \\
\textbf{60+   responses} & { 22} & { 30} & { 5} & { 9} &  & \textbf{60+   responses} & 5 & 11 & \multicolumn{1}{r}{6} & \multicolumn{1}{r}{9} \\
 & \multicolumn{1}{l}{} & \multicolumn{1}{l}{} & \multicolumn{1}{l}{} & \multicolumn{1}{l}{} &  &  & \multicolumn{1}{l}{} & \multicolumn{1}{l}{} &  &  \\
\textbf{} & \multicolumn{2}{l}{\textbf{}} & \multicolumn{2}{l}{\textbf{}} &  & \cellcolor{lightgray}\textbf{Minsk   city} & \multicolumn{2}{c}{\cellcolor{lightgray}\textbf{Urban}} & \multicolumn{2}{c}{\cellcolor{lightgray}\textbf{Rural}} \\
\textbf{} & \multicolumn{1}{l}{\textbf{}} & \multicolumn{1}{l}{\textbf{}} & \multicolumn{1}{l}{\textbf{}} & \multicolumn{1}{l}{\textbf{}} &  & \cellcolor{lightgray}\textbf{Age} & \cellcolor{lightgray}\textbf{Male} & \cellcolor{lightgray}\textbf{Female} & \cellcolor{lightgray}\textbf{Male} & \cellcolor{lightgray}\textbf{Female} \\
\textbf{} & \multicolumn{1}{l}{} & \multicolumn{1}{l}{} & \multicolumn{1}{l}{} & \multicolumn{1}{l}{} &  & \textbf{18-30   quotas} & 12 & 12 & 0 & 0 \\
\textbf{} & \multicolumn{1}{l}{} & \multicolumn{1}{l}{} & \multicolumn{1}{l}{} & \multicolumn{1}{l}{} &  & \textbf{18-30   responses} & 15 & 14 & 0 & 0 \\
\textbf{} & \multicolumn{1}{l}{} & \multicolumn{1}{l}{} & \multicolumn{1}{l}{} & \multicolumn{1}{l}{} &  & \textbf{31-40   quotas} & 12 & 13 & 0 & 0 \\
\textbf{} & \multicolumn{1}{l}{} & \multicolumn{1}{l}{} & \multicolumn{1}{l}{} & \multicolumn{1}{l}{} &  & \textbf{31-40   responses} & 16 & 13 & 0 & 0 \\
\textbf{} & \multicolumn{1}{l}{} & \multicolumn{1}{l}{} & \multicolumn{1}{l}{} & \multicolumn{1}{l}{} &  & \textbf{41-50   quotas} & 8 & 9 & 0 & 0 \\
\textbf{} & \multicolumn{1}{l}{} & \multicolumn{1}{l}{} & \multicolumn{1}{l}{} & \multicolumn{1}{l}{} &  & \textbf{41-50   responses} & 13 & 14 & 0 & 0 \\
\textbf{} & \multicolumn{1}{l}{} & \multicolumn{1}{l}{} & \multicolumn{1}{l}{} & \multicolumn{1}{l}{} &  & \textbf{51-60   quotas} & 7 & 9 & 0 & 0 \\
\textbf{} & \multicolumn{1}{l}{} & \multicolumn{1}{l}{} & \multicolumn{1}{l}{} & \multicolumn{1}{l}{} &  & \textbf{51-60   responses} & 17 & 13 & 0 & 0 \\
\textbf{} & \multicolumn{1}{l}{} & \multicolumn{1}{l}{} & \multicolumn{1}{l}{} & \multicolumn{1}{l}{} &  & \textbf{60+   quotas} & 9 & 17 & 0 & 0 \\
\textbf{} & \multicolumn{1}{l}{} & \multicolumn{1}{l}{} & \multicolumn{1}{l}{} & \multicolumn{1}{l}{} &  & \textbf{60+   responses} & 15 & 39 & 0 & 0 \\
 & \multicolumn{1}{l}{} & \multicolumn{1}{l}{} & \multicolumn{1}{l}{} & \multicolumn{1}{l}{} &  &  & \multicolumn{1}{l}{} & \multicolumn{1}{l}{} &  &  \\
\textbf{} & \multicolumn{2}{c}{\textbf{}} & \multicolumn{2}{c}{\textbf{}} &  &  & \multicolumn{1}{l}{} & \multicolumn{1}{l}{} &  &  \\
\textbf{} & \textbf{} & \textbf{} & \textbf{} & \textbf{} &  &  & \multicolumn{1}{l}{} & \multicolumn{1}{l}{} &  &  \\
\textbf{} &  &  &  &  &  &  & \multicolumn{1}{l}{} & \multicolumn{1}{l}{} &  &  \\
\textbf{} &  &  &  &  &  &  & \multicolumn{1}{l}{} & \multicolumn{1}{l}{} &  &  \\
\textbf{} &  &  &  &  &  &  & \multicolumn{1}{l}{} & \multicolumn{1}{l}{} &  &  \\
\textbf{} &  &  &  &  &  &  & \multicolumn{1}{l}{} & \multicolumn{1}{l}{} &  &  \\
\textbf{} &  &  &  &  &  &  & \multicolumn{1}{l}{} & \multicolumn{1}{l}{} &  &  \\
\textbf{} &  &  &  &  &  &  & \multicolumn{1}{l}{} & \multicolumn{1}{l}{} &  &  \\
\textbf{} &  &  &  &  &  &  & \multicolumn{1}{l}{} & \multicolumn{1}{l}{} &  &  \\
\textbf{} &  &  &  &  &  &  & \multicolumn{1}{l}{} & \multicolumn{1}{l}{} &  &  \\
\textbf{} &  &  &  &  &  &  & \multicolumn{1}{l}{} & \multicolumn{1}{l}{} &  &  \\
\textbf{} &  &  &  &  &  &  & \multicolumn{1}{l}{} & \multicolumn{1}{l}{} &  &  \\
 & \multicolumn{1}{l}{} & \multicolumn{1}{l}{} & \multicolumn{1}{l}{} & \multicolumn{1}{l}{} &  &  & \multicolumn{1}{l}{} & \multicolumn{1}{l}{} &  &  \\
\textbf{} & \multicolumn{2}{c}{\textbf{}} & \multicolumn{2}{c}{\textbf{}} &  &  & \multicolumn{1}{l}{} & \multicolumn{1}{l}{} &  &  \\
\textbf{} & \textbf{} & \textbf{} & \textbf{} & \textbf{} &  &  & \multicolumn{1}{l}{} & \multicolumn{1}{l}{} &  &  \\
\textbf{} &  &  &  &  &  &  & \multicolumn{1}{l}{} & \multicolumn{1}{l}{} &  &  \\
\textbf{} &  &  &  &  &  &  & \multicolumn{1}{l}{} & \multicolumn{1}{l}{} &  &  \\
\textbf{} &  &  &  &  &  &  & \multicolumn{1}{l}{} & \multicolumn{1}{l}{} &  &  \\
\textbf{} &  &  &  &  &  &  & \multicolumn{1}{l}{} & \multicolumn{1}{l}{} &  &  \\
\textbf{} &  &  &  &  &  &  & \multicolumn{1}{l}{} & \multicolumn{1}{l}{} &  &  \\
\textbf{} &  &  &  &  &  &  & \multicolumn{1}{l}{} & \multicolumn{1}{l}{} &  &  \\
\textbf{} &  &  &  &  &  &  & \multicolumn{1}{l}{} & \multicolumn{1}{l}{} &  &  \\
\textbf{} &  &  &  &  &  &  & \multicolumn{1}{l}{} & \multicolumn{1}{l}{} &  &  \\
\textbf{} &  &  &  &  &  &  & \multicolumn{1}{l}{} & \multicolumn{1}{l}{} &  &  \\
\textbf{} &  &  &  &  &  &  & \multicolumn{1}{l}{} & \multicolumn{1}{l}{} &  &  \\
 & \multicolumn{1}{l}{} & \multicolumn{1}{l}{} & \multicolumn{1}{l}{} & \multicolumn{1}{l}{} &  &  & \multicolumn{1}{l}{} & \multicolumn{1}{l}{} &  &  \\
\textbf{} & \multicolumn{2}{c}{\textbf{}} & \multicolumn{1}{l}{} & \multicolumn{1}{l}{} &  &  & \multicolumn{1}{l}{} & \multicolumn{1}{l}{} &  &  \\
\textbf{} & \textbf{} & \textbf{} & \multicolumn{1}{l}{} & \multicolumn{1}{l}{} &  &  & \multicolumn{1}{l}{} & \multicolumn{1}{l}{} &  &  \\
\textbf{} &  &  & \multicolumn{1}{l}{} & \multicolumn{1}{l}{} &  &  & \multicolumn{1}{l}{} & \multicolumn{1}{l}{} &  &  \\
\textbf{} &  &  & \multicolumn{1}{l}{} & \multicolumn{1}{l}{} &  &  & \multicolumn{1}{l}{} & \multicolumn{1}{l}{} &  &  \\
\textbf{} &  &  & \multicolumn{1}{l}{} & \multicolumn{1}{l}{} &  &  & \multicolumn{1}{l}{} & \multicolumn{1}{l}{} &  &  \\
\textbf{} &  &  & \multicolumn{1}{l}{} & \multicolumn{1}{l}{} &  &  & \multicolumn{1}{l}{} & \multicolumn{1}{l}{} &  &  \\
\textbf{} &  &  & \multicolumn{1}{l}{} & \multicolumn{1}{l}{} &  &  & \multicolumn{1}{l}{} & \multicolumn{1}{l}{} &  &  \\
\textbf{} &  &  & \multicolumn{1}{l}{} & \multicolumn{1}{l}{} &  &  & \multicolumn{1}{l}{} & \multicolumn{1}{l}{} &  &  \\
\textbf{} &  &  & \multicolumn{1}{l}{} & \multicolumn{1}{l}{} &  &  & \multicolumn{1}{l}{} & \multicolumn{1}{l}{} &  &  \\
\textbf{} &  &  & \multicolumn{1}{l}{} & \multicolumn{1}{l}{} &  &  & \multicolumn{1}{l}{} & \multicolumn{1}{l}{} &  &  \\
\textbf{} &  &  & \multicolumn{1}{l}{} & \multicolumn{1}{l}{} &  &  & \multicolumn{1}{l}{} & \multicolumn{1}{l}{} &  &  \\
\textbf{} &  &  & \multicolumn{1}{l}{} & \multicolumn{1}{l}{} &  &  & \multicolumn{1}{l}{} & \multicolumn{1}{l}{} &  &  \\ \cline{1-5}
\end{tabular}
\label{tab:quotastab}
\end{table}

\subsubsection{Volunteers instructions}\label{sec:volinstrappendix}
Below, the instructions for volunteers are presented: \\

\begin{itemize}
\item Fill Google form to record the results - one for each respondent.
\item Follow quotas (see Table \ref{tab:quotastab}).
\item Select random people - for example one in three.
\item If you approach a group of people, choose one, for example, who has the closest birthday
\item Survey in places with maximum heterogeneity of the audience:
    \begin{itemize}
        \item near supermarkets;
        \item along the street;
        \item parks;
        \item change place during the surveying.
    \end{itemize}

\item Record the number of people who refuse to answer. It is important that there are as few such people as possible. Make an emphasis on anonymity and safety. Try to start a dialog in any way, after this, people are more likely to respond to the questions.
\item Optionally, make an audio record of the interviews, this will make the data more reliable.
\item Try to carry the conversation out politely and tactfully. You can start it with distracted topics to make it easier for the respondents. Introduce yourself as an independent civic initiative or say that you want to conduct a survey on your own in your town.

\end{itemize}

As a result, we expected from volunteers:
\begin{itemize}
    \item Completed forms (questionnaires) - 30 people, taking into account quotas (it was very desirable to collect such a number but it was allowed to collect more. Less was also allowed if the volunteer had no possibility to reach the quota).
    \item Number of people who refused to answer.
    \item Time and place of the survey (for example, Minsk, 2nd of August, from 13 to 15).

\end{itemize}

\clearpage

\section{Further results}\label{sec:resultsappendix}

\begin{table}[!htb]
 \caption{Official election results and posterior statistics of interest for the poststratified distribution on the population level for the alternative model (II). Here, $g_x$ means $x$'s quantile of the focus posterior distribution. For the election results, the notation ${}^{***}$ means outside the 99.9\% credible interval, ${}^{**}$ means outside the 99\% credible interval, ${}^{*}$ means outside the 95\% credible interval, and ${}^{.}$ means outside the 90\% credible interval  of the parameter with respect to its posterior distribution. The anomalies in the official results found on either of the addressed credible levels are bold.}
  \centering
    \addtolength{\tabcolsep}{-3pt}
  \small
  \begin{tabular}{llrrrrrrrrrrr}
    \toprule
Event&Official&Mean&Mode&Median&$q_{0.0500}$&$q_{0.9500}$&$q_{0.0250}$&$q_{0.9750}$&$q_{0.0050}$&$q_{0.9950}$&$q_{0.0005}$&$q_{0.9995}$\\\hline
Lukashenka&\textbf{0.8010}${}^{***}$&0.1560&0.1561&0.1560&0.1378&0.1753&0.1344&0.1791&0.1280&0.1866&0.1207&0.1954\\
Tsikhanouskaia&\textbf{0.1012}${}^{***}$&0.7754&0.7753&0.7753&0.7556&0.7944&0.7517&0.7979&0.7441&0.8048&0.7352&0.8127\\
Dmitriyeu&\textbf{0.0120}${}^{**}$&0.0041&0.0043&0.0042&0.0022&0.0077&0.0019&0.0087&0.0015&0.0108&0.0011&0.0139\\
Cherachen&0.0114&0.0077&0.0079&0.0078&0.0047&0.0127&0.0042&0.0139&0.0035&0.0165&0.0027&0.0202\\
Kanapatskaya&\textbf{0.0167}${}^{***}$&0.0011&0.0013&0.0012&0.0004&0.0034&0.0003&0.0041&0.0002&0.0060&0.0001&0.0094\\
Against all&\textbf{0.0459}${}^{.}$&0.0577&0.0579&0.0578&0.0473&0.0700&0.0454&0.0726&0.0420&0.0778&0.0383&0.0842\\
Early voting&\textbf{0.4954}${}^{***}$&0.1069&0.1071&0.1070&0.0929&0.1225&0.0904&0.1256&0.0856&0.1319&0.0802&0.1395\\
    \bottomrule
  \end{tabular}
  \label{tab:other_models_preds1}
\end{table}

\begin{figure}[!htb]
  \centering
  \includegraphics[scale=0.58]{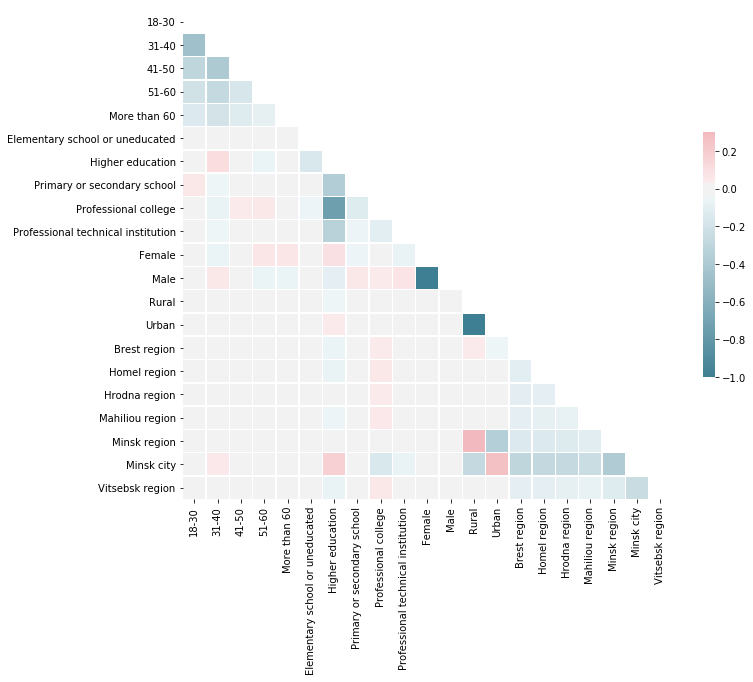}
  \caption{Correlations between all levels of categorical variables in the merged sample.}
  \label{fig:corplot_mult}
\end{figure}

\begin{table}[!htb]
 \caption{Official election results and posterior statistics of interest for the poststratified distribution on the population level for the alternative model (III). Here, $g_x$ means $x$'s quantile of the focus posterior distribution. For the election results, the notation ${}^{***}$ means outside the 99.9\% credible interval, ${}^{**}$ means outside the 99\% credible interval, ${}^{*}$ means outside the 95\% credible interval, and ${}^{.}$ means outside the 90\% credible interval  of the parameter with respect to its posterior distribution. The anomalies in the official results found on either of the addressed credible levels are bold.}
  \centering
    \addtolength{\tabcolsep}{-3pt}
  \small
  \begin{tabular}{llrrrrrrrrrrr}
    \toprule
Event&Official&Mean&Mode&Median&$q_{0.0500}$&$q_{0.9500}$&$q_{0.0250}$&$q_{0.9750}$&$q_{0.0050}$&$q_{0.9950}$&$q_{0.0005}$&$q_{0.9995}$\\\hline
Lukashenka&\textbf{0.8010}${}^{***}$&0.1559&0.1561&0.1560&0.1378&0.1752&0.1344&0.1790&0.1280&0.1865&0.1207&0.1953\\
Tsikhanouskaya&\textbf{0.1012}${}^{***}$&0.7754&0.7753&0.7754&0.7556&0.7944&0.7518&0.7980&0.7442&0.8049&0.7353&0.8127\\
Dmitriyeu&\textbf{0.0120}${}^{**}$&0.0041&0.0043&0.0042&0.0022&0.0077&0.0019&0.0086&0.0015&0.0108&0.0011&0.0139\\
Cherachen&0.0114&0.0077&0.0079&0.0078&0.0047&0.0127&0.0042&0.0139&0.0035&0.0166&0.0027&0.0203\\
Kanapatskaya&\textbf{0.0167}${}^{***}$&0.0011&0.0013&0.0012&0.0004&0.0034&0.0003&0.0041&0.0002&0.0060&0.0001&0.0095\\
Against all&\textbf{0.0459}${}^{.}$&0.0577&0.0579&0.0578&0.0473&0.0701&0.0454&0.0726&0.0420&0.0778&0.0383&0.0842\\
Early voting&\textbf{0.4954}${}^{***}$&0.1069&0.1071&0.1070&0.0929&0.1225&0.0904&0.1256&0.0855&0.1319&0.0802&0.1394\\
    \bottomrule
  \end{tabular}
  \label{tab:other_models_preds2}
\end{table}

\begin{figure}[!htb]
  \centering
  \includegraphics[scale=0.39]{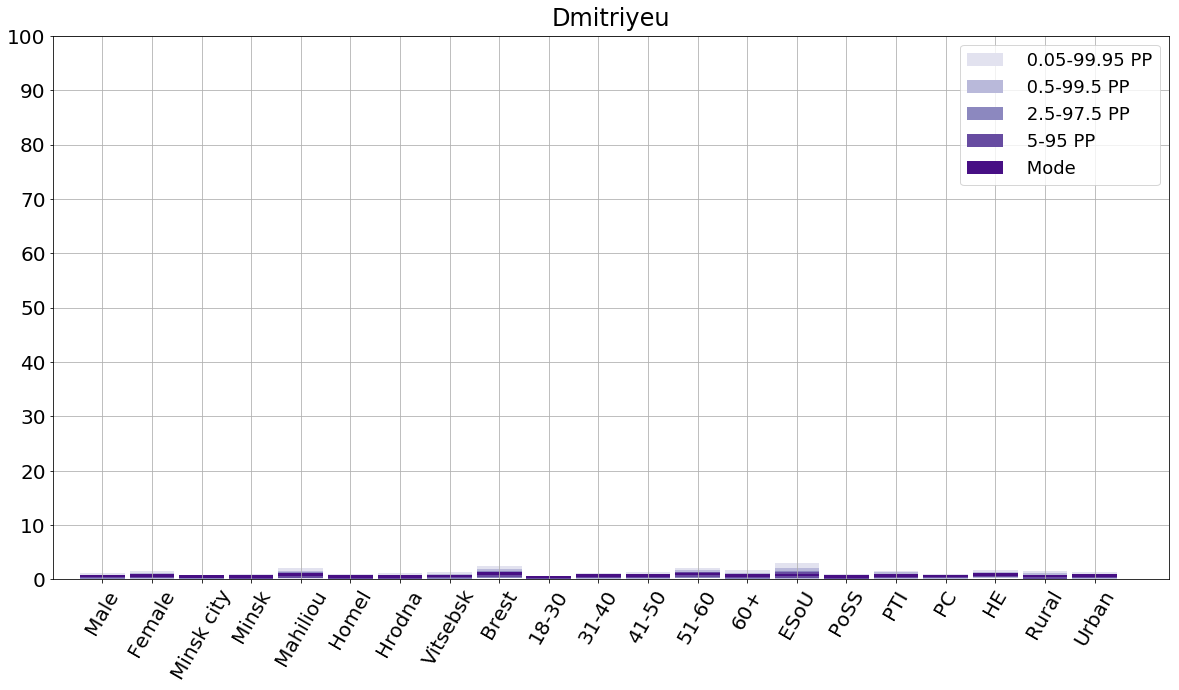}
  \caption{Posterior modes and quantiles for Dmitiyeu across all marginal subpopulations.}
  \label{fig:ratings_dimtr}
\end{figure}

\begin{figure}[!htb]
  \centering
  \includegraphics[scale=0.39]{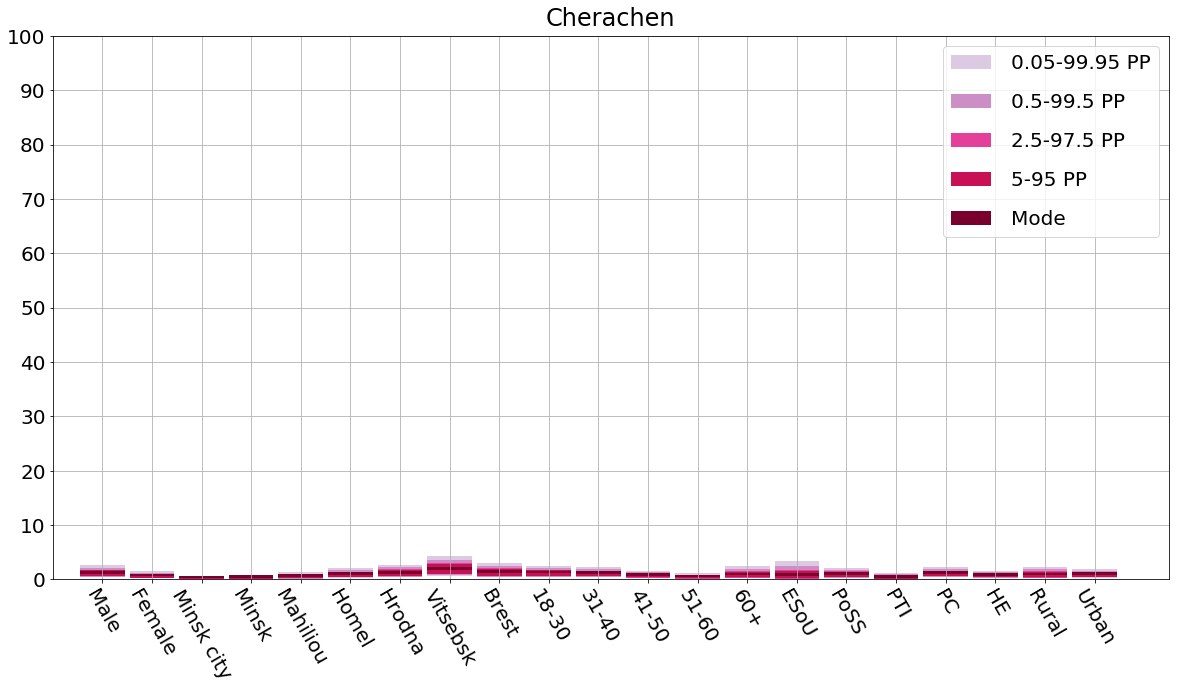}
  \caption{Posterior modes and quantiles for Cherachen and across all marginal subpopulations.}
  \label{fig:ratings_chera}
\end{figure}

\begin{figure}[!htb]
  \centering
    \includegraphics[scale=0.39]{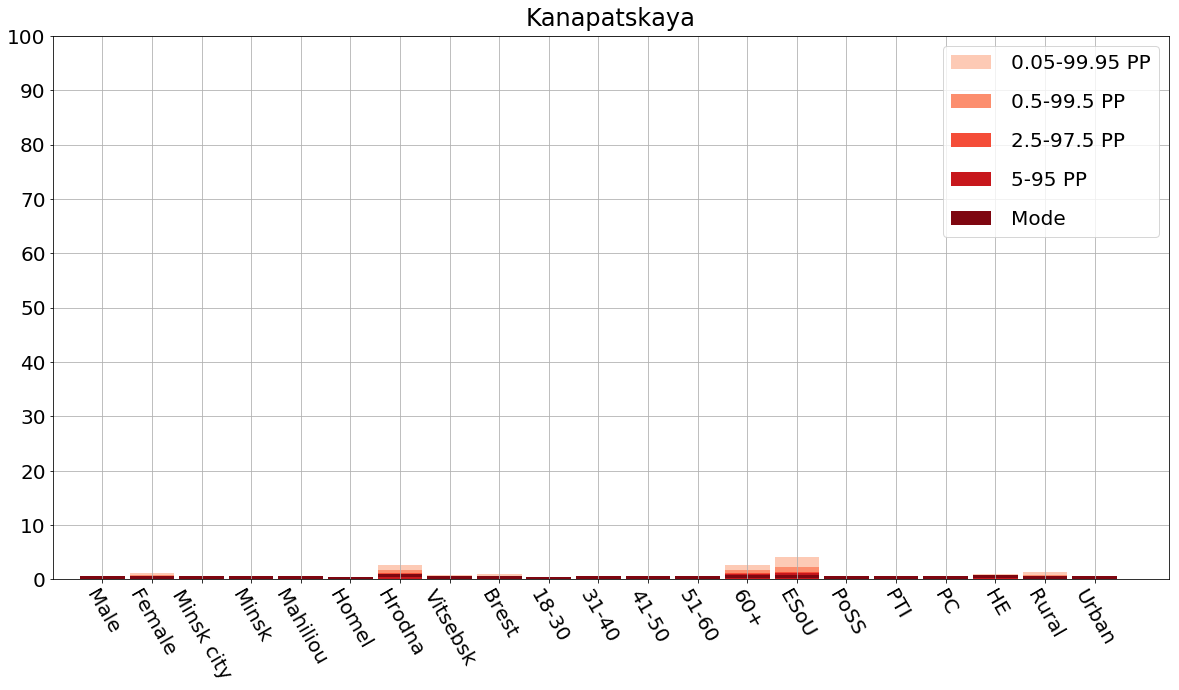}
  \caption{Posterior modes and quantiles for Kanapatskaya across all marginal subpopulations.}
  \label{fig:ratings_kana}
\end{figure}

\begin{figure}[!htb]
  \centering
  \includegraphics[scale=0.39]{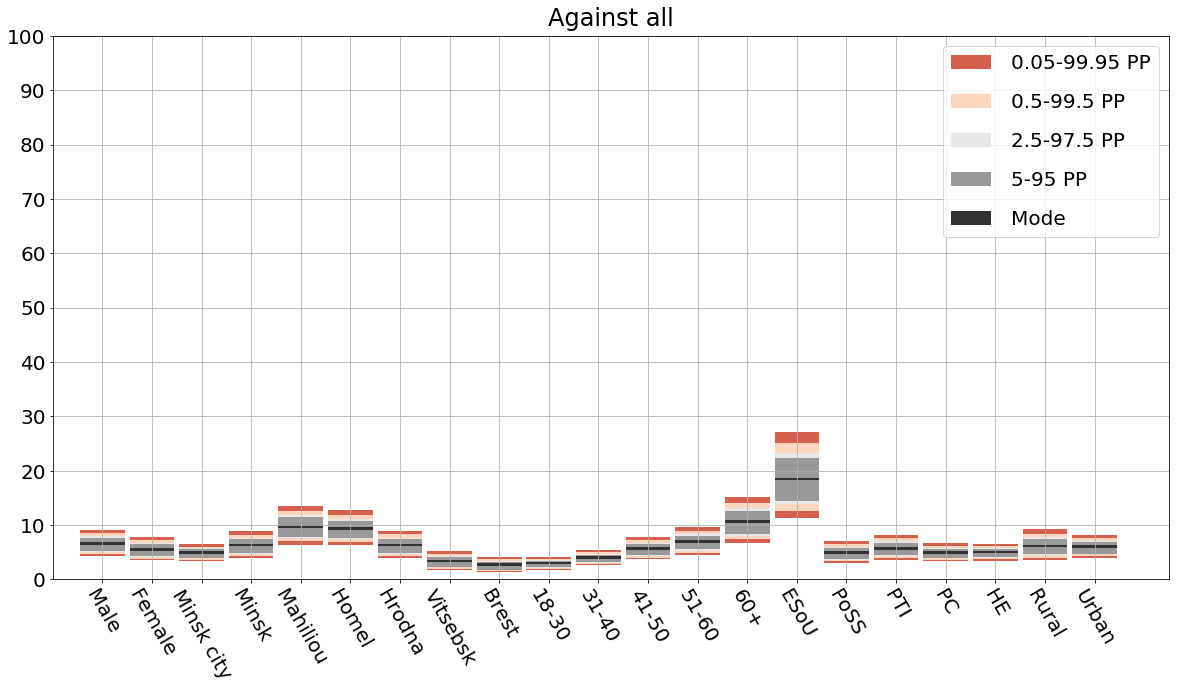}
  \caption{Posterior modes and quantiles for the votes {``}Against all{''}  across all marginal subpopulations.}
  \label{fig:ratings_nega}
\end{figure}

\begin{figure}[!htb]
  \centering
  \includegraphics[scale=0.39]{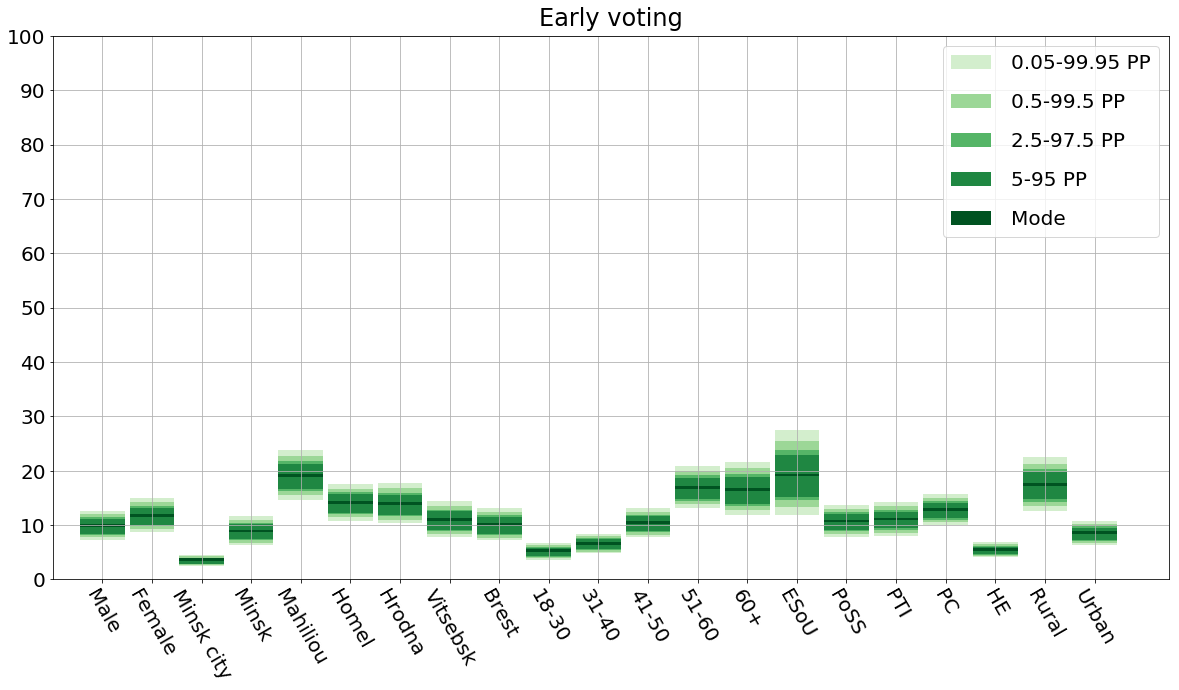}
  \caption{Posterior modes and quantiles for the early voting across all marginal subpopulations.}
  \label{fig:ratings_earl}
\end{figure}

\begin{figure}[!htb]
  \centering
  \includegraphics[scale=0.39]{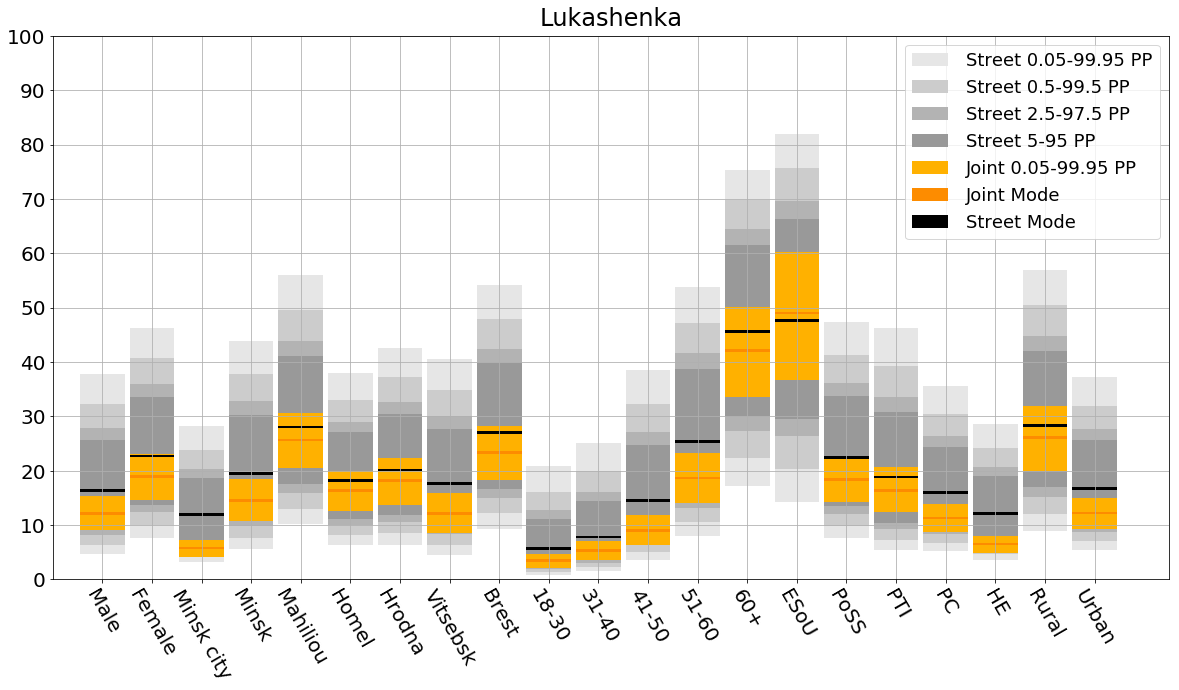}
  \caption{Posterior modes and quantiles for Lukashenka across all marginal subpopulations on the street and merged data.}
  \label{fig:ratings_luka_str}
\end{figure}

\begin{figure}[!htb]
  \centering
  \includegraphics[scale=0.39]{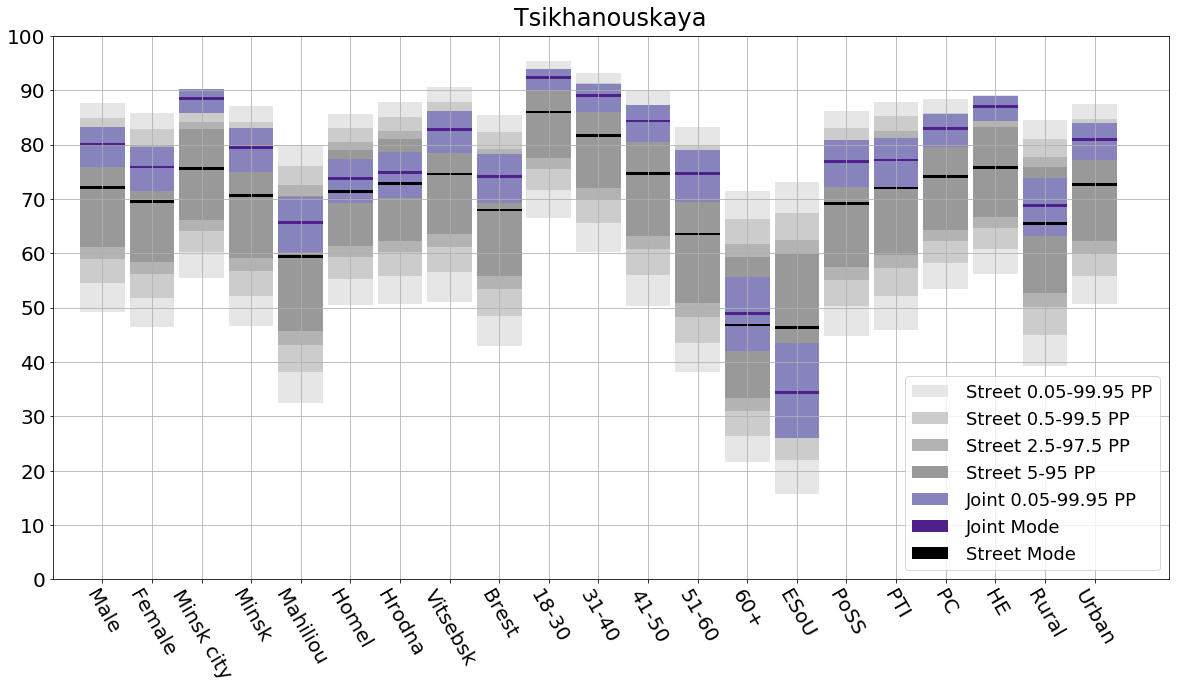}
  \caption{Posterior modes and quantiles for Tsikhanouskaya across all marginal subpopulations on the street and merged data.}
  \label{fig:ratings_tiha_str}
\end{figure}

\begin{figure}[!htb]
  \centering
  \includegraphics[scale=0.39]{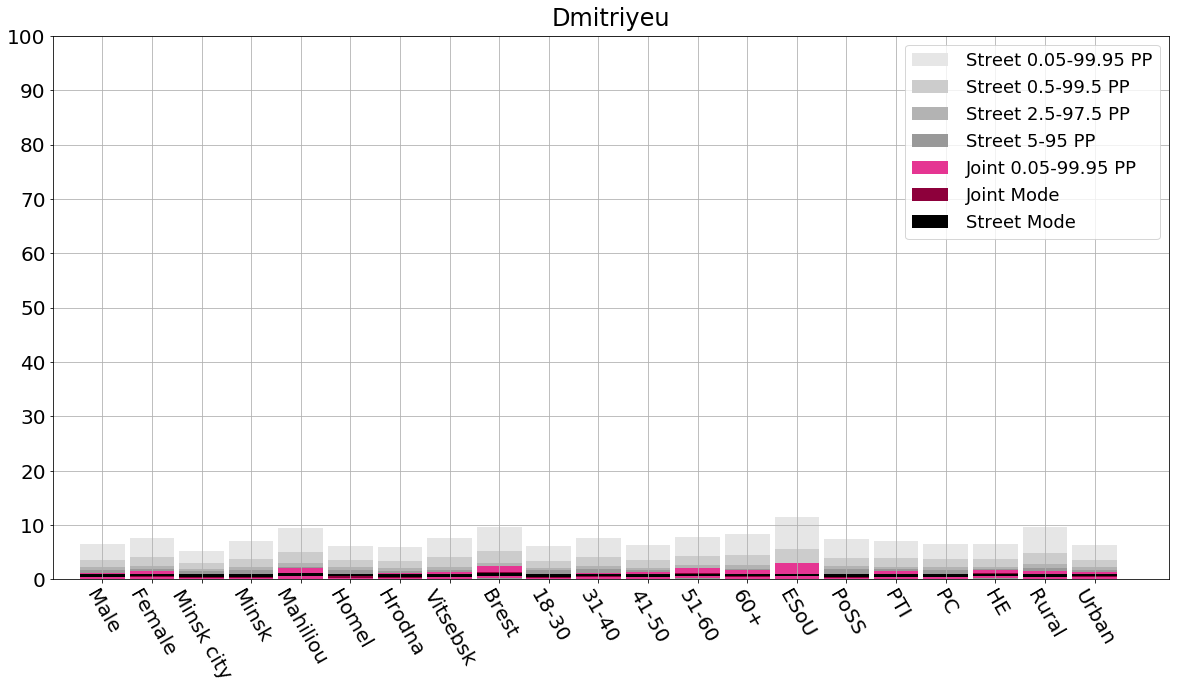}
  \caption{Posterior modes and quantiles for Dmitiyeu across all marginal subpopulations on the street and merged data.}
  \label{fig:ratings_dimtr_str}
\end{figure}

\begin{figure}[!htb]
  \centering
  \includegraphics[scale=0.39]{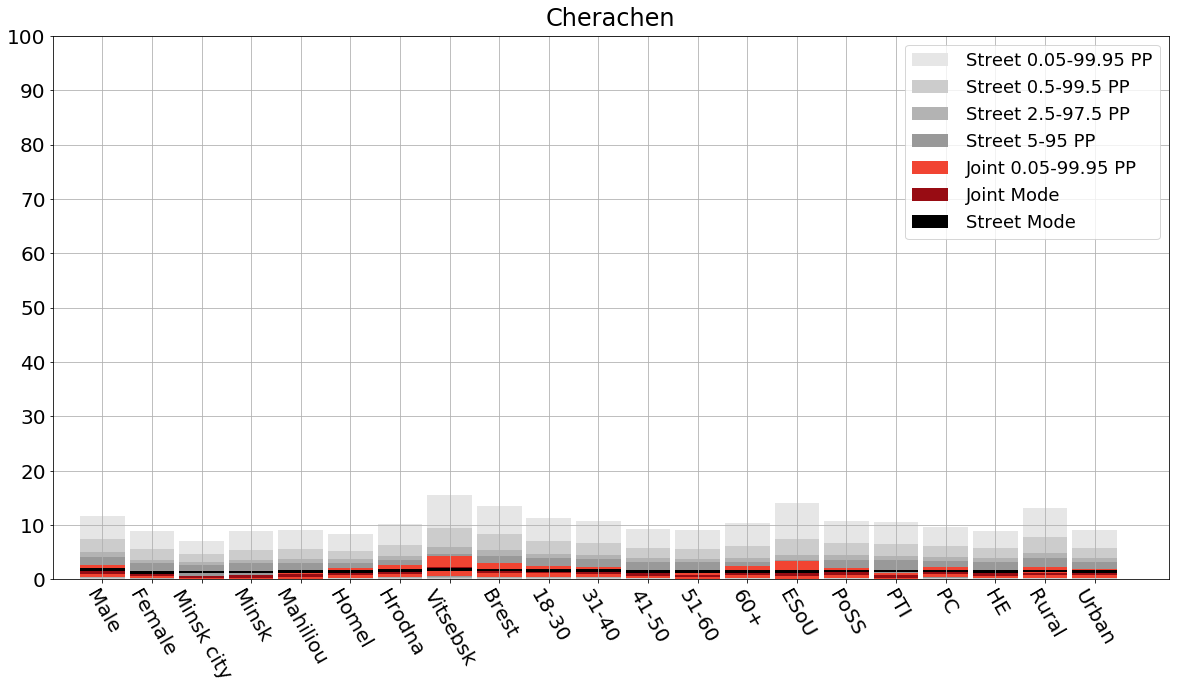}
  \caption{Posterior modes and quantiles for Cherachen and across all marginal subpopulations on the street and merged data.}
  \label{fig:ratings_chera_str}
\end{figure}

\begin{figure}[!htb]
  \centering
    \includegraphics[scale=0.39]{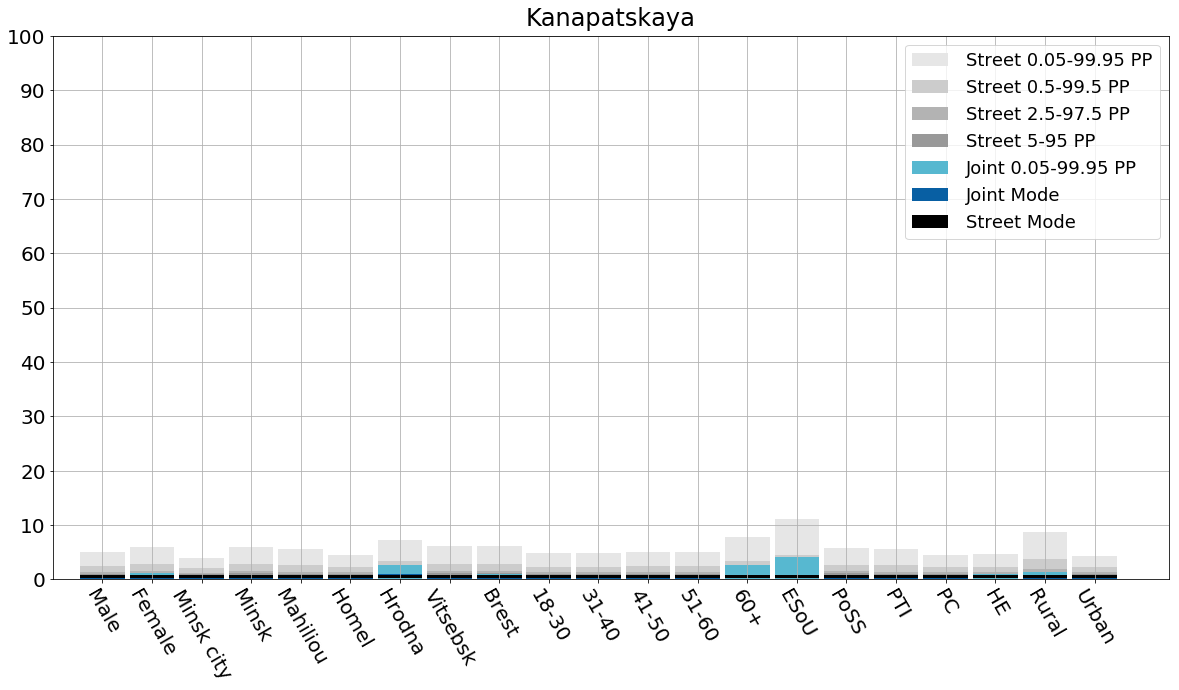}
  \caption{Posterior modes and quantiles for Kanapatskaya across all marginal subpopulations on the street and merged data.}
  \label{fig:ratings_kana_str}
\end{figure}

\begin{figure}[!htb]
  \centering
  \includegraphics[scale=0.39]{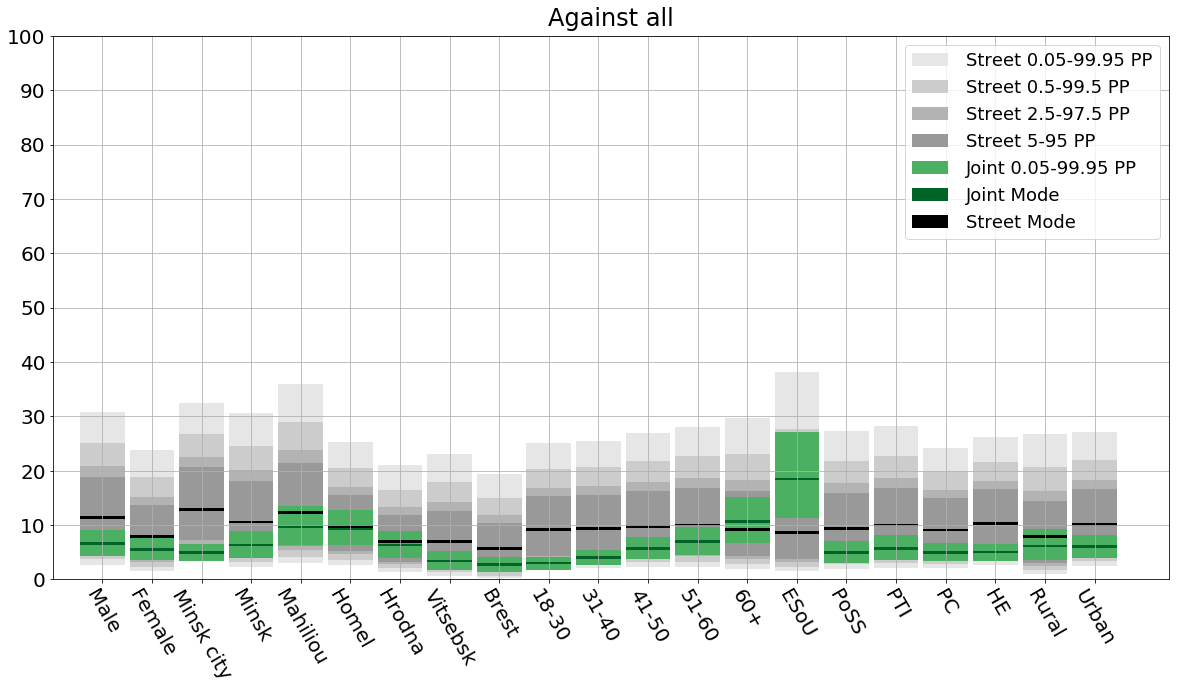}
  \caption{Posterior modes and quantiles for the votes Against all  across all marginal subpopulations on the street and merged data.}
  \label{fig:ratings_nega_str}
\end{figure}

\begin{figure}[!htb]
  \centering
  \includegraphics[scale=0.39]{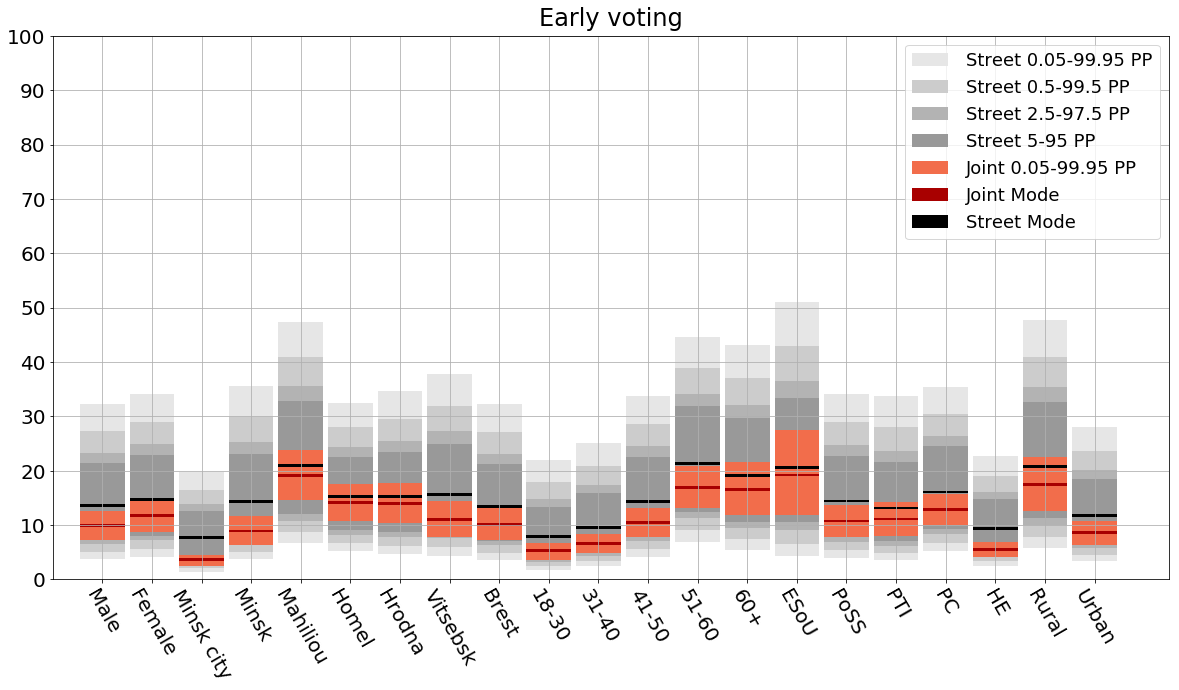}
  \caption{Posterior modes and quantiles for the early voting across all marginal subpopulations on the street and merged data.}
  \label{fig:ratings_earl_str}
\end{figure}

\begin{figure}[!htb]
  \centering
  \includegraphics[scale=0.28]{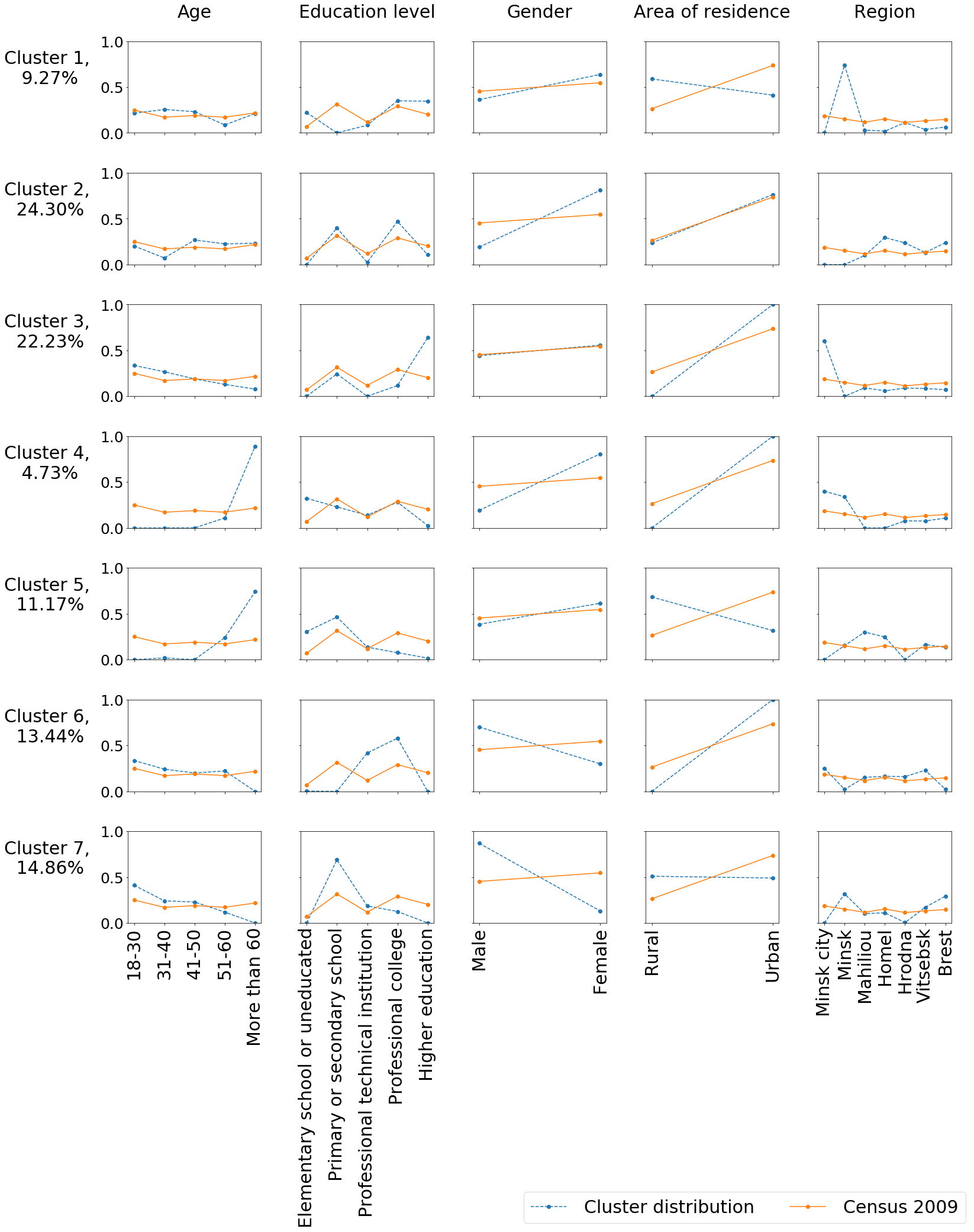}
  \caption{Subpopulations corresponding to the seven clusters obtained by the mixtures of multinomials. The poststratified shares of the population belonging to these clusters are shown on the right of each row.}
  \label{fig:democlust}
\end{figure}

 \begin{figure}[!htb]
 \centering
   \includegraphics[scale=0.28]{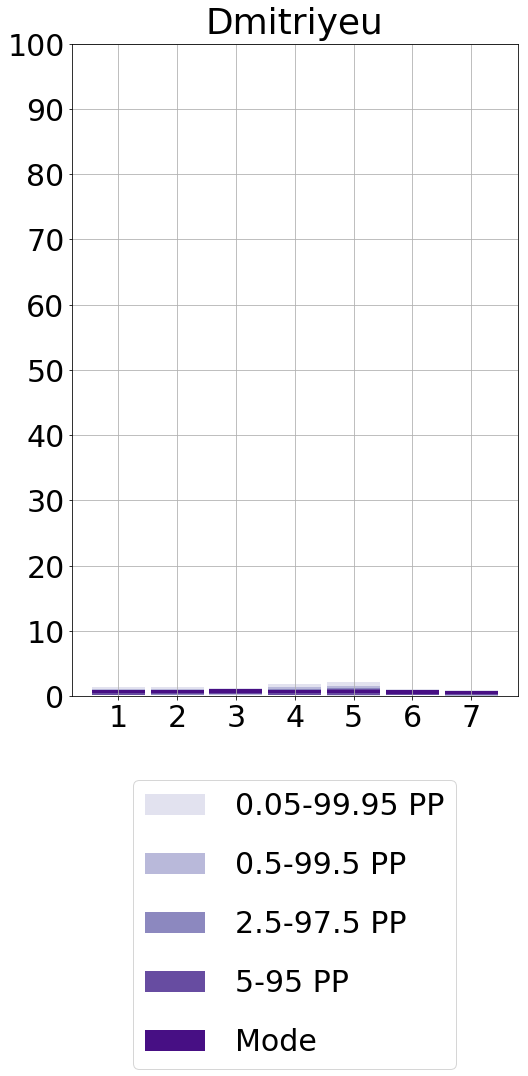}
   \includegraphics[scale = 0.28]{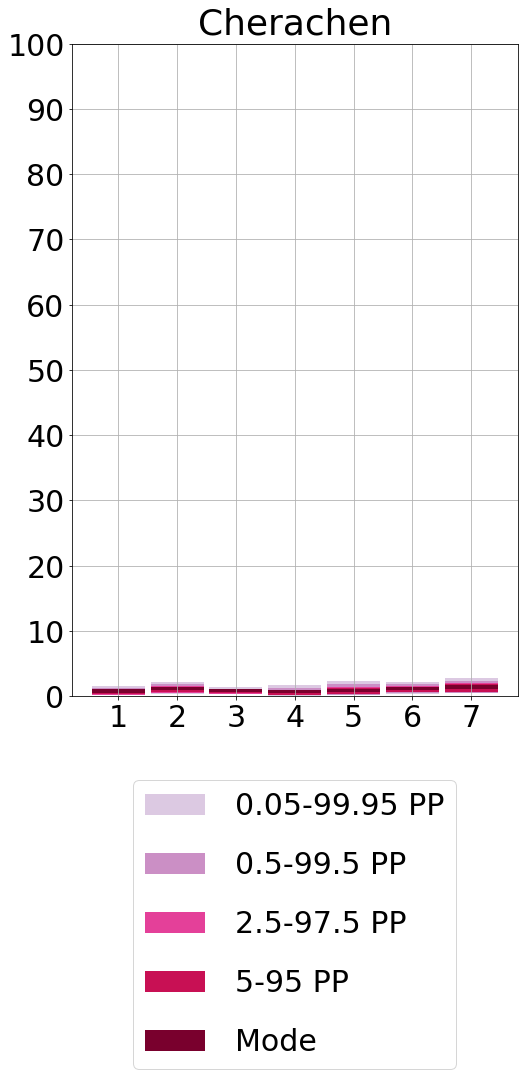}
   \includegraphics[scale=0.28]{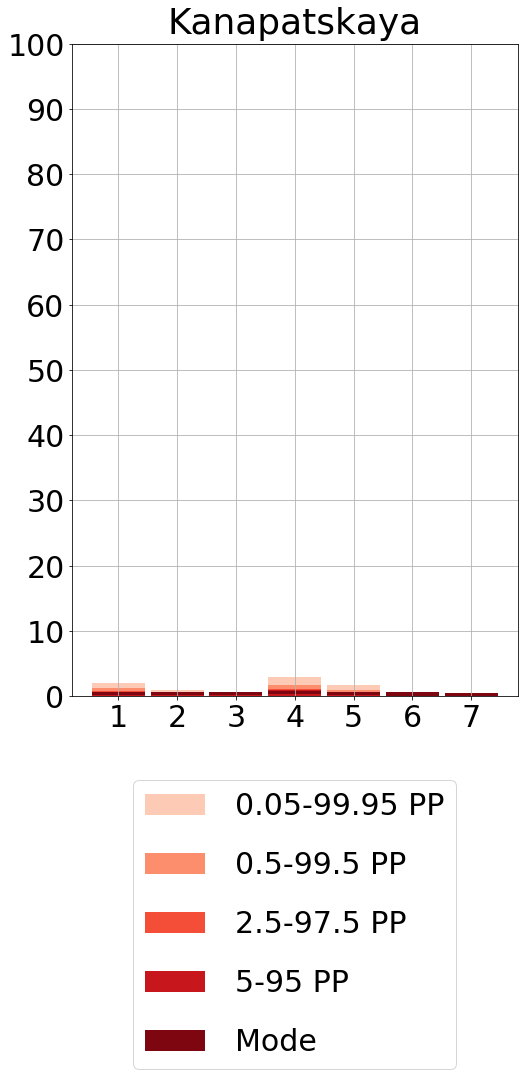}
   \includegraphics[scale = 0.28]{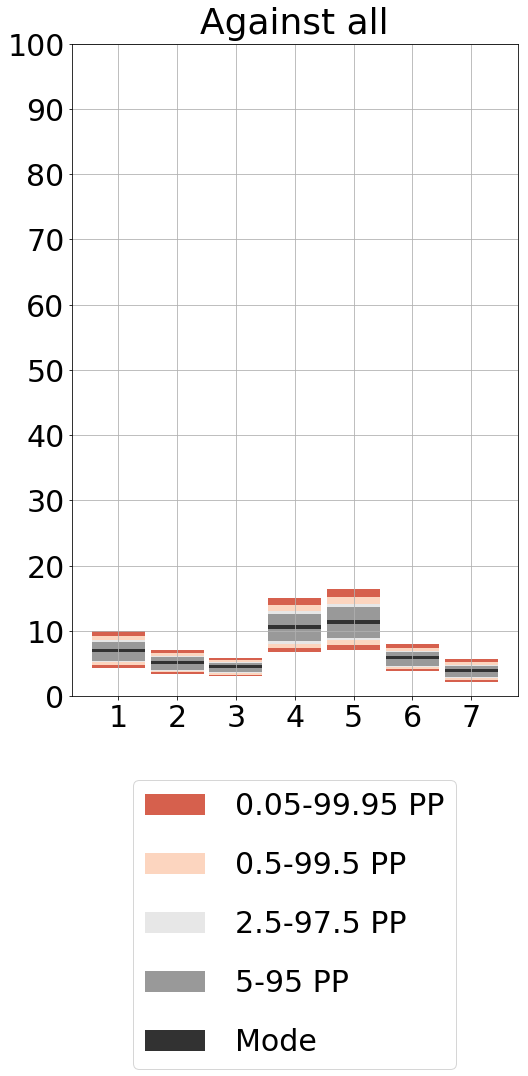}
   \includegraphics[scale = 0.28]{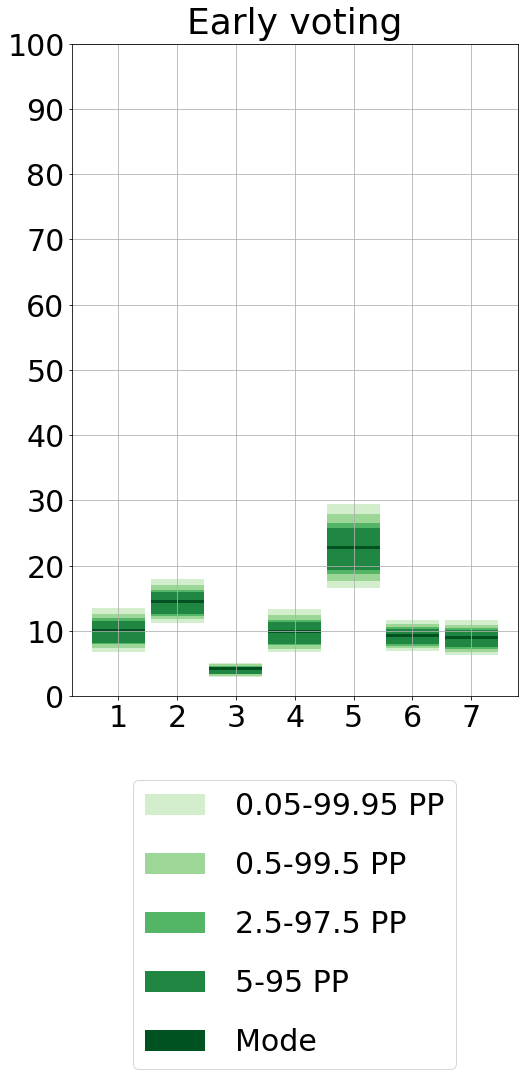}
   \caption{Posterior modes and quantiles for the support of Dmitriyeu, Cherachen, Kanapatskaya (top row) and Agaist all, Early voting (bottom row) on the level on subpopulations corresponding to the 7 found clusters.}
   \label{fig:othersclust}
 \end{figure}

\clearpage
\begin{table}[!htb]

\caption{Random effects for Lukashenka’s model}

\begin{adjustbox}{max width=\textwidth}
\begin{tabular}{lrrrrrrrrrrrrrr}
  \hline
 & mean & sd & 0.025quant & 0.5quant & 0.975quant & 5e-04quant & 0.9995quant & 0.005quant & 0.995quant & 0.05quant & 0.95quant & mode \\ 
  \hline
BYM Brest region 1 & 0.14 & 0.27 & -0.32 & 0.09 & 0.76 & -0.90 & 1.35 & -0.56 & 1.01 & -0.22 & 0.64 & 0.02 \\ 
  BYM Homel region 1 & -0.01 & 0.24 & -0.52 & -0.00 & 0.49 & -1.16 & 1.07 & -0.79 & 0.74 & -0.41 & 0.38 & 0.00 \\ 
  BYM Hrodna region 1 & 0.00 & 0.24 & -0.51 & 0.00 & 0.50 & -1.13 & 1.09 & -0.77 & 0.75 & -0.39 & 0.39 & 0.00 \\ 
  BYM Mahiliou region 1 & 0.21 & 0.30 & -0.26 & 0.13 & 0.91 & -0.81 & 1.50 & -0.49 & 1.16 & -0.17 & 0.79 & 0.02 \\ 
  BYM Minsk region 1 & -0.24 & 0.29 & -0.91 & -0.17 & 0.16 & -1.52 & 0.60 & -1.15 & 0.33 & -0.79 & 0.11 & -0.02 \\ 
  BYM Vitsebsk region 1 & -0.17 & 0.29 & -0.84 & -0.11 & 0.29 & -1.51 & 0.83 & -1.12 & 0.51 & -0.72 & 0.20 & -0.02 \\ 
  BYM Brest region 2 & 0.26 & 0.99 & -1.75 & 0.29 & 2.16 & -3.17 & 3.50 & -2.41 & 2.77 & -1.41 & 1.84 & 0.37 \\ 
  BYM Homel region 2 & 0.07 & 0.97 & -1.86 & 0.07 & 1.99 & -3.26 & 3.37 & -2.50 & 2.63 & -1.54 & 1.67 & 0.06 \\ 
  BYM Hrodna region 2 & -0.03 & 0.97 & -1.96 & -0.03 & 1.89 & -3.34 & 3.28 & -2.59 & 2.54 & -1.63 & 1.57 & -0.03 \\ 
  BYM Mahiliou region 2 & 0.34 & 1.01 & -1.71 & 0.37 & 2.24 & -3.15 & 3.55 & -2.38 & 2.84 & -1.37 & 1.94 & 0.48 \\ 
  BYM Minsk region 2 & -0.32 & 0.76 & -1.77 & -0.34 & 1.22 & -2.73 & 2.27 & -2.21 & 1.71 & -1.55 & 0.96 & -0.40 \\ 
  BYM Vitsebsk region 2 & -0.31 & 1.00 & -2.21 & -0.34 & 1.72 & -3.54 & 3.15 & -2.82 & 2.38 & -1.90 & 1.38 & -0.43 \\ 
  id & -0.45 & 0.72 & -2.24 & -0.20 & 0.56 & -3.75 & 2.00 & -2.92 & 1.14 & -1.90 & 0.35 & -0.03 \\ 
  18-30 & -2.40 & 0.85 & -4.05 & -2.40 & -0.80 & -5.56 & 0.48 & -4.67 & -0.29 & -3.78 & -1.05 & -2.37 \\ 
  31-40 & -1.93 & 0.85 & -3.57 & -1.93 & -0.33 & -5.08 & 0.96 & -4.19 & 0.18 & -3.30 & -0.58 & -1.90 \\ 
  41-50 & -1.36 & 0.85 & -3.00 & -1.35 & 0.24 & -4.51 & 1.53 & -3.62 & 0.75 & -2.73 & -0.01 & -1.33 \\ 
  51-60 & -0.46 & 0.85 & -2.10 & -0.46 & 1.13 & -3.61 & 2.42 & -2.72 & 1.65 & -1.83 & 0.89 & -0.44 \\ 
  More than 60 & 0.60 & 0.84 & -1.04 & 0.60 & 2.19 & -2.55 & 3.48 & -1.66 & 2.70 & -0.77 & 1.95 & 0.62 \\ 
  Elementary school or uneducated & 0.01 & 0.56 & -1.46 & 0.10 & 0.91 & -2.84 & 2.18 & -2.12 & 1.40 & -1.10 & 0.73 & 0.16 \\ 
  Primary or secondary school & 0.01 & 0.56 & -1.47 & 0.10 & 0.89 & -2.84 & 2.17 & -2.13 & 1.39 & -1.11 & 0.71 & 0.16 \\ 
  Professional technical institution & 0.31 & 0.55 & -1.16 & 0.40 & 1.19 & -2.53 & 2.47 & -1.82 & 1.68 & -0.81 & 1.01 & 0.46 \\ 
  Professional college & -0.35 & 0.55 & -1.83 & -0.26 & 0.53 & -3.20 & 1.80 & -2.49 & 1.02 & -1.47 & 0.34 & -0.19 \\ 
  Higher education & -0.95 & 0.56 & -2.43 & -0.86 & -0.08 & -3.81 & 1.19 & -3.09 & 0.42 & -2.07 & -0.26 & -0.79 \\ 
  Female & 0.06 & 0.36 & -0.88 & 0.11 & 0.64 & -2.13 & 1.52 & -1.44 & 0.98 & -0.62 & 0.51 & 0.15 \\ 
  Male & -0.30 & 0.36 & -1.25 & -0.24 & 0.28 & -2.50 & 1.16 & -1.81 & 0.62 & -0.99 & 0.15 & -0.19 \\ 
  Urban & -0.35 & 0.38 & -1.32 & -0.28 & 0.30 & -2.14 & 1.15 & -1.70 & 0.68 & -1.12 & 0.14 & -0.23 \\ 
  Rural & 0.07 & 0.38 & -0.89 & 0.14 & 0.73 & -1.71 & 1.58 & -1.27 & 1.11 & -0.68 & 0.57 & 0.18 \\ 
  Brest region & 0.34 & 0.33 & -0.34 & 0.35 & 0.97 & -1.03 & 1.54 & -0.62 & 1.20 & -0.22 & 0.86 & 0.39 \\ 
  Vitsebsk region & -0.36 & 0.33 & -1.03 & -0.36 & 0.29 & -1.81 & 0.89 & -1.33 & 0.54 & -0.90 & 0.18 & -0.37 \\ 
  Homel region & -0.04 & 0.30 & -0.66 & -0.03 & 0.54 & -1.40 & 1.14 & -0.96 & 0.79 & -0.53 & 0.43 & -0.02 \\ 
  Hrodna region & 0.05 & 0.30 & -0.58 & 0.05 & 0.63 & -1.31 & 1.23 & -0.87 & 0.88 & -0.45 & 0.52 & 0.06 \\ 
  Minsk city & -0.68 & 0.32 & -1.32 & -0.68 & -0.08 & -2.08 & 0.38 & -1.59 & 0.11 & -1.20 & -0.17 & -0.70 \\ 
  Minsk region & -0.11 & 0.32 & -0.74 & -0.10 & 0.48 & -1.50 & 0.95 & -1.01 & 0.67 & -0.63 & 0.39 & -0.11 \\ 
  Mahiliou region & 0.51 & 0.37 & -0.23 & 0.53 & 1.17 & -0.90 & 1.75 & -0.50 & 1.41 & -0.11 & 1.07 & 0.62 \\ 
   \hline
\end{tabular}
\end{adjustbox}
\end{table}

\begin{table}[!htb]

\caption{Random effects for Tsikhanouskaya’s model}

\begin{adjustbox}{max width=\textwidth}
\begin{tabular}{lrrrrrrrrrrrrrr}
 \hline
 & mean & sd & 0.025quant & 0.5quant & 0.975quant & 5e-04quant & 0.9995quant & 0.005quant & 0.995quant & 0.05quant & 0.95quant & mode \\ 
  \hline
BYM Brest region 1 & -0.04 & 0.21 & -0.47 & -0.03 & 0.39 & -0.96 & 0.92 & -0.67 & 0.59 & -0.39 & 0.30 & -0.02 \\ 
  BYM Homel region 1 & -0.08 & 0.21 & -0.53 & -0.07 & 0.34 & -1.01 & 0.87 & -0.72 & 0.54 & -0.44 & 0.25 & -0.03 \\ 
  BYM Hrodna region 1 & 0.00 & 0.21 & -0.42 & 0.00 & 0.44 & -0.91 & 0.97 & -0.61 & 0.64 & -0.34 & 0.34 & 0.00 \\ 
  BYM Mahiliou region 1 & -0.25 & 0.27 & -0.83 & -0.22 & 0.19 & -1.30 & 0.71 & -1.01 & 0.39 & -0.73 & 0.12 & -0.05 \\ 
  BYM Minsk region 1 & 0.22 & 0.23 & -0.15 & 0.19 & 0.70 & -0.56 & 1.23 & -0.31 & 0.90 & -0.10 & 0.62 & 0.05 \\ 
  BYM Vitsebsk region 1 & 0.21 & 0.26 & -0.23 & 0.17 & 0.77 & -0.72 & 1.33 & -0.42 & 0.98 & -0.15 & 0.67 & 0.05 \\ 
  BYM Brest region 2 & -0.09 & 0.93 & -1.94 & -0.10 & 1.80 & -3.30 & 3.20 & -2.56 & 2.44 & -1.62 & 1.47 & -0.13 \\ 
  BYM Homel region 2 & -0.26 & 0.94 & -2.09 & -0.28 & 1.68 & -3.43 & 3.10 & -2.71 & 2.33 & -1.79 & 1.34 & -0.32 \\ 
  BYM Hrodna region 2 & 0.08 & 0.93 & -1.78 & 0.08 & 1.95 & -3.18 & 3.32 & -2.43 & 2.58 & -1.46 & 1.63 & 0.06 \\ 
  BYM Mahiliou region 2 & -0.54 & 1.01 & -2.41 & -0.60 & 1.54 & -3.67 & 3.01 & -2.98 & 2.22 & -2.11 & 1.19 & -0.79 \\ 
  BYM Minsk region 2 & 0.37 & 0.75 & -1.17 & 0.40 & 1.76 & -2.23 & 2.70 & -1.66 & 2.19 & -0.91 & 1.54 & 0.50 \\ 
  BYM Vitsebsk region 2 & 0.45 & 0.98 & -1.59 & 0.49 & 2.28 & -3.04 & 3.57 & -2.26 & 2.87 & -1.24 & 1.99 & 0.64 \\ 
  id & 0.30 & 0.54 & -0.55 & 0.17 & 1.61 & -1.72 & 2.96 & -0.99 & 2.16 & -0.38 & 1.35 & 0.02 \\ 
  18-30 & 1.61 & 0.61 & 0.45 & 1.62 & 2.80 & -0.75 & 4.06 & -0.01 & 3.31 & 0.64 & 2.59 & 1.66 \\ 
  31-40 & 1.22 & 0.61 & 0.05 & 1.22 & 2.41 & -1.15 & 3.66 & -0.41 & 2.92 & 0.25 & 2.20 & 1.27 \\ 
  41-50 & 0.83 & 0.61 & -0.34 & 0.83 & 2.01 & -1.54 & 3.27 & -0.80 & 2.52 & -0.15 & 1.80 & 0.87 \\ 
  51-60 & 0.21 & 0.61 & -0.96 & 0.21 & 1.40 & -2.16 & 2.65 & -1.42 & 1.91 & -0.76 & 1.18 & 0.26 \\ 
  More than 60 & -0.68 & 0.61 & -1.85 & -0.68 & 0.50 & -3.05 & 1.75 & -2.31 & 1.01 & -1.66 & 0.29 & -0.64 \\ 
  Elementary school or uneducated & -0.48 & 0.47 & -1.26 & -0.54 & 0.69 & -2.35 & 1.84 & -1.63 & 1.20 & -1.12 & 0.42 & -0.61 \\ 
  Primary or secondary school & 0.29 & 0.47 & -0.47 & 0.22 & 1.46 & -1.56 & 2.63 & -0.86 & 1.99 & -0.33 & 1.19 & 0.13 \\ 
  Professional technical institution & 0.04 & 0.47 & -0.73 & -0.03 & 1.21 & -1.81 & 2.37 & -1.11 & 1.73 & -0.59 & 0.94 & -0.13 \\ 
  Professional college & 0.50 & 0.47 & -0.25 & 0.44 & 1.68 & -1.34 & 2.84 & -0.64 & 2.20 & -0.11 & 1.41 & 0.33 \\ 
  Higher education & 0.80 & 0.47 & 0.04 & 0.73 & 1.97 & -1.05 & 3.13 & -0.34 & 2.49 & 0.18 & 1.70 & 0.62 \\ 
  Female & 0.01 & 0.13 & -0.27 & 0.00 & 0.32 & -0.54 & 0.58 & -0.39 & 0.44 & -0.20 & 0.25 & -0.00 \\ 
  Male & 0.02 & 0.13 & -0.25 & 0.01 & 0.34 & -0.52 & 0.60 & -0.38 & 0.46 & -0.18 & 0.27 & 0.00 \\ 
  Urban & 0.11 & 0.22 & -0.25 & 0.07 & 0.69 & -0.92 & 1.40 & -0.53 & 1.04 & -0.16 & 0.51 & 0.04 \\ 
  Rural & -0.01 & 0.21 & -0.38 & -0.03 & 0.56 & -1.05 & 1.27 & -0.66 & 0.91 & -0.28 & 0.37 & -0.03 \\ 
  Brest region & -0.07 & 0.25 & -0.57 & -0.07 & 0.43 & -1.09 & 1.06 & -0.78 & 0.67 & -0.47 & 0.33 & -0.05 \\ 
  Vitsebsk region & 0.31 & 0.29 & -0.21 & 0.30 & 0.89 & -0.69 & 1.54 & -0.40 & 1.13 & -0.13 & 0.78 & 0.30 \\ 
  Homel region & -0.10 & 0.25 & -0.60 & -0.09 & 0.41 & -1.13 & 1.03 & -0.81 & 0.64 & -0.51 & 0.31 & -0.07 \\ 
  Hrodna region & -0.05 & 0.25 & -0.54 & -0.05 & 0.45 & -1.07 & 1.09 & -0.75 & 0.69 & -0.45 & 0.35 & -0.04 \\ 
  Minsk city & 0.41 & 0.24 & -0.01 & 0.40 & 0.93 & -0.40 & 1.51 & -0.16 & 1.14 & 0.05 & 0.83 & 0.37 \\ 
  Minsk region & 0.07 & 0.24 & -0.35 & 0.05 & 0.58 & -0.75 & 1.16 & -0.51 & 0.79 & -0.28 & 0.49 & -0.01 \\ 
  Mahiliou region & -0.40 & 0.31 & -1.01 & -0.41 & 0.17 & -1.51 & 0.70 & -1.21 & 0.36 & -0.91 & 0.09 & -0.46 \\ 
   \hline
\end{tabular}
\end{adjustbox}
\end{table}

\begin{table}[!htb]

\caption{Random effects for Dmitriyeu’s model}

\begin{adjustbox}{max width=\textwidth}
\begin{tabular}{lrrrrrrrrrrrrrr}
  \hline
 & mean & sd & 0.025quant & 0.5quant & 0.975quant & 5e-04quant & 0.9995quant & 0.005quant & 0.995quant & 0.05quant & 0.95quant & mode \\ 
  \hline
BYM Brest region 1 & 0.25 & 0.30 & -0.21 & 0.20 & 0.91 & -0.71 & 1.45 & -0.40 & 1.14 & -0.13 & 0.80 & 0.03 \\ 
  BYM Homel region 1 & -0.11 & 0.25 & -0.67 & -0.07 & 0.33 & -1.30 & 0.84 & -0.93 & 0.53 & -0.56 & 0.24 & -0.02 \\ 
  BYM Hrodna region 1 & -0.11 & 0.25 & -0.67 & -0.07 & 0.33 & -1.30 & 0.85 & -0.92 & 0.54 & -0.55 & 0.24 & -0.02 \\ 
  BYM Mahiliou region 1 & 0.15 & 0.26 & -0.29 & 0.11 & 0.74 & -0.82 & 1.28 & -0.49 & 0.97 & -0.20 & 0.63 & 0.02 \\ 
  BYM Minsk region 1 & -0.19 & 0.24 & -0.72 & -0.14 & 0.19 & -1.31 & 0.61 & -0.95 & 0.35 & -0.62 & 0.12 & -0.03 \\ 
  BYM Vitsebsk region 1 & -0.06 & 0.24 & -0.61 & -0.04 & 0.40 & -1.25 & 0.93 & -0.87 & 0.62 & -0.49 & 0.31 & -0.01 \\ 
  BYM Brest region 2 & 0.52 & 1.00 & -1.54 & 0.56 & 2.39 & -3.00 & 3.67 & -2.22 & 2.97 & -1.20 & 2.10 & 0.67 \\ 
  BYM Homel region 2 & -0.14 & 0.96 & -2.00 & -0.16 & 1.80 & -3.34 & 3.20 & -2.61 & 2.44 & -1.69 & 1.47 & -0.21 \\ 
  BYM Hrodna region 2 & -0.21 & 0.96 & -2.07 & -0.22 & 1.74 & -3.41 & 3.15 & -2.68 & 2.39 & -1.76 & 1.41 & -0.27 \\ 
  BYM Mahiliou region 2 & 0.29 & 0.97 & -1.69 & 0.32 & 2.14 & -3.11 & 3.46 & -2.35 & 2.74 & -1.35 & 1.84 & 0.39 \\ 
  BYM Minsk region 2 & -0.29 & 0.73 & -1.69 & -0.31 & 1.19 & -2.65 & 2.24 & -2.13 & 1.67 & -1.47 & 0.94 & -0.36 \\ 
  BYM Vitsebsk region 2 & -0.15 & 0.96 & -2.04 & -0.16 & 1.78 & -3.39 & 3.18 & -2.66 & 2.43 & -1.73 & 1.46 & -0.17 \\ 
  id & -4.87 & 0.97 & -6.48 & -5.01 & -2.62 & -7.93 & -0.89 & -7.07 & -1.81 & -6.22 & -3.05 & -5.26 \\ 
  18-30 & -0.84 & 0.65 & -2.61 & -0.71 & 0.10 & -4.32 & 1.36 & -3.41 & 0.58 & -2.17 & -0.08 & -0.60 \\ 
  31-40 & -0.27 & 0.62 & -1.97 & -0.15 & 0.65 & -3.73 & 2.00 & -2.79 & 1.19 & -1.53 & 0.46 & -0.08 \\ 
  41-50 & -0.08 & 0.61 & -1.74 & 0.02 & 0.89 & -3.50 & 2.24 & -2.56 & 1.42 & -1.30 & 0.68 & 0.06 \\ 
  51-60 & 0.39 & 0.61 & -1.25 & 0.49 & 1.35 & -3.05 & 2.73 & -2.08 & 1.90 & -0.82 & 1.14 & 0.53 \\ 
  More than 60 & -0.28 & 0.62 & -1.95 & -0.19 & 0.71 & -3.67 & 2.01 & -2.75 & 1.22 & -1.51 & 0.51 & -0.14 \\ 
  Elementary school or uneducated & 0.14 & 0.56 & -1.27 & 0.24 & 1.06 & -2.58 & 1.88 & -1.89 & 1.39 & -0.95 & 0.91 & 0.31 \\ 
  Primary or secondary school & -0.66 & 0.56 & -2.15 & -0.50 & 0.07 & -3.39 & 0.87 & -2.72 & 0.35 & -1.85 & -0.04 & -0.36 \\ 
  Professional technical institution & -0.07 & 0.58 & -1.59 & 0.11 & 0.64 & -2.85 & 1.41 & -2.18 & 0.89 & -1.28 & 0.55 & 0.29 \\ 
  Professional college & -0.40 & 0.45 & -1.63 & -0.33 & 0.30 & -2.82 & 1.25 & -2.18 & 0.70 & -1.34 & 0.15 & -0.31 \\ 
  Higher education & 0.27 & 0.44 & -0.94 & 0.35 & 0.93 & -2.12 & 1.88 & -1.49 & 1.34 & -0.65 & 0.79 & 0.37 \\ 
  Female & -0.01 & 0.42 & -1.22 & 0.06 & 0.63 & -2.59 & 1.73 & -1.86 & 1.07 & -0.87 & 0.49 & 0.04 \\ 
  Male & -0.29 & 0.45 & -1.59 & -0.17 & 0.30 & -2.96 & 1.36 & -2.23 & 0.71 & -1.24 & 0.17 & -0.08 \\ 
  Urban & -0.03 & 0.42 & -1.12 & 0.02 & 0.76 & -2.45 & 1.71 & -1.71 & 1.18 & -0.84 & 0.57 & 0.02 \\ 
  Rural & -0.30 & 0.48 & -1.57 & -0.14 & 0.36 & -2.93 & 1.27 & -2.18 & 0.75 & -1.28 & 0.21 & -0.05 \\ 
  Brest region & 0.37 & 0.34 & -0.20 & 0.35 & 1.06 & -0.77 & 1.67 & -0.41 & 1.31 & -0.12 & 0.94 & 0.15 \\ 
  Vitsebsk region & -0.07 & 0.29 & -0.70 & -0.05 & 0.48 & -1.48 & 1.04 & -1.00 & 0.72 & -0.57 & 0.37 & -0.03 \\ 
  Homel region & -0.17 & 0.29 & -0.80 & -0.15 & 0.33 & -1.58 & 0.87 & -1.10 & 0.56 & -0.68 & 0.24 & -0.09 \\ 
  Hrodna region & -0.14 & 0.29 & -0.77 & -0.12 & 0.37 & -1.56 & 0.92 & -1.07 & 0.60 & -0.64 & 0.28 & -0.07 \\ 
  Minsk city & -0.17 & 0.26 & -0.74 & -0.14 & 0.29 & -1.46 & 0.76 & -1.00 & 0.48 & -0.63 & 0.21 & -0.08 \\ 
  Minsk region & -0.21 & 0.29 & -0.86 & -0.17 & 0.28 & -1.64 & 0.73 & -1.16 & 0.46 & -0.73 & 0.20 & -0.09 \\ 
  Mahiliou region & 0.24 & 0.30 & -0.30 & 0.21 & 0.88 & -0.92 & 1.49 & -0.54 & 1.13 & -0.20 & 0.76 & 0.11 \\ 
   \hline
\end{tabular}
\end{adjustbox}
\end{table}

\begin{table}[!htb]

\caption{Random effects for Cherachen’s model}

\begin{adjustbox}{max width=\textwidth}
\begin{tabular}{lrrrrrrrrrrrrrr}
  \hline
 & mean & sd & 0.025quant & 0.5quant & 0.975quant & 5e-04quant & 0.9995quant & 0.005quant & 0.995quant & 0.05quant & 0.95quant & mode \\ 
  \hline
BYM Brest region 1 & 0.21 & 0.38 & -0.49 & 0.19 & 0.99 & -1.28 & 1.72 & -0.80 & 1.28 & -0.36 & 0.85 & 0.12 \\ 
  BYM Homel region 1 & 0.03 & 0.35 & -0.68 & 0.03 & 0.73 & -1.50 & 1.48 & -1.01 & 1.04 & -0.54 & 0.60 & 0.02 \\ 
  BYM Hrodna region 1 & 0.17 & 0.36 & -0.53 & 0.15 & 0.91 & -1.34 & 1.66 & -0.85 & 1.21 & -0.40 & 0.77 & 0.10 \\ 
  BYM Mahiliou region 1 & -0.30 & 0.42 & -1.21 & -0.26 & 0.44 & -2.07 & 1.14 & -1.56 & 0.72 & -1.05 & 0.32 & -0.12 \\ 
  BYM Minsk region 1 & -0.66 & 0.53 & -1.68 & -0.63 & 0.18 & -2.48 & 0.63 & -2.00 & 0.36 & -1.53 & 0.09 & -0.16 \\ 
  BYM Vitsebsk region 1 & 0.38 & 0.43 & -0.38 & 0.34 & 1.25 & -1.11 & 1.96 & -0.66 & 1.54 & -0.27 & 1.12 & 0.14 \\ 
  BYM Brest region 2 & 0.27 & 0.95 & -1.66 & 0.30 & 2.10 & -3.07 & 3.41 & -2.31 & 2.70 & -1.34 & 1.79 & 0.40 \\ 
  BYM Homel region 2 & -0.03 & 0.93 & -1.90 & -0.02 & 1.82 & -3.26 & 3.19 & -2.52 & 2.45 & -1.58 & 1.51 & 0.02 \\ 
  BYM Hrodna region 2 & 0.28 & 0.94 & -1.64 & 0.31 & 2.12 & -3.04 & 3.44 & -2.29 & 2.73 & -1.31 & 1.81 & 0.37 \\ 
  BYM Mahiliou region 2 & -0.37 & 0.97 & -2.22 & -0.41 & 1.60 & -3.53 & 3.02 & -2.82 & 2.25 & -1.92 & 1.27 & -0.50 \\ 
  BYM Minsk region 2 & -0.58 & 0.80 & -2.06 & -0.60 & 1.01 & -2.98 & 2.09 & -2.48 & 1.51 & -1.85 & 0.76 & -0.62 \\ 
  BYM Vitsebsk region 2 & 0.43 & 0.97 & -1.57 & 0.47 & 2.25 & -3.00 & 3.54 & -2.23 & 2.84 & -1.23 & 1.96 & 0.64 \\ 
  id & -4.17 & 1.24 & -6.17 & -4.36 & -1.27 & -7.53 & 0.76 & -6.75 & -0.29 & -5.89 & -1.81 & -4.71 \\ 
  18-30 & -0.13 & 0.97 & -2.98 & 0.17 & 0.97 & -5.17 & 2.12 & -4.08 & 1.37 & -2.30 & 0.83 & 0.34 \\ 
  31-40 & -0.29 & 0.98 & -3.16 & 0.01 & 0.82 & -5.35 & 1.96 & -4.26 & 1.22 & -2.48 & 0.67 & 0.19 \\ 
  41-50 & -0.72 & 1.00 & -3.62 & -0.41 & 0.40 & -5.83 & 1.53 & -4.73 & 0.79 & -2.94 & 0.27 & -0.20 \\ 
  51-60 & -1.29 & 1.05 & -4.32 & -0.98 & -0.09 & -6.55 & 0.98 & -5.44 & 0.29 & -3.63 & -0.23 & -0.73 \\ 
  More than 60 & -0.47 & 0.99 & -3.33 & -0.17 & 0.67 & -5.54 & 1.76 & -4.43 & 1.03 & -2.66 & 0.52 & 0.03 \\ 
  Elementary school or uneducated & -0.19 & 0.60 & -1.58 & -0.14 & 0.90 & -3.08 & 1.86 & -2.26 & 1.31 & -1.26 & 0.71 & -0.09 \\ 
  Primary or secondary school & 0.09 & 0.52 & -1.18 & 0.15 & 0.95 & -2.72 & 1.88 & -1.88 & 1.33 & -0.86 & 0.79 & 0.21 \\ 
  Professional technical institution & -1.21 & 0.68 & -2.88 & -1.09 & -0.19 & -4.38 & 0.60 & -3.57 & 0.08 & -2.53 & -0.31 & -0.88 \\ 
  Professional college & 0.33 & 0.49 & -0.88 & 0.38 & 1.17 & -2.41 & 2.13 & -1.58 & 1.57 & -0.56 & 1.01 & 0.40 \\ 
  Higher education & 0.13 & 0.48 & -1.07 & 0.17 & 0.96 & -2.56 & 1.91 & -1.75 & 1.36 & -0.75 & 0.80 & 0.17 \\ 
  Female & -0.57 & 0.57 & -2.10 & -0.45 & 0.26 & -3.80 & 1.27 & -2.87 & 0.64 & -1.72 & 0.11 & -0.34 \\ 
  Male & 0.06 & 0.56 & -1.43 & 0.16 & 0.90 & -3.13 & 1.93 & -2.20 & 1.30 & -1.06 & 0.75 & 0.23 \\ 
  Urban & -0.05 & 0.34 & -1.04 & 0.00 & 0.48 & -2.41 & 1.40 & -1.67 & 0.87 & -0.73 & 0.33 & 0.00 \\ 
  Rural & -0.14 & 0.37 & -1.21 & -0.05 & 0.33 & -2.58 & 1.23 & -1.84 & 0.70 & -0.90 & 0.21 & -0.01 \\ 
  Brest region & 0.26 & 0.44 & -0.54 & 0.23 & 1.18 & -1.46 & 1.94 & -0.90 & 1.51 & -0.39 & 1.02 & 0.11 \\ 
  Vitsebsk region & 0.48 & 0.50 & -0.38 & 0.43 & 1.51 & -1.23 & 2.25 & -0.70 & 1.84 & -0.25 & 1.34 & 0.16 \\ 
  Homel region & 0.09 & 0.41 & -0.72 & 0.08 & 0.92 & -1.66 & 1.70 & -1.09 & 1.25 & -0.56 & 0.77 & 0.05 \\ 
  Hrodna region & 0.17 & 0.42 & -0.63 & 0.15 & 1.05 & -1.57 & 1.81 & -1.00 & 1.38 & -0.47 & 0.90 & 0.07 \\ 
  Minsk city & -0.80 & 0.57 & -1.99 & -0.74 & 0.09 & -2.87 & 0.55 & -2.36 & 0.27 & -1.80 & 0.00 & -0.27 \\ 
  Minsk region & -0.33 & 0.50 & -1.44 & -0.25 & 0.47 & -2.32 & 1.06 & -1.81 & 0.71 & -1.25 & 0.36 & -0.01 \\ 
  Mahiliou region & -0.33 & 0.47 & -1.36 & -0.28 & 0.48 & -2.32 & 1.23 & -1.76 & 0.78 & -1.17 & 0.34 & -0.13 \\ 
   \hline
\end{tabular}
\end{adjustbox}
\end{table}

\begin{table}[!htb]

\caption{Random effects for Kanapatskaya’s model}

\begin{adjustbox}{max width=\textwidth}
\begin{tabular}{lrrrrrrrrrrrrrr}
  \hline
 & mean & sd & 0.025quant & 0.5quant & 0.975quant & 5e-04quant & 0.9995quant & 0.005quant & 0.995quant & 0.05quant & 0.95quant & mode \\ 
  \hline
BYM Brest region 1 & 0.06 & 0.41 & -0.77 & 0.04 & 0.94 & -1.52 & 1.66 & -1.11 & 1.28 & -0.60 & 0.77 & 0.01 \\ 
  BYM Homel region 1 & -0.43 & 0.51 & -1.63 & -0.33 & 0.34 & -2.52 & 0.99 & -2.05 & 0.63 & -1.41 & 0.22 & -0.04 \\ 
  BYM Hrodna region 1 & 0.62 & 0.55 & -0.19 & 0.56 & 1.80 & -0.82 & 2.50 & -0.45 & 2.13 & -0.11 & 1.62 & 0.06 \\ 
  BYM Mahiliou region 1 & -0.21 & 0.44 & -1.24 & -0.14 & 0.58 & -2.10 & 1.27 & -1.64 & 0.89 & -1.03 & 0.42 & -0.02 \\ 
  BYM Minsk region 1 & -0.06 & 0.34 & -0.79 & -0.04 & 0.62 & -1.41 & 1.24 & -1.08 & 0.90 & -0.65 & 0.49 & -0.01 \\ 
  BYM Vitsebsk region 1 & -0.08 & 0.42 & -1.03 & -0.05 & 0.75 & -1.87 & 1.47 & -1.42 & 1.08 & -0.84 & 0.59 & -0.00 \\ 
  BYM Brest region 2 & 0.15 & 0.96 & -1.76 & 0.16 & 2.03 & -3.15 & 3.35 & -2.39 & 2.63 & -1.44 & 1.73 & 0.17 \\ 
  BYM Homel region 2 & -0.62 & 1.01 & -2.56 & -0.64 & 1.43 & -3.87 & 2.91 & -3.16 & 2.12 & -2.26 & 1.09 & -0.68 \\ 
  BYM Hrodna region 2 & 0.92 & 1.05 & -1.24 & 0.96 & 2.87 & -2.79 & 4.15 & -1.97 & 3.46 & -0.87 & 2.58 & 1.05 \\ 
  BYM Mahiliou region 2 & -0.38 & 0.98 & -2.29 & -0.39 & 1.58 & -3.60 & 3.00 & -2.90 & 2.24 & -1.99 & 1.25 & -0.41 \\ 
  BYM Minsk region 2 & -0.04 & 0.72 & -1.45 & -0.04 & 1.36 & -2.43 & 2.35 & -1.90 & 1.82 & -1.22 & 1.13 & -0.04 \\ 
  BYM Vitsebsk region 2 & -0.01 & 0.97 & -1.92 & -0.01 & 1.90 & -3.27 & 3.25 & -2.54 & 2.52 & -1.61 & 1.59 & -0.01 \\ 
  id & -5.36 & 2.06 & -8.35 & -5.82 & -0.38 & -10.12 & 1.98 & -9.13 & 0.78 & -7.97 & -1.09 & -6.28 \\ 
  18-30 & -1.98 & 1.63 & -6.64 & -1.61 & 0.06 & -9.69 & 1.22 & -8.16 & 0.55 & -5.65 & -0.17 & -1.27 \\ 
  31-40 & -1.85 & 1.62 & -6.47 & -1.48 & 0.18 & -9.51 & 1.38 & -7.98 & 0.69 & -5.49 & -0.05 & -1.10 \\ 
  41-50 & -1.40 & 1.62 & -6.00 & -1.02 & 0.65 & -9.04 & 1.87 & -7.52 & 1.17 & -5.02 & 0.41 & -0.63 \\ 
  51-60 & -1.24 & 1.60 & -5.83 & -0.86 & 0.82 & -8.88 & 2.08 & -7.35 & 1.35 & -4.85 & 0.55 & -0.51 \\ 
  More than 60 & 0.31 & 1.57 & -4.20 & 0.69 & 2.29 & -7.17 & 3.51 & -5.68 & 2.81 & -3.24 & 2.05 & 1.06 \\ 
  Elementary school or uneducated & -1.07 & 1.72 & -6.57 & -0.71 & 0.93 & -10.90 & 2.04 & -8.82 & 1.40 & -4.86 & 0.69 & -0.45 \\ 
  Primary or secondary school & -1.47 & 1.62 & -6.74 & -1.08 & 0.32 & -10.69 & 1.39 & -8.82 & 0.75 & -5.12 & 0.11 & -0.67 \\ 
  Professional technical institution & -0.76 & 1.53 & -5.70 & -0.44 & 1.02 & -9.52 & 2.28 & -7.72 & 1.51 & -4.11 & 0.81 & -0.16 \\ 
  Professional college & -0.89 & 1.41 & -5.54 & -0.60 & 0.82 & -9.21 & 2.13 & -7.48 & 1.35 & -4.00 & 0.56 & -0.50 \\ 
  Higher education & 0.37 & 1.37 & -4.18 & 0.65 & 2.03 & -7.75 & 3.40 & -6.06 & 2.57 & -2.72 & 1.78 & 0.73 \\ 
  Female & 0.10 & 0.49 & -1.07 & 0.12 & 1.02 & -2.22 & 2.06 & -1.62 & 1.47 & -0.78 & 0.84 & 0.01 \\ 
  Male & -0.48 & 0.53 & -1.81 & -0.39 & 0.36 & -2.96 & 1.32 & -2.36 & 0.75 & -1.51 & 0.19 & -0.24 \\ 
  Urban & -0.05 & 0.39 & -1.06 & -0.01 & 0.56 & -2.87 & 1.71 & -1.98 & 0.90 & -0.61 & 0.45 & -0.00 \\ 
  Rural & -0.14 & 0.42 & -1.36 & -0.07 & 0.44 & -3.23 & 1.38 & -2.30 & 0.70 & -0.80 & 0.34 & -0.01 \\ 
  Brest region & 0.01 & 0.47 & -0.93 & 0.02 & 0.95 & -1.93 & 1.79 & -1.32 & 1.30 & -0.76 & 0.78 & 0.02 \\ 
  Vitsebsk region & -0.20 & 0.50 & -1.26 & -0.17 & 0.74 & -2.57 & 1.50 & -1.78 & 1.07 & -1.04 & 0.59 & -0.14 \\ 
  Homel region & -0.54 & 0.58 & -1.87 & -0.48 & 0.46 & -3.62 & 1.11 & -2.58 & 0.75 & -1.58 & 0.31 & -0.40 \\ 
  Hrodna region & 0.75 & 0.58 & -0.31 & 0.72 & 1.91 & -0.97 & 2.84 & -0.62 & 2.31 & -0.16 & 1.72 & 0.61 \\ 
  Minsk city & -0.07 & 0.41 & -0.90 & -0.07 & 0.73 & -1.86 & 1.44 & -1.26 & 1.03 & -0.75 & 0.59 & -0.06 \\ 
  Minsk region & -0.07 & 0.45 & -1.00 & -0.06 & 0.77 & -2.07 & 1.50 & -1.42 & 1.08 & -0.82 & 0.63 & -0.04 \\ 
  Mahiliou region & -0.21 & 0.51 & -1.31 & -0.19 & 0.72 & -2.67 & 1.46 & -1.85 & 1.04 & -1.08 & 0.57 & -0.15 \\
   \hline
\end{tabular}
\end{adjustbox}
\end{table}

\begin{table}[!htb]

\caption{Random effects for {``}Against all{''} model}

\begin{adjustbox}{max width=\textwidth}
\begin{tabular}{lrrrrrrrrrrrrrr}
  \hline
 & mean & sd & 0.025quant & 0.5quant & 0.975quant & 5e-04quant & 0.9995quant & 0.005quant & 0.995quant & 0.05quant & 0.95quant & mode \\ 
  \hline
BYM Brest region 1 & -0.34 & 0.37 & -1.14 & -0.28 & 0.19 & -1.76 & 0.70 & -1.42 & 0.39 & -1.01 & 0.12 & -0.02 \\ 
  BYM Homel region 1 & 0.23 & 0.30 & -0.26 & 0.17 & 0.87 & -0.84 & 1.41 & -0.50 & 1.10 & -0.17 & 0.76 & 0.02 \\ 
  BYM Hrodna region 1 & -0.01 & 0.25 & -0.54 & -0.01 & 0.51 & -1.14 & 1.08 & -0.79 & 0.75 & -0.42 & 0.40 & -0.00 \\ 
  BYM Mahiliou region 1 & 0.25 & 0.31 & -0.24 & 0.20 & 0.92 & -0.81 & 1.46 & -0.47 & 1.15 & -0.15 & 0.81 & 0.02 \\ 
  BYM Minsk region 1 & 0.01 & 0.21 & -0.44 & 0.01 & 0.45 & -0.98 & 0.94 & -0.67 & 0.66 & -0.34 & 0.36 & 0.00 \\ 
  BYM Vitsebsk region 1 & -0.24 & 0.32 & -0.95 & -0.17 & 0.26 & -1.56 & 0.81 & -1.21 & 0.49 & -0.82 & 0.17 & -0.02 \\ 
  BYM Brest region 2 & -0.59 & 1.03 & -2.50 & -0.63 & 1.53 & -3.72 & 3.00 & -3.06 & 2.21 & -2.22 & 1.17 & -0.74 \\ 
  BYM Homel region 2 & 0.48 & 0.99 & -1.56 & 0.52 & 2.33 & -3.03 & 3.59 & -2.24 & 2.90 & -1.22 & 2.05 & 0.63 \\ 
  BYM Hrodna region 2 & -0.12 & 0.94 & -1.98 & -0.12 & 1.75 & -3.34 & 3.15 & -2.61 & 2.39 & -1.67 & 1.43 & -0.12 \\ 
  BYM Mahiliou region 2 & 0.56 & 1.01 & -1.52 & 0.60 & 2.42 & -2.99 & 3.67 & -2.21 & 3.00 & -1.17 & 2.14 & 0.72 \\ 
  BYM Minsk region 2 & 0.04 & 0.70 & -1.37 & 0.04 & 1.42 & -2.38 & 2.42 & -1.83 & 1.88 & -1.13 & 1.19 & 0.05 \\ 
  BYM Vitsebsk region 2 & -0.36 & 0.98 & -2.21 & -0.40 & 1.65 & -3.48 & 3.09 & -2.79 & 2.31 & -1.92 & 1.31 & -0.49 \\ 
  id & -1.26 & 1.06 & -3.39 & -1.17 & 0.54 & -4.86 & 1.84 & -4.05 & 1.03 & -3.07 & 0.31 & -0.56 \\ 
  18-30 & -1.12 & 0.81 & -3.21 & -0.92 & 0.03 & -4.90 & 1.21 & -4.03 & 0.49 & -2.78 & -0.15 & -0.71 \\ 
  31-40 & -0.77 & 0.81 & -2.86 & -0.57 & 0.38 & -4.54 & 1.56 & -3.67 & 0.84 & -2.43 & 0.20 & -0.37 \\ 
  41-50 & -0.37 & 0.81 & -2.47 & -0.17 & 0.78 & -4.14 & 1.95 & -3.27 & 1.23 & -2.04 & 0.60 & 0.02 \\ 
  51-60 & -0.15 & 0.81 & -2.24 & 0.05 & 1.00 & -3.92 & 2.18 & -3.05 & 1.46 & -1.81 & 0.82 & 0.25 \\ 
  More than 60 & -0.04 & 0.81 & -2.13 & 0.16 & 1.11 & -3.81 & 2.28 & -2.94 & 1.56 & -1.70 & 0.92 & 0.35 \\ 
  Elementary school or uneducated & 0.41 & 0.91 & -1.88 & 0.69 & 1.58 & -3.51 & 2.60 & -2.61 & 1.94 & -1.48 & 1.43 & 0.97 \\ 
  Primary or secondary school & -0.99 & 0.91 & -3.29 & -0.70 & 0.16 & -4.93 & 1.19 & -4.03 & 0.53 & -2.89 & 0.02 & -0.41 \\ 
  Professional technical institution & -0.84 & 0.92 & -3.15 & -0.55 & 0.30 & -4.79 & 1.32 & -3.89 & 0.67 & -2.75 & 0.17 & -0.26 \\ 
  Professional college & -0.96 & 0.91 & -3.26 & -0.67 & 0.17 & -4.90 & 1.20 & -4.00 & 0.54 & -2.87 & 0.04 & -0.38 \\ 
  Higher education & -0.97 & 0.91 & -3.27 & -0.68 & 0.16 & -4.91 & 1.19 & -4.01 & 0.53 & -2.87 & 0.03 & -0.38 \\ 
  Female & -0.41 & 0.50 & -1.81 & -0.30 & 0.29 & -3.14 & 1.26 & -2.47 & 0.66 & -1.43 & 0.15 & -0.22 \\ 
  Male & -0.01 & 0.49 & -1.40 & 0.10 & 0.69 & -2.73 & 1.67 & -2.06 & 1.06 & -1.03 & 0.55 & 0.16 \\ 
  Urban & -0.03 & 0.53 & -1.54 & 0.10 & 0.70 & -2.88 & 1.76 & -2.20 & 1.10 & -1.17 & 0.56 & 0.16 \\ 
  Rural & -0.45 & 0.54 & -1.99 & -0.31 & 0.27 & -3.32 & 1.32 & -2.64 & 0.66 & -1.61 & 0.14 & -0.22 \\ 
  Brest region & -0.57 & 0.42 & -1.40 & -0.56 & 0.18 & -2.24 & 0.70 & -1.74 & 0.39 & -1.26 & 0.08 & -0.38 \\ 
  Vitsebsk region & -0.41 & 0.37 & -1.17 & -0.40 & 0.26 & -1.99 & 0.81 & -1.50 & 0.48 & -1.03 & 0.16 & -0.36 \\ 
  Homel region & 0.31 & 0.33 & -0.35 & 0.31 & 0.95 & -1.01 & 1.53 & -0.61 & 1.19 & -0.24 & 0.84 & 0.31 \\ 
  Hrodna region & 0.04 & 0.30 & -0.58 & 0.04 & 0.63 & -1.32 & 1.25 & -0.89 & 0.89 & -0.45 & 0.52 & 0.05 \\ 
  Minsk city & -0.10 & 0.25 & -0.63 & -0.10 & 0.38 & -1.34 & 0.88 & -0.91 & 0.59 & -0.52 & 0.29 & -0.09 \\ 
  Minsk region & 0.10 & 0.25 & -0.43 & 0.10 & 0.59 & -1.14 & 1.09 & -0.71 & 0.80 & -0.32 & 0.50 & 0.11 \\ 
  Mahiliou region & 0.33 & 0.34 & -0.33 & 0.33 & 1.00 & -0.99 & 1.57 & -0.59 & 1.23 & -0.22 & 0.89 & 0.32 \\ 
   \hline
\end{tabular}
\end{adjustbox}
\end{table}

\begin{table}[!htb]

\caption{Random effects for Early voting turnout model}

\begin{adjustbox}{max width=\textwidth}
\begin{tabular}{lrrrrrrrrrrrrrr}
  \hline
 & mean & sd & 0.025quant & 0.5quant & 0.975quant & 5e-04quant & 0.9995quant & 0.005quant & 0.995quant & 0.05quant & 0.95quant & mode \\ 
  \hline
BYM Brest region 1 & -0.09 & 0.24 & -0.62 & -0.07 & 0.35 & -1.21 & 0.89 & -0.87 & 0.57 & -0.51 & 0.26 & -0.03 \\ 
  BYM Homel region 1 & 0.09 & 0.24 & -0.37 & 0.07 & 0.60 & -0.94 & 1.15 & -0.60 & 0.83 & -0.27 & 0.50 & 0.04 \\ 
  BYM Hrodna region 1 & 0.05 & 0.23 & -0.42 & 0.04 & 0.54 & -0.99 & 1.09 & -0.65 & 0.77 & -0.32 & 0.44 & 0.02 \\ 
  BYM Mahiliou region 1 & 0.23 & 0.29 & -0.23 & 0.18 & 0.85 & -0.75 & 1.39 & -0.42 & 1.08 & -0.16 & 0.75 & 0.06 \\ 
  BYM Minsk region 1 & -0.32 & 0.30 & -0.94 & -0.27 & 0.13 & -1.47 & 0.47 & -1.17 & 0.25 & -0.84 & 0.08 & -0.08 \\ 
  BYM Vitsebsk region 1 & -0.04 & 0.23 & -0.54 & -0.03 & 0.42 & -1.12 & 0.96 & -0.79 & 0.64 & -0.43 & 0.33 & -0.01 \\ 
  BYM Brest region 2 & -0.19 & 0.96 & -2.06 & -0.21 & 1.75 & -3.42 & 3.16 & -2.68 & 2.40 & -1.74 & 1.42 & -0.26 \\ 
  BYM Homel region 2 & 0.19 & 0.96 & -1.76 & 0.21 & 2.05 & -3.16 & 3.41 & -2.41 & 2.67 & -1.42 & 1.73 & 0.26 \\ 
  BYM Hrodna region 2 & 0.02 & 0.95 & -1.90 & 0.03 & 1.89 & -3.29 & 3.27 & -2.54 & 2.53 & -1.57 & 1.57 & 0.07 \\ 
  BYM Mahiliou region 2 & 0.47 & 1.00 & -1.58 & 0.51 & 2.33 & -3.02 & 3.64 & -2.25 & 2.92 & -1.24 & 2.04 & 0.63 \\ 
  BYM Minsk region 2 & -0.45 & 0.78 & -1.90 & -0.47 & 1.12 & -2.81 & 2.18 & -2.32 & 1.61 & -1.69 & 0.86 & -0.53 \\ 
  BYM Vitsebsk region 2 & -0.03 & 0.95 & -1.92 & -0.04 & 1.88 & -3.29 & 3.27 & -2.55 & 2.52 & -1.59 & 1.55 & -0.06 \\ 
  id & -1.12 & 0.85 & -2.73 & -1.09 & 0.37 & -3.92 & 1.48 & -3.25 & 0.77 & -2.50 & 0.18 & -0.58 \\ 
  18-30 & -1.23 & 0.74 & -2.98 & -1.07 & -0.09 & -4.37 & 0.95 & -3.59 & 0.34 & -2.66 & -0.27 & -0.78 \\ 
  31-40 & -1.04 & 0.74 & -2.79 & -0.88 & 0.10 & -4.17 & 1.15 & -3.39 & 0.54 & -2.47 & -0.08 & -0.59 \\ 
  41-50 & -0.56 & 0.74 & -2.31 & -0.41 & 0.58 & -3.69 & 1.63 & -2.91 & 1.02 & -1.99 & 0.40 & -0.11 \\ 
  51-60 & 0.07 & 0.74 & -1.68 & 0.22 & 1.20 & -3.06 & 2.25 & -2.29 & 1.64 & -1.37 & 1.02 & 0.51 \\ 
  More than 60 & -0.05 & 0.74 & -1.79 & 0.11 & 1.09 & -3.17 & 2.14 & -2.40 & 1.53 & -1.48 & 0.91 & 0.40 \\ 
  Elementary school or uneducated & -0.05 & 0.20 & -0.48 & -0.04 & 0.33 & -1.04 & 0.77 & -0.70 & 0.52 & -0.38 & 0.26 & -0.03 \\ 
  Primary or secondary school & -0.06 & 0.17 & -0.44 & -0.06 & 0.29 & -1.00 & 0.74 & -0.66 & 0.49 & -0.34 & 0.20 & -0.06 \\ 
  Professional technical institution & 0.09 & 0.17 & -0.28 & 0.09 & 0.45 & -0.84 & 0.91 & -0.50 & 0.66 & -0.18 & 0.36 & 0.08 \\ 
  Professional college & 0.31 & 0.17 & -0.06 & 0.31 & 0.65 & -0.62 & 1.10 & -0.28 & 0.86 & 0.04 & 0.56 & 0.31 \\ 
  Higher education & -0.52 & 0.17 & -0.89 & -0.51 & -0.17 & -1.45 & 0.28 & -1.11 & 0.03 & -0.79 & -0.26 & -0.51 \\ 
  Female & -0.02 & 0.23 & -0.60 & 0.02 & 0.41 & -1.20 & 0.93 & -0.87 & 0.63 & -0.46 & 0.31 & 0.04 \\ 
  Male & -0.12 & 0.23 & -0.71 & -0.07 & 0.30 & -1.32 & 0.82 & -0.98 & 0.52 & -0.57 & 0.20 & -0.04 \\ 
  Urban & -0.62 & 0.53 & -1.91 & -0.49 & 0.18 & -2.87 & 1.15 & -2.35 & 0.56 & -1.68 & 0.04 & -0.37 \\ 
  Rural & -0.00 & 0.53 & -1.29 & 0.12 & 0.80 & -2.24 & 1.77 & -1.73 & 1.19 & -1.06 & 0.66 & 0.24 \\ 
  Brest region & -0.13 & 0.28 & -0.70 & -0.12 & 0.42 & -1.38 & 0.97 & -0.97 & 0.65 & -0.59 & 0.32 & -0.11 \\ 
  Vitsebsk region & -0.06 & 0.28 & -0.63 & -0.06 & 0.48 & -1.30 & 1.02 & -0.90 & 0.70 & -0.52 & 0.38 & -0.05 \\ 
  Homel region & 0.16 & 0.29 & -0.43 & 0.16 & 0.71 & -1.07 & 1.23 & -0.70 & 0.92 & -0.31 & 0.61 & 0.16 \\ 
  Hrodna region & 0.13 & 0.28 & -0.45 & 0.13 & 0.67 & -1.10 & 1.20 & -0.71 & 0.89 & -0.33 & 0.58 & 0.13 \\ 
  Minsk city & -0.65 & 0.31 & -1.27 & -0.65 & -0.10 & -1.94 & 0.31 & -1.53 & 0.07 & -1.17 & -0.17 & -0.50 \\ 
  Minsk region & -0.12 & 0.30 & -0.73 & -0.11 & 0.42 & -1.39 & 0.85 & -0.98 & 0.60 & -0.62 & 0.34 & 0.01 \\ 
  Mahiliou region & 0.40 & 0.34 & -0.26 & 0.41 & 1.02 & -0.88 & 1.54 & -0.51 & 1.22 & -0.15 & 0.93 & 0.47 \\ 
   \hline
\end{tabular}
\end{adjustbox}
\end{table}

\clearpage

\bibliographystyle{plainnat}
\bibliography{references}  

\end{document}